\documentclass[footinbib,a4paper,aps,prx,superscriptaddress,reprint,twocolumn,preprintnumbers,amsmath,amssymb,nobalancelastpage,10pt]{revtex4-1}
\usepackage{amsmath}
\usepackage{verbatim}
\usepackage{graphicx}
\usepackage{color}
\usepackage{tikz}
\usepackage{siunitx}
\usepackage{physics}
\usepackage{subfigure}
\usepackage{float}

\usepackage{mathptmx}
\usepackage{amssymb}
\usepackage{amsmath}
\usepackage{amsfonts}
\usepackage{array}

\usepackage{bm}
\usepackage{xcolor}
\usepackage{braket}

\definecolor{mygrey}{gray}{0.35}
\definecolor{myblue}{rgb}{0.2,0.2,0.8}
\definecolor{myzard}{cmyk}{0,0,0.05,0}
\definecolor{mywhite}{rgb}{1,1,1}
\definecolor{myred}{rgb}{1,0.,0.3}

\usepackage[colorlinks=true,citecolor=myblue,linkcolor=myred]{hyperref}

 \def\ee{\mathord{\rm e}}
 
 \def\ii{\mathord{\rm i}}

\def\half{\textstyle\frac{1}{2}}

\renewcommand{\ii}{{\rm i}}
\renewcommand{\ee}{{\rm e}}
\def\beq{\begin{equation}}
\def\eeq{\end{equation}}

\def\barray{\begin{eqnarray}}
\def\earray{\end{eqnarray}}

\begin{document}


\title{The rotor Jackiw-Rebbi model: a cold-atom approach to \\chiral symmetry restoration and quark confinement}



\author{D. Gonz\'{a}lez-Cuadra}\email{daniel.gonzalez@icfo.eu}
\affiliation{ICFO - Institut de Ci\`encies Fot\`oniques, The Barcelona Institute of Science and Technology, Av. Carl Friedrich Gauss 3, 08860 Castelldefels (Barcelona), Spain}

\author{A. Dauphin}
\affiliation{ICFO - Institut de Ci\`encies Fot\`oniques, The Barcelona Institute of Science and Technology, Av. Carl Friedrich Gauss 3, 08860 Castelldefels (Barcelona), Spain}
%

\author{M. Aidelsburger}
\affiliation{Fakult\"{a}t f\"{u}r Physik, Ludwig-Maximilians-Universit\"{a}t, Schellingstr. 4, D-80799 M\"{u}nchen, Germany}
\affiliation{Munich Center for Quantum Science and Technology (MCQST), Schellingstr. 4, D-80799 M\"{u}nchen, Germany}

\author{M. Lewenstein}
\affiliation{ICFO - Institut de Ci\`encies Fot\`oniques, The Barcelona Institute of Science and Technology, Av. Carl Friedrich Gauss 3, 08860 Castelldefels (Barcelona), Spain} 
\affiliation{ICREA, Lluis Companys 23, 08010 Barcelona, Spain}

\author{A. Bermudez}
\affiliation{Departamento de F\'{i}sica Te\'{o}rica, Universidad Complutense, 28040 Madrid, Spain}

\begin{abstract}
Understanding the nature of confinement, as well as its relation with the spontaneous breaking of chiral symmetry, remains one of the long-standing questions in high-energy physics. The difficulty of this task stems from the limitations of current analytical and numerical techniques to address non-perturbative phenomena in non-Abelian gauge theories. In this work, we show how similar phenomena emerge in simpler models, and how these can be further investigated using state-of-the-art cold-atom quantum simulators. More specifically, we introduce the rotor Jackiw-Rebbi  model, a (1+1)-dimensional quantum field theory where interactions between Dirac fermions are mediated by quantum rotors. Starting from a mixture of ultracold atoms in an optical lattice, we show how this quantum field  theory emerges in the long-wavelength limit. For a wide and experimentally-relevant parameter regime, the Dirac fermions acquire a dynamical mass via the spontaneous breakdown of chiral symmetry. Moreover, we study the effect of both quantum and  thermal fluctuations, which lead to the phenomenon of chiral symmetry restoration. Finally, we uncover a confinement-deconfinement quantum phase transition, where meson-like fermions fractionalise into quark-like quasi-particles bound to topological solitons of the rotor field. The proliferation of these solitons at  finite chemical potentials again serves to  restore  the chiral symmetry, yielding a clear analogy with the quark-gluon plasma in quantum chromodynamics, where  this symmetry coexists with the deconfined fractional charges. Our results show how the interplay between these phenomena could be analyse in realistic atomic experiments.
\end{abstract}


\maketitle

{\bf Introduction.--} Quantum field theory (QFT) provides a unifying  framework to understand    many-body  systems  at widely different   scales. At the highest energies reached so far, the standard model of particle physics explains  all  observed phenomena by means of  a relativistic QFT of fermions  coupled to  scalar and gauge bosons~\cite{Peskin:1995ev}. At  lower energies, non-relativistic QFTs of interacting fermionic and bosonic particles form the core of the standard model of condensed-matter physics~\cite{Xiao:803748}, which explains a wide variety of phases via Landau's seminal contributions of spontaneous symmetry-breaking (SSB)~\cite{landau} and quasiparticle renormalisation~\cite{landau_fl}.  In the vicinity of certain SSB phase transitions, the quasiparticles  governing the long-wavelength phenomena  can be completely different from the original non-relativistic  constituents~\cite{Anderson393}, and even be described by relativistic models analogous to those of particle physics. In fact, it is the careful  understanding of this quasiparticle renormalisation, which yields the very definition of a relativistic QFT~\cite{WILSON197475,RevModPhys.47.773,hollowood_2013}, and  sets the basis for the non-perturbative approach to lattice gauge theories~\cite{PhysRevD.10.2445,RevModPhys.51.659}. 

More recently, the range of applications of relativistic theories  has been extended  to much lower energies, as they also appear   in experiments dealing  with the coldest type of quantum matter controlled in a laboratory:  ultracold neutral atoms~\cite{Tarruell2012,Duca288,flaschner_experimental_2016,Schweizer2019,Mil1128,2003.08945} and trapped atomic ions~\cite{Gerritsma2010,PhysRevLett.106.060503,Martinez2016}. In contrast to condensed matter and high-energy  physics, cold-atom/trapped-ion experiments deal with quantum many-body systems that can be accurately initialised, controlled, and measured, even  at the single-particle level, turning Feynman's idea of a quantum simulator (QS)~\cite{Feynman_1982,doi:10.1080/00018730701223200} into a practical reality~\cite{Bloch2012,Blatt2012}. One of the unique properties of  cold-atom QSs is the possibility of controlling the effective dimensionality of the model in an experiment. This is particularly important in a QFT context, where  interactions  tend to be  more relevant as the dimensionality   is lowered~\cite{WILSON197475}, bringing in an increased richness in the form of non-perturbative effects.  Moreover, the reduced dimensionality sometimes  captures  the essence of these  non-perturbative effects, characteristic of higher-dimensional non-Abelian gauge theories,  in a  much simpler   arena. Some paradigmatic examples of this trend  are the  axial anomaly of the Schwinger model~\cite{PhysRev.128.2425,MANTON1985220}, the strong-weak  duality of the Thirring model~\cite{THIRRING195891,PhysRevD.11.2088}, asymptotic freedom and dynamical mass generation  in the Gross-Neveu model~\cite{PhysRevD.10.3235}, and the  fractionalization of charge by solitons in the Jackiw-Rebbi model~\cite{PhysRevD.13.3398}. Therefore, the first QSs  of QFTs are  targeting models in low dimensions~\cite{1911.00003,Zohar_2015,doi:10.1080/00107514.2016.1151199,doi:10.1002/andp.201300104,Carmen_Ba_uls_2020,2006.01258}.

We note that the flexibility of  these platforms offers an exceptional alternative: rather than using the QS to target a  QFT already studied in the realm of high-energy physics or condensed matter, one can design the QS to realise new QFTs which, although inspired by  phenomena first considered in these disciplines, lead to partially-uncharted territory  and give an alternative take on long-standing open problems in these fields.   For instance, despite the huge success of the standard model of particle physics,   the absence of fractionally-charged quarks from the spectrum  still presents unsolved questions in quantum chromodynamics (QCD), such as understanding  the specific microscopic  mechanism for the  {\it confinement} of quarks into  mesons/hadrons with integer electric charges~\cite{greensite_2020}. One related problem that remains open is the nature of the confinement-deconfinement transitions at finite temperatures/densities~\cite{kogut_stephanov_2003} that lead to phases  with   isolated quarks and gluons\textemdash as well as its relation with the restoration of chiral symmetry~\cite{Bazanov_2019}.  Unfortunately, gauge theories  in  (1+1) dimensions are all confining  regardless of the Abelian or non-Abelian nature of the gauge group. In higher dimensions, deconfinement is usually driven by four-body plaquette interactions, which are challenging to implement in cold-atom experiments~\cite{dai_four-body_2017,1911.00003}. Therefore, rather than looking for QSs of gauge theories, one can  exploit the aforementioned flexibility of QSs to design new QFTs where characteristic phenomena of higher-dimensional non-Abelian gauge theories emerge in strongly-correlated phases. In this article, we follow this route, and identify a simple lattice model in (1+1) dimensions that regularises a relativistic QFTs where the interplay between dynamical mass generation and charge fractionalization leads to confinement-deconfinement transitions of quark-like quasi-particles, the  mechanism of  which can be neatly understood at the microscopic level. Although various mechanisms of confinement in QSs have been discussed in the literature~\cite{PhysRevLett.107.275301, PhysRevLett.109.125302, TAGLIACOZZO2013160,Tagliacozzo2013, PhysRevLett.117.240504, RICO2018466, PhysRevA.100.013629,PhysRevLett.122.150601,1912.11117, PhysRevLett.124.120503,PhysRevResearch.2.013288,PhysRevX.10.021041, Magnifico2020realtimedynamics,2002.06013, PhysRevLett.124.180602, 2007.07258}, to the best of our knowledge, our work identifies for the first time a confinement-deconfinement transition between fractionally-charged quasi-particles. Moreover, the feasibility of our QS proposal with state-of-the-art cold-atom experiments indicates the possibility of experimentally observing this transition, together with other QCD-like phenomena, such as chiral symmetry restoration at finite densities.



\vspace{0.5ex}
{\bf The lattice model.--} To motivate the nature of  our model, we note that  imposing non-linear constraints in QFTs  is also a source of non-perturbative phenomena  that  resemble the phenomenology of non-Abelian gauge theories. The  $O(N)$ non-linear sigma model, where  a vector field is constrained to take values on the  $(N-1)$-sphere,   also displays asymptotic freedom and dynamical mass generation~\cite{POLYAKOV197579,RevModPhys.51.659}, although the latter cannot be accompanied by SSB in (1+1) dimensions~\cite{Coleman1973}. Remarkably,  the $O(3)$ non-linear sigma model with an additional topological term   arises as the long-wavelength description of Heisenberg antiferromagnetic  spin chains~\cite{HALDANE1983464,AFFLECK1985397}.  For leading antiferromagnetic correlations, these   systems are effectively described by quantum rotor models, which consist  of  particles  rotating in the surface of a sphere, such that their angular momentum  competes with the  interactions  that  favour a collective orientation~\cite{HALDANE1983464,AFFLECK1985397,sachdev_2011}. This  motivates our study  of constrained  QFTs involving rotors as mediators of interactions between   Dirac fermions.


\begin{figure}[t]
  \centering
  \includegraphics[width=0.8\linewidth]{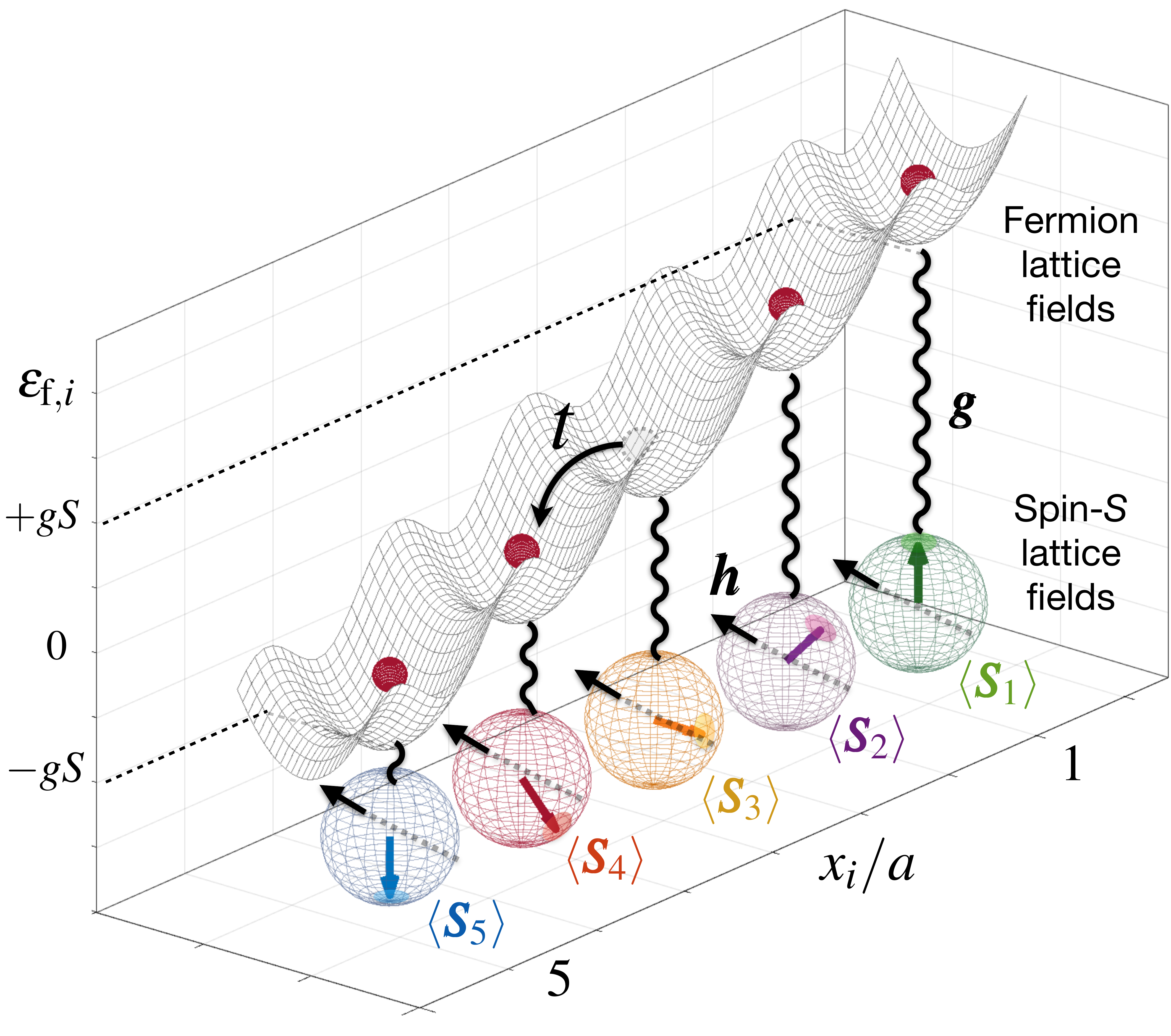}   
\caption{\label{fig:scheme} {\bf Discretised rotor Jackiw-Rebbi model:} Fermions tunnel with strength $t$  against an  energy landscape set by the  lattice spins $\epsilon_{{\rm f},i}=\boldsymbol{g}\cdot\langle\boldsymbol{S}_i\rangle$, which  additionally precess under a magnetic field $\boldsymbol{h}$. In the figure, we show a possible configuration of the spins that lead to an energy landscape with a simple gradient. In the text, we show that other configurations that break translational symmetry appear directly in the equilibrium states of the model. }
\end{figure}

Let us introduce a (1+1)-dimensional  lattice  model that  can display SSB of a chiral symmetry, leading to dynamical mass generation, which allows, as we will see, the emergence of confinement. We consider a Hamiltonian lattice field theory of fermions $c_i^{\phantom{\dagger}},c_i^{\dagger}$ and spins $\boldsymbol{S}_i$, both residing at the sites $x_i=ia$ of a 1D chain of $N_{\rm s}$ sites and length $L=N_{\rm s}a$ (see Fig.~\ref{fig:scheme}).  The fermions  hop  between neighbouring  sites with tunnelling strength $t$ across a potential landscape set by  a spin-fermion coupling $\boldsymbol{g}=g{\bf e}_z$,  and determined by the  spin along the quantisation axis 
\beq
\label{eq:spin_fermion_lattice}
H=\sum_i\left(-t\left(c_i^{\dagger}c_{i+1}^{\phantom{\dagger}}+c_{i+1}^{\dagger}c_{i}^{\phantom{\dagger}}\right)+\boldsymbol{g}\cdot\boldsymbol{S}^{\phantom{\dagger}}_ic_i^{\dagger}c_{i}^{\phantom{\dagger}}-\boldsymbol{h}\cdot \boldsymbol{S}_i^{\phantom{\dagger}}\right).
\eeq
  Additionally, the spins precess under an external  field  with both longitudinal and transverse components $\boldsymbol{h}=h_{\ell}{\bf e}_z+h_{\mathsf{t}}{\bf e}_x$, the latter controlling the quantum fluctuations of the spins. Although Eq.~\eqref{eq:spin_fermion_lattice} bares a certain  resemblance to the Kondo model of fermions coupled to magnetic impurities~\cite{10.1143/PTP.32.37,RevModPhys.51.659}, the fermions are spinless in the present case, and there is no continuous $SU(2)$ symmetry in the coupling. 
  
\begin{figure}[t]
  \centering
  \includegraphics[width=0.95\linewidth]{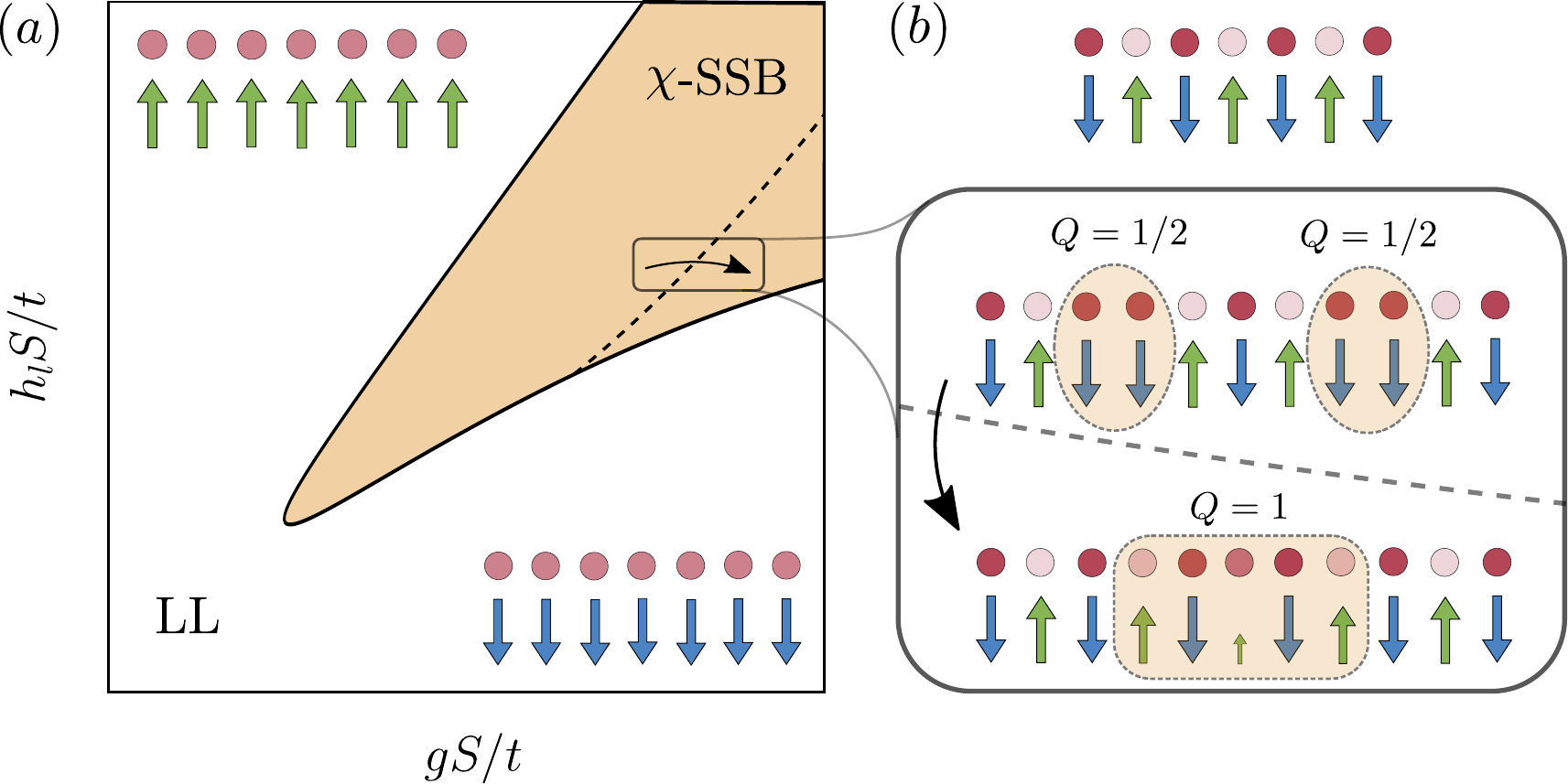}
\caption{\label{fig:scheme_phase_diagram} {\bf Phase diagram of the rotor Jackiw-Rebbi model:} \textbf{(a)} In the figure, we represent qualitatively the different phases that appear in the half-filled vacuum in terms of the interaction $g$ and the longitudinal field $h_\ell$, for fixed values of the transverse field $h_\mathsf{t}$ and temperature $T$. Chiral symmetry is spontaneously broken in the shaded region ($\chi$-SSB), where the fermions develop a dynamical mass and the spins display  N\'eel long-range order, as depicted in the upper panel of {\bf (b)}. This region is surrounded by a  chiral-symmetric phase with a longitudinal paramagnet for the spins, such that the interacting massless fermions form a Luttinger liquid (LL). \textbf{(b)} Within the ordered region, we find two different quasi-particle regimes, separated in the figure by a dashed line. In the first one, deconfined topological defects in the spins bound  repulsive quark-like fermions with fractional charges. In the second one, the quasi-particles attract each other, forming meson-like fermion bags with integer charge. For a finite doping density, the two regimes are separated by a first-order confinement-deconfinement phase transition, which coincides with a chiral symmetry restoration due to the proliferation of defects.}
\end{figure}

Let us briefly summarise our findings. Dispensing with the continuous O(3) symmetry of  the non-linear sigma model~\cite{HALDANE1983464,AFFLECK1985397}, we show that long-range antiferromagnetic order can take place even in (1+1) dimensions, as it is the result of  the breakdown of a discrete chiral symmetry (Fig.~\ref{fig:scheme_phase_diagram}{\bf (a)}). These properties can be neatly understood in the long-wavelength limit, where we find an effective Jackiw-Rebbi-type QFT with  rotor fields playing the role of the self-interacting scalar field: a {\it rotor Jackiw-Rebbi  model}. Interestingly, in contrast to the standard Jackiw-Rebbi model~\cite{PhysRevD.13.3398}, the SSB  does not take place at the classical level, but   requires genuine quantum effects that lead to the non-perturbative dynamical generation of a  fermion mass. We also explore how different rotor profiles can arise in the groundstate by varying the fermion density, which either leads to fractionally-charged or fermion-bag quasi-particles, and  proof their stability even in the ultimate quantum limit of spin $S=1/2$. In fact, we show that the fermion-bag quasi-particles can be understood as confined pairs of fractional charges, resembling the confinement of fractionally-charged quarks in mesons that occurs in the standard model of particle physics (Fig.~\ref{fig:scheme_phase_diagram}{\bf (b)}). Interestingly, we find that a confinement-deconfinement transition can be controlled by tuning a single microscopic parameter, and that this quantum phase transition is associated to the restoration of chiral symmetry by soliton proliferation.

Similarly to lattice gauge theories, in our lattice model~\eqref{eq:spin_fermion_lattice}, fermion-fermion interactions  are mediated by bosonic fields. However, contrary to the former, our model does not possess gauge invariance, a challenging feature to simulate with atomic resources~\cite{1911.00003}. We show how this simplification allows one to implement the lattice model using state-of-the-art cold-atom QSs in a large regime of realistic experimental parameters. Our results thus show how non-perturbative high energy phenomena, such as charge confinement, could be investigated using minimal experimental resources. Let us start with the proposed scheme for the cold-atom QS.

\begin{figure}[t]
  \centering
  \includegraphics[width=1\linewidth]{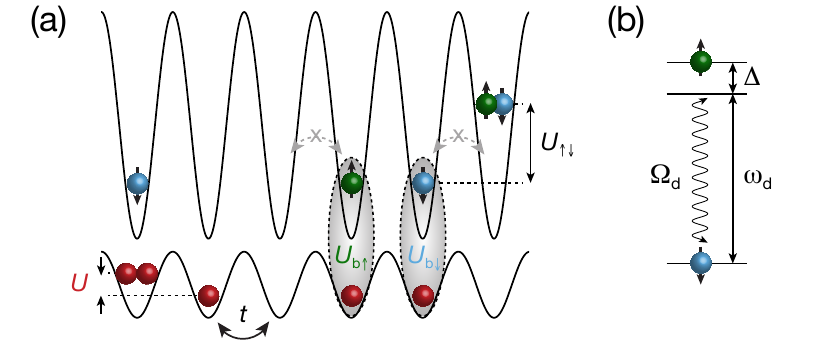}
\caption{\label{fig:scheme_ol} {\bf Bose-Fermi quantum simulator:} {\bf (a)} Optical lattice potentials for both the bosonic   atoms, here represented by red spheres, and the  two-state fermionic atoms, here represented by green (blue) spheres for the $\uparrow$ ($\downarrow$) states. The  bosons have a tunnelling amplitude $t,$ and the Bose-Hubbard on-site interaction $U$. Conversely, the fermionic tunnelling is suppressed by the very deep lattice, and the Fermi-Hubbard ons-site interaction strength is $U_{\uparrow\downarrow}$.   When residing on the same lattice site, bosons and fermions  interact with different strengths  $U_{b\uparrow}\neq U_{b\downarrow}$. {\bf (b)}  Driving that induces  Rabi oscillations between the  fermionic states with Rabi frequency $\Omega_{\rm d}$ and detuning $\Delta_{\rm d}$.}
\end{figure}

\vspace{0.5ex}
{\bf Cold-atom quantum simulation.--}
The lattice Hamiltonian~\eqref{eq:spin_fermion_lattice} can be realised with a Bose-Fermi mixture of ultra-cold atoms confined in spin-independent optical lattices.  We will focus on the $S=\half$ case, as it presents fewer requirements and, moreover, all the relevant phenomena already appear in this limit of maximal quantum fluctuations. We discuss here a generic situation where the bosonic atoms populate a single internal state, whereas the fermionic atoms can be in two possible states, here labelled as $\sigma\in\{\uparrow,\downarrow\}$. In App.~\ref{app:implcoldatoms}, we present a detailed account for a particular choice of atomic species: $^{87}$Rb atoms in their absolute groundstate for the bosons, and  $^{40}$K atoms in two Zeeman sub-levels of the hyperfine groundstate manifold for the fermions. The bosonic and fermionic  species will be individually trapped by a single one-dimensional optical lattice, such that the relative strength can be adjusted by tuning the corresponding lattice wavelength. The particular choice of  atoms allows for a much deeper lattice for the fermionic species (see App.~\ref{app:implcoldatoms}), such that the dynamics can be described by a Bose-Fermi Hubbard model~\cite{PhysRevLett.81.3108,PhysRevLett.89.220407,PhysRevA.68.023606} with mobile bosons and effectively immobile fermions. This is   described by the Hamiltonian $H=H_{\rm m}+H_{\rm i}$  depicted  in Fig.~\ref{fig:scheme_ol}, where the external motional degrees of freedom are governed by the grand-canonical Hamiltonian
\beq
\begin{split}
\label{eq:BFH_model}
H_{\rm m}&=-t\sum_i\left(\hat{b}^\dagger_i \hat{b}^{\phantom{\dagger}}_{i+1} + \text{H.c.} \right) + \frac{U}{2} \sum_i \hat{b}^{\dagger}_i \hat{b}_i^{\dagger}\hat{b}^{\phantom{\dagger}}_i\hat{b}^{\phantom{\dagger}}_i-\sum_i \mu_i \hat{b}_i^{\dagger}\hat{b}^{\phantom{\dagger}}_i\\
&+\sum_i\!\bigg(\!U_{\uparrow\downarrow} \hat{f}^{\dagger}_{i\uparrow} \hat{f}^\dagger_{i\downarrow}\hat{f}^{\phantom{\dagger}}_{i\downarrow}\hat{f}^{\phantom{\dagger}}_{i\uparrow}+\sum_{\sigma}\!U_{b\sigma} \hat{f}^{\dagger}_{i\sigma} \hat{f}^{\phantom{\dagger}}_{i\sigma}\hat{b}^{\dagger}_i \hat{b}^{\phantom{\dagger}}_i \!\!\bigg)\!-\! \sum_{i,\sigma} \mu_{ i,\sigma}\hat{f}_{i\sigma}^{\dagger}\hat{f}^{\phantom{\dagger}}_{i\sigma}.
\end{split}
\eeq
Here, $\mu_i$ ($\mu_{i,\sigma}$) is the local chemical potential for the bosons (fermions),   $U$ ($U_{\uparrow\downarrow}$) is the on-site Hubbard interaction due to  $s$-wave collisions between a pair of bosons (fermions), and $U_{b \uparrow},U_{b\downarrow}$ stem from the corresponding fermion-boson scattering. In addition, the Hamiltonian for the internal degrees of freedom of the fermionic atoms reads
 \beq
\label{eq:BFH_model_int}
H_{\rm int}=\sum_i \!\bigg( \sum_{\sigma} \frac{\epsilon_{ \sigma}}{2}\hat{f}_{i\sigma}^{\dagger}\hat{f}^{\phantom{\dagger}}_{i\sigma}+\frac{\Omega_{\rm d}}{2}\ee^{-\ii\omega_{\rm d}t}\hat{f}_{i\uparrow}^{\dagger}\hat{f}^{\phantom{\dagger}}_{i\downarrow}+{\rm H.c.}\bigg),
\eeq
which includes the  atomic energies for the fermions $\epsilon_{ \uparrow},\epsilon_{ \uparrow}$, and a local term  e.g. a radio-frequency  $\omega_{\rm d}$ that induces  oscillations  between the two fermionic hyperfine states with the Rabi frequency $\Omega_{\rm d}$ (see Fig.~\ref{fig:scheme_ol}{\bf (b)}).

 We consider an average filling of one fermion per lattice site,  effectively reducing the fermionic Hamiltonian in a rotating frame to a sum of on-site terms, whose dynamics can be expressed in terms of spin-$\half$ operators $S^z_i=\half(\hat{f}_{i\uparrow}^{\dagger}\hat{f}^{\phantom{\dagger}}_{i\uparrow}-\hat{f}_{i\downarrow}^{\dagger}\hat{f}^{\phantom{\dagger}}_{i\downarrow})$,  $ S^x_i=\half(\hat{f}_{i\uparrow}^{\dagger}\hat{f}^{\phantom{\dagger}}_{i\downarrow}-\hat{f}_{i\downarrow}^{\dagger}\hat{f}^{\phantom{\dagger}}_{i\uparrow})$. As discussed in App.~\ref{app:implcoldatoms}, the strength of the local external driving  leads  to the transverse field $\boldsymbol{h}_\mathsf{t}$, and its detuning from  resonance  $\Delta_{\rm d}=(\epsilon_{ f\uparrow}-\epsilon_{ f\downarrow})-\omega_{\rm d}$ naturally realizes the longitudinal field $\boldsymbol{h}_\ell$ in a rotating frame
\beq
\boldsymbol{h}_{\ell}=-\Delta_{\rm d}{\bf e}_z, \hspace{2ex}\boldsymbol{h}_\mathsf{t}=-\Omega_{\rm d}{\bf e}_x.
\eeq
The key of the present scheme is the possibility of controlling the different Hubbard interactions in Eq.~\eqref{eq:BFH_model} with a single magnetic field using suitable Feshbach resonances. As detailed in App.~\ref{app:implcoldatoms}, the difference  between the bose-fermi scattering channels   readily implements
\beq
\boldsymbol{g}=2(U_{b\uparrow} - U_{b\downarrow}){\bf e}_z,
\eeq
 which can be adjusted using a single Feshbach resonance. Finally, in the  limit $U\gg t $, the Hamiltonian coincides with the targeted lattice model~\eqref{eq:spin_fermion_lattice} by means of a Jordan-Wigner transformation~\cite{Jordan1928}. In this limit, it becomes clear that we use hardcore bosonic atoms to simulate the fermion fields, and spinfull fermionic atoms to simulate the rotor fields. We note, however, that  the relevant physics  survives   away from this hardcore limit, e.g. we explicitly show a spontaneous breaking of chiral symmetry at a finite value of $U/t$, giving rise to the generation of a dynamical mass for strongly-correlated bosons (see  App.~\ref{app:implcoldatoms}).

 In this Appendix, we also present the detailed calculations of all the relevant parameters for the Rb-K mixture, demonstrating a wide range of experimentally tunability. We would like to emphasise that the above ingredients have been realised in current cold-atom experiments, and the QS of the model~\eqref{eq:spin_fermion_lattice} is thus realistic. We believe that the main experimental challenge will likely be the preparation of extended regions in the lattice with the proper filling for the bosons and fermions. 
 
 In the sections below, we  give a detailed account of the results summarised above about the equilibrium states and low-energy quasi-particles of the Hamiltonian~\eqref{eq:spin_fermion_lattice}.  When discussing these particular phenomena, we will comment on specific probing techniques  and the  temperatures that   would be required to observe them. Let us start by describing these effects by means of an effective long-wavelength theory.


\vspace{0.5ex}
{\bf The rotor Jackiw-Rebbi QFT.--} We now derive the continuum Hamiltonian QFT  that captures the low-energy properties of the lattice model~\eqref{eq:spin_fermion_lattice} at half filling, this is, with fermionic density $\rho = \frac{1}{N_s}\sum_i \langle{c}^\dagger_i {c}^{\vphantom{\dagger}}_i \rangle=  1 / 2$, which will be  considered as the vacuum state in the following. As customary in various (1+1)-dimensional systems~\cite{Haldane_1981}, one starts by linearising the  dispersion relation to obtain a Tomonaga-Luttinger model~\cite{10.1143/ptp/5.4.544,doi:10.1063/1.1704046}. In this case, this amounts to a long-wavelength approximation around  the Fermi points $\pm k_{F}=\pm\pi/2a$, such that 
$c_i\approx\ee^{-\ii k_{ F}x_i}\sqrt{a}\psi_+(x)+\ee^{+\ii k_{ F}x_i}\sqrt{a}\psi_-(x)$ is expressed in terms of slowly-varying right-  and left-moving fermion fields  $\psi_\pm(x)$. As detailed in  Appendix~\ref{app:H_con}, this can be recast in terms of the staggered-fermion discretization of lattice gauge theories~\cite{PhysRevD.11.395}, where one  identifies  the Fermi velocity $v_{\rm F}=2ta$ as the effective speed of light $c=v_{\rm F}$. To proceed with the continuum limit, the spin operators are also expressed in terms of  slowly-varying fields $\boldsymbol{S}_i\approx\cos(k_{\rm  N}x_i)S\boldsymbol{n}(x)+a\boldsymbol{\ell}(x)$ with $k_{N}=\pi/a$, corresponding to the   N\'eel $\boldsymbol{n}(x) $ and canting $\boldsymbol{\ell}(x)$ fields~\cite{HALDANE1983464,AFFLECK1985397}. Performing a gradient expansion and neglecting rapidly-oscillating terms yields  $ H=\!\!\int\!{\rm d}x\mathcal{H}(x)$ with
 \beq
 \label{eq:rotor_JR}
 \mathcal{H}=\overline{\Psi}(x)\left(-\ii c\gamma^1\partial_1+\boldsymbol{g}_s\cdot\boldsymbol{n}(x)\right)\Psi(x)+(\boldsymbol{g}j_0(x)-\boldsymbol{h})\cdot\boldsymbol{\ell}(x),
 \eeq
where $\overline{\Psi}(x)={\Psi}^\dagger(x)\gamma^0$ is the adjoint   for the spinor ${\Psi}(x)=(\psi_+(x),\psi_-(x))^{\rm t}$,  the    gamma matrices are $\gamma^0=\sigma^x$, $\gamma^1=-\ii\sigma^y$,   and the charge-density  is $j_0(x)=\overline{\Psi}(x)\gamma_0{\Psi}(x)$. Additionally, we have introduced the  coupling $\boldsymbol{g}_{s}=gS{\bf e}_z$, and $\partial_1=\partial/\partial x$.

This relativistic QFT~\eqref{eq:rotor_JR} can be understood as a  Jackiw-Rebbi-type model~\cite{PhysRevD.13.3398}, in which a  Yukawa term couples the fermion-mass bi-linear $\overline{\Psi}(x)\Psi(x)$ to the  N\'eel field $\boldsymbol{n}(x)$, instead of the standard Yukawa coupling to a scalar field $\phi(x)$. Additionally, the rotor dynamics is determined by   the precession of the canting field under $\boldsymbol{g}_s j_0(x) / S - \boldsymbol{h}$, instead of the more familiar $\lambda\phi^4$ term of the Jackiw-Rebbi QFT. Hence,  this precession includes the back-action of the matter field onto the mediating fields via the  charge density. In the large-$S$ limit, and in phases dominated by N\'eel correlations, $\boldsymbol{n}(x)$ and $\boldsymbol{\ell}(x)$ represent, respectively,  the orientation and angular momentum of a quantum rotor lying on the unit sphere,  such that this QFT~\eqref{eq:rotor_JR} can be understood as  a rotor Jackiw-Rebbi (rJR) model. In contrast to the standard rotor model~\cite{sachdev_2011}, neighbouring rotor fields are not coupled via $O(N)$-symmetric interactions, but rotor-rotor couplings will  instead be generated through their coupling to  the Dirac fields, and viceversa. As shown below, the lack of a continuous symmetry in Eq.~\eqref{eq:rotor_JR}  plays a crucial role, and makes the physics of the rJR model very different form these $O(N)$ counterparts.

If the spins order according to a N\'eel pattern $\langle n_{z}(x)\rangle\neq 0$, fermions will acquire a mass by the SSB of the discrete chiral symmetry $\Psi(x)\to\gamma^5\Psi(x)$ with $\gamma^5=\sigma^z$, and $n_z(x)\to-n_z(x)$. However, in contrast to the JR model~\cite{PhysRevD.13.3398}, this chiral SSB cannot be predicted classically by looking at the  symmetry-broken sectors of the scalar field, i.e. the double-well minima of the classical potential in the $\lambda\phi^4$ theory. In our case, the bare rotor term only describes precession of the rotor angular momentum, and does not include a collective coupling of the rotors  that would  induce N\'eel order already  classically. Instead, in closer similarity to the Gross-Neveu model~\cite{PhysRevD.10.3235}, chiral SSB shall occur by a {\it dynamical mass generation} that can only be accounted for by including quantum effects (i.e. rotor-fermion loops). On the other hand, there are important differences, as the Gross-Neveu model uses an auxiliary Hubbard-Stratonovich field~\cite{coleman_1985}, whereas our rotor fields represent real degrees of freedom with their intrinsic quantum dynamics. As discussed below, this will lead to crucial differences for the chiral SSB, the quasiparticle spectrum, and confinement.

\vspace{0.5ex}
{\bf Dynamical mass generation and large-$S$ limit.--} Using a coherent-state basis~\cite{Radcliffe_1971,fradkin_2013}, as discussed in App.~\ref{app:S_con}, the partition function   $
\mathsf{Z}=\int{\rm D}[\overline{\Psi}, \Psi,\boldsymbol{n},\boldsymbol{\ell}]\ee^{-\mathsf{S}_{\rm E}}$ can be expressed in terms of an  action $\mathsf{S}_{\rm E}=\int{\rm d}^2x\mathcal{L}(\boldsymbol{x})$  with  the Lagrangian
\beq
\label{eq:euc_L}
\mathcal{L}=\overline{\Psi}(\boldsymbol{x})\left(\hat{\gamma}^\mu\partial_\mu+\boldsymbol{g}_s\cdot\boldsymbol{n}(\boldsymbol{x})\right){\Psi}(\boldsymbol{x})+(\boldsymbol{g}j_0(\boldsymbol{x})-\boldsymbol{h})\cdot\boldsymbol{\ell}(\boldsymbol{x}).
\eeq
Here, $\boldsymbol{x}$ is a 2-dimensional Euclidean  space with imaginary time $x^0=c\tau= c(\ii t)$, and $x^1=x$, and  the Euclidean gamma matrices are  $\hat{\gamma}^0=\gamma^0, \hat{\gamma}^1=-\ii\gamma^1$. The Dirac and adjoint spinors   are Grassmann-valued fields ${\Psi}(\boldsymbol{x})=(\psi_+(\tau,x),\psi_-(\tau,x))^{\rm t}$, ${\overline\Psi}(\boldsymbol{x})=(\psi_-(\tau,x),\psi_+(\tau,x))$, and the charge density is the Grassmann bi-linear $j_0(\boldsymbol{x})=\overline{\Psi}(\tau,x)\hat{\gamma}_0{\Psi}(\tau,x)$. Likewise,  the   spins  lead to constrained vector fields which, in the large-$S$ and  N\'eel-dominated limits, correspond to the position $\boldsymbol{n}(\boldsymbol{x})$ and angular momentum $\boldsymbol{\ell}(\boldsymbol{x})$ of a collection  of  rotors with
\beq
\label{eq:constraints}
\boldsymbol{n}(\boldsymbol{x})\cdot\boldsymbol{n}(\boldsymbol{x})=1,\hspace{2ex}\boldsymbol{n}(\boldsymbol{x})\cdot\boldsymbol{\ell}(\boldsymbol{x})=0.
\eeq

In the continuum limit, 
Since the Lagrangian~\eqref{eq:euc_L}
 is quadratic in Grassmann fields, the fermions can  be integrated out  to obtain an effective action for the rotor fields $\mathsf{S}_{\rm eff}=-\log\left(\int{\rm D}[\overline{\Psi},\Psi]\ee^{-\mathsf{S}_{\rm E}}\right)$. In analogy to the Gross-Neveu model~\cite{PhysRevD.10.3235},  where  dynamical mass generation can be derived by assuming a homogeneous auxiliary field, we consider  $\boldsymbol{n}(\boldsymbol{x})=\boldsymbol{n}$, $\boldsymbol{\ell}(\boldsymbol{x})=\boldsymbol{\ell}$, $\forall \boldsymbol{x}\in(0,\beta]\times(0,L]$. As shown in App.~\ref{app:S_mf},  the functional integral leads to  
  \beq
  \label{eq:eff_action}
  \mathsf{S}_{\rm eff}=\beta L\!\left(\!\left(\frac{\boldsymbol{g}}{2}-\boldsymbol{h}\right)\cdot\boldsymbol{\ell}-\frac{(\boldsymbol{g}_{ s}\cdot\boldsymbol{n})^2}{4\pi c}\left(\log\left(\frac{\Lambda_{\rm c}}{\boldsymbol{g}_{ s}\cdot\boldsymbol{n}}\right)^{\!\!2}+1\!\right)\!\!\right),
 \eeq 
 where we have introduced the UV cutoff $\Lambda_{\rm c}=2 t$ set by the bandwith of the bare fermion dispersion on the lattice~\eqref{eq:spin_fermion_lattice}.

 Exploiting the  analogy to the Gross-Neveu model~\cite{PhysRevD.10.3235}, where one deals with  $N$ flavours of Dirac fermions and finds the non-perturbative dynamical mass generation in the $N\to\infty$ limit, we will send   $S\to\infty$    to determine the non-perturbative phase diagram quantitatively. Since  phases  not governed by N\'eel correlations can also appear, and we are interested in possible quantum phase transitions transitions thereof, we must relax  the first  constraint in Eq.~\eqref{eq:constraints}. By considering the explicit construction of the rotor fields in terms of the spin coherent states, we find  the   N\'eel and canting  fields 
 \beq
 \label{eq:fields_angles}
 \begin{split}
 \boldsymbol{n}&=\frac{1}{\phantom{.}2\phantom{.}}(\sin\theta_A-\sin\theta_B){\bf e}_{z}-\frac{1}{\phantom{.}2\phantom{.}}(\cos\theta_A-\cos\theta_B){\bf e}_{x},\\
 \boldsymbol{\ell}&=\frac{S}{2a}(\sin\theta_A+\sin\theta_B){\bf e}_{z}-\frac{S}{2a}(\cos\theta_A+\cos\theta_B){\bf e}_{x}.
  \end{split}
 \eeq  
The fields are thus parametrized by  $\theta_A,\theta_{B}\in[0,\pi]$, each of which represents the angle of  the spin coherent state associated to the $A$ (odd sites) and  $B$ (even sites) sub-lattice, pointing along  the great circle contained in the $xz$ plane (see Fig.~\ref{fig:scheme}). One can  check that the second constraint in Eq.~\eqref{eq:constraints} is  readily satisfied $\boldsymbol{n}\cdot\boldsymbol{\ell}=0$, whereas the first one will only be recovered in N\'eel-dominated phases $\theta_{A}\approx-\theta_{B}=\pm\pi/2$, where  $\boldsymbol{n}\cdot\boldsymbol{n}\approx1$.


\begin{figure*}[t]
  \centering
  \includegraphics[width=1.0\linewidth]{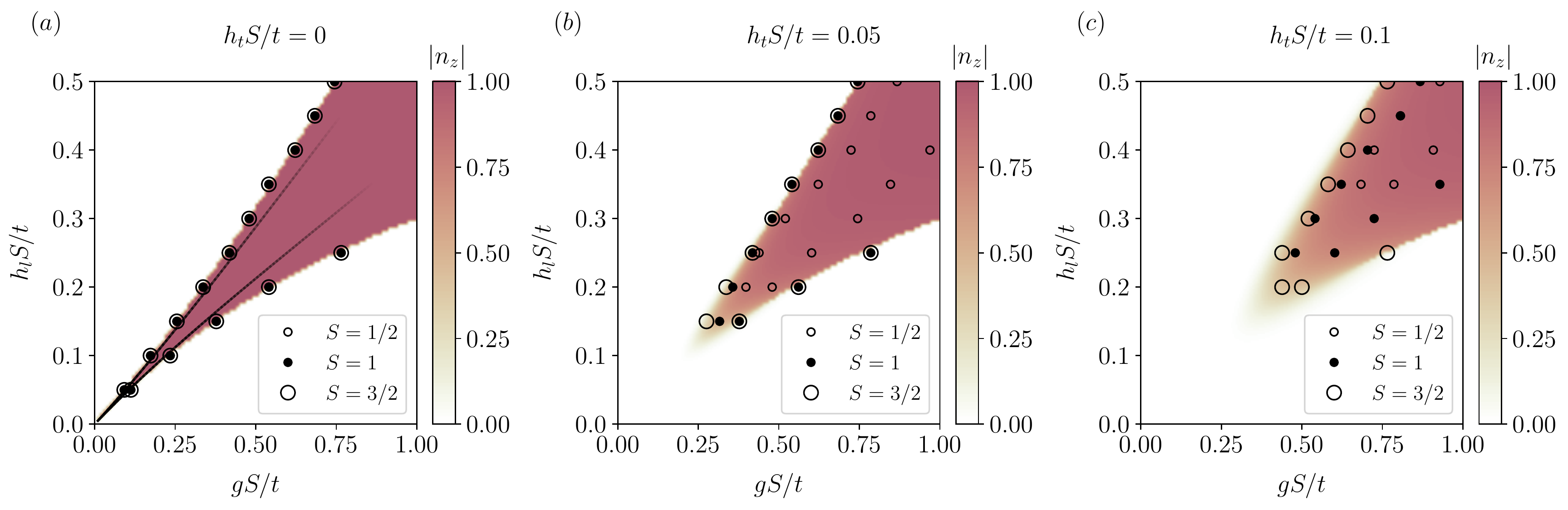}
\caption{\label{fig:phase_diagram} {\textbf{Dynamical mass generation at finite $S$:}  In the background of the different phase diagrams represented in the figure, we show the absolute value of $n_z$ calculated with the HF method for a periodic chain of $N_s = 64$ sites and $N_f = 32$ fermions. The different dots correspond to the critical points separating the AF from the LL phase, obtained using iDMRG for various values of $S$. {\bf (a)} When quantum fluctuations are absent, all the points coincide with the HF solution and the large-$S$ predictions for  the whole phase diagram. The solid lines correspond to the solution obtained using the large-$S$ expansion in the continuum model~\eqref{eq:critical_lines},  agreeing with the lattice critical lines for small values of $gS/t$. {\bf (b)} In the presence of finite quantum fluctuations, the HF solution departs from the correct ground state, although it provides a good approximation even for small values of $S$ in the case of small fluctuations. {\bf (c)} The AF region shrinks as $h_{\mathsf{t}}S/t$ increases, and it does it faster for smaller values of $S$. However, this phase can always be found, even in the limit $S=1/2$, provided that $gS/t$ and $h_{\ell}S/t$ are sufficiently large.}}
\end{figure*}

 By inspecting the effective action~\eqref{eq:eff_action}, one finds that it is proportional  to $ \mathsf{S}_{\rm eff}=  \beta LS\mathsf{V}_{\rm eff}(\theta_A,\theta_B)$, such that the large-$S$ limit is obtained through  the saddle-point equations $\left.\boldsymbol{\nabla}_{\theta} \mathsf{V}_{\rm eff}\right|_{\boldsymbol{\theta}^\star}=\boldsymbol{0}$. The features of the phase diagram can be understood in two complementary regimes: {\it (a)} For $h_{\mathsf{t}}/t\gg h_{\ell}/t, g/t$, the saddle-point equations will be solved by $\theta^\star_A=\theta^\star_{\rm B}=\pi$, such that all spins are maximally polarised in the direction of the leading transverse field $\ket{\mathsf{g}_{\mathsf{tP}}}=\otimes_i\ket{S,S}_{x,i}$, where $\ket{S,m}_{\alpha,i}$ is the common eigenstate of $\boldsymbol{S}_i^2,S^\alpha_i$  with eigenvalues $S(S+1)$, and $m\in\{-S,-S+1,\cdots,S\}$, respectively. This state can  thus be understood as a disordered {\it transverse paramagnet}, with all spins aligned along the equator of Fig.~\ref{fig:scheme}. Since $\langle n_z\rangle=0$, there is no mass generation, and the chiral symmetry remains intact. Therefore, the fermionic sector will correspond to a {\it metallic Luttinger liquid}. {\it (b)} For $h_{\rm t}/t\ll h_{\ell}/t, g/t$, we find a competition between two distinct phases. If $h_{\ell}> h_{\ell}^{+}$ or $h_{\ell}< h_{\ell}^{-}$, where we have introduced 
 \beq
 \label{eq:critical_lines}
 h_{\ell}^{\pm}=\frac{g}{2}\pm\frac{t}{\pi S}\left(\frac{gS}{2t}\right)^{\!\!2}\!\left(\log\left(\frac{2t}{gS}\right)+\frac{1}{2}\right),
 \eeq
the saddle-point equations are solved by $\theta_A^\star=\theta_{\rm B}^\star=\pi/2$, or $\theta_A^\star=\theta_{\rm B}^\star=-\pi/2$, respectively. These correspond to  disordered {\it longitudinal paramagnets} with groundstates $\ket{\mathsf{g}_{\ell\mathsf{P}}^+}=\otimes_i\ket{S,S}_{z,i}$, or  $\ket{\mathsf{g}_{\ell\mathsf{P}}^-}=\otimes_i\ket{S,-S}_{z,i}$, with all spins pointing towards the north or south poles of Fig.~\ref{fig:scheme}, respectively. Once again,  since $\bra{\mathsf{g}_{\ell\mathsf{P}}^\pm} n_z\ket{\mathsf{g}_{\ell\mathsf{P}}^\pm}=0$, there is no mass generation and the fermions  are described by a  massless Luttinger liquid. On the contrary, if $h_{\ell}^{-}\!<h_{\ell}< h_{\ell}^{+}$, the saddle points correspond to $\theta_A^\star=-\theta_{\rm B}^\star=\pm\pi/2$, which yields  two  {\it N\'eel antiferromagnets} $\ket{\mathsf{g}_{\mathsf{N}}^\pm}=\otimes_{i\in A}\ket{S,\pm S}_{z,i}\otimes_{i\in B}\ket{S,\mp S}_{z,i}$, in which chiral symmetry is spontaneously broken, yielding   $\bra{\mathsf{g}_{\mathsf{N}}^\pm} n_z\ket{\mathsf{g}_{\mathsf{N}}^\pm}=\pm1$ respectively. In both cases, the spins of Fig.~\ref{fig:scheme} alternate between the north and south poles, and the Dirac fermions acquire a mass dynamically,  accompanied by a so-called scalar condensate
\begin{equation}
\langle\overline{\Psi}\Psi\rangle=\langle\psi^\dagger_+(x)\psi_-(x)+\psi^\dagger_-(x)\psi_+(x)\rangle=\Sigma_0.
\end{equation}
Since the spinor components $\psi_\pm(x)$ correspond to the long-wavelength excitations around  momenta $\pm k_{\rm F}=\pm\pi/2a$, the  scalar condensate  of the continuum QFT leads to  a periodic modulation of the lattice density $\langle c_i^\dagger c^{\phantom{\dagger}}_i\rangle=\half+(-1)^{i}\Sigma_0 a $. This phase is reminiscent of a  { charge-density wave insulator} in electron-phonon  systems. However,  in these systems,  Peierls' argument shows that the 1D metal is always unstable towards the insulator~\cite{peierls}, instead of  showing different phases separated by quantum critical lines, as occurs for our rotor Jackiw-Rebbi QFT~\eqref{eq:critical_lines}. Let us remark once more  that, while in the standard JR model~\cite{PhysRevD.13.3398} chiral SSB can be understood by means of perturbation theory  about the classically SSB sectors in $g\ll1$, the mass is dynamically generated in our case, and cannot be understood perturbatively, which  can be appreciated by the fact that the $\log(St/g)$ dependence in Eq.~\eqref{eq:critical_lines} cannot be Taylor expanded  for small $g$.

\vspace{0.5ex}
{\bf Quantum fluctuations and Ising universality class.--} Let us now benchmark the above large-$S$ calculations with numerical results based on a Hartree-Fock (HF)~\cite{giuliani_vignale_2008} self-consistent mean-field method, and a matrix-product-state (MPS)~\cite{SCHOLLWOCK201196,10.21468/SciPostPhysLectNotes.5} formulation of the density-matrix renormalization group (DMRG)~\cite{PhysRevLett.69.2863}. This serves a two-fold purpose: on the one hand, both methods are directly applied to the discretised  model on the lattice~\eqref{eq:spin_fermion_lattice}, and can thus be used to identify the  parameter regime  where the continuum QFT predictions~\eqref{eq:eff_action} are recovered. On the other hand, the quasi-exact MPS method  gives direct access to  corrections of the large-$S$  predictions for finite values of $S\in\{\frac{3}{2},1,\frac{1}{2}\}$,   testing  the dynamical mass generation in  the regime of large quantum fluctuations.  Likewise,  we can adapt the HF method  to non-zero temperatures $T$ and chemical potentials $\mu$  in order to explore the role of   thermal fluctuations and finite densities (see Appendix~\ref{app:MF}).

 Figure~\ref{fig:phase_diagram} contains our results for  the   zero-temperature half-filling phase diagram as a function of $(gS/t,h_{\ell}S/t)$ for various values of the transverse field $h_{\mathsf{t}}S/t$. In the background, we represent the N\'eel order parameter, $n_z=\frac{2}{N_{\rm s}}\sum_x{\bf e}_z\cdot\boldsymbol{n}(x)$, obtained by averaging over the HF estimate of the N\'eel field  $\boldsymbol{n}(x)=\frac{1}{2S}\left(\langle \boldsymbol{S}_{2i}\rangle-\langle \boldsymbol{S}_{2i-1}\rangle\right)$. In Fig.~\ref{fig:phase_diagram}{\bf (a)}, one can see how the HF predicts an intermediate  region, here depicted in red,  displaying antiferromagnetic long-range order $n_z\approx1$ due to the SSB of the discrete chiral symmetry. In order to test the validity of our QFT predictions based on the phenomenon of dynamical mass generation, we benchmark these numerical results against the critical lines of Eq.~\eqref{eq:critical_lines}, which are depicted as solid black lines in Fig.~\ref{fig:phase_diagram}{\bf (a)}. In analogy to the large-$N$ limit of other strongly-coupled QFTs~\cite{coleman_1985}, we must rescale the coupling to obtain physical results for  $S\to\infty$, such that $gS$ remains finite. From the comparison of the HF and large-$S$ results, we understand that it is the regime of $gS\ll t$ (i.e. couplings much smaller than the UV cutoff), where we can recover the continuum QFT from the lattice discretization, as typically occurs in asymptotically-free lattice field theories. 
 
 Note that, in fact, the lattice theory already agrees with  the continuum predictions for intermediate couplings $gS\sim0.25t$, which is a sensible fraction of the UV cutoff, and signals the wide validity of the aforementioned scheme of dynamical mass generation. It is worth mentioning, however, that  the SSB mechanism that yields the  N\'eel phase is valid for an even wider range of parameters. Indeed, around any of the critical lines of Fig.~\ref{fig:phase_diagram}{\bf (a)}, there will be an effective continuum QFT, albeit with different renormalised parameters that require us to rely on numerical methods or  experimental QSs. This wider parameter regime is useful in light of the cold-atom realization presented above, which will be able to probe antiferromagnetic correlations in a regime more favourable than the one set by  super-exchange mechanisms~\cite{PhysRevLett.91.090402,Trotzky295,Greif1307,Boll1257,mazurenko_cold-atom_2017,PhysRevLett.124.043204}. In particular, for larger values of $gS$, the order survives to larger temperatures, as we will also show below.
 
The generation of a dynamical mass described in this section can be implemented following the previous cold-atom scheme, and experimentally probed  using standard detection techniques. In particular, the scalar condensate (i.e. charge-density-wave ordering) can be readily probed by measuring the  imbalance   $I=(n_A-n_B)/(n_A+n_B)$ between the occupation of Rb atoms on $A$- and $B$-sublattices by using superlattices~\cite{sebby-strabley_lattice_2006,folling_direct_2007,trotzky_probing_2012} or via noise correlations~\cite{folling_spatial_2005,rom_free_2006,2003.08945}. The antiferromagnetic N\'eel ordering can be further revealed by evaluating the imbalance observable, or the noise correlations, for the fermionic K atoms in a spin-resolved manner, which  requires separating the two hyperfine states during the detection by means of a Stern-Gerlach sequence~\cite{Trotzky295}, or other similar techniques~\cite{trotzky_controlling_2010,Greif1307}. Since in the SBB phase, however, there are two energetically-degenerate configurations shifted by one lattice site, and the experiment will consist of many independent copies of the one-dimensional chains, the ensemble-averaged observables may fail to signal the phase transition without an additional term that weakly breaks the symmetry between the configurations. If a small symmetry-breaking field cannot be globally introduced in all these copies, one may resort to a combination with quantum gas microscopy~\cite{bakr_probing_2010,sherson_single-atom-resolved_2010,mazurenko_cold-atom_2017, Koepsell_2020, Hartke_2020}, which now also enables full spin and charge read-out. 
 
 Let us now explore the effect of  finite $S$   and non-zero  $h_{\mathsf{t}}$. In Fig.~\ref{fig:phase_diagram}{\bf (a)}, the circles represent the critical values of the SSB phase transition for different values of $S$, and are obtained with the MPS method based on the iDMRG scheme for an infinite chain~\cite{10.21468/SciPostPhysLectNotes.5}. These critical points are estimated by localising the divergence of the spin susceptibility, $\chi_S = \partial n_z / \partial (gS/t)$, where we use bond dimension $D=200$ and a four-sites repeating unit cell. As can be observed in the figure, for a vanishing transverse field, the critical points for different $S$ are all arranged along the same critical line which, furthermore, delimits  the N\'eel-ordered phase obtained with the HF method, and agrees with the  large-$S$ predictions~\eqref{eq:critical_lines} in the  regime where we expect to recover the continuum QFT  from the lattice regularisation. We note that, in our model, changing the value of the spin $S$ for a vanishing  $h_{\mathsf{t}}=0$, does not modify the quantum fluctuations, such that the large-$S$ prediction works equally well for any value of $S$. This contrasts  the typical situation in models with $O(3)$ symmetry, where the the classical limit is associated with $S\to\infty$, and quantum fluctuations appear as soon as $S$ is finite. Moreover, as outlined in the appendix,  we confirm that there is no qualitative distinction in the underlying physics for integer or half-integer spins, as occurs for   models with a continuous  $O(3)$ rotational  symmetry.

 \begin{figure}[t]
  \centering
  \includegraphics[width=0.9\linewidth]{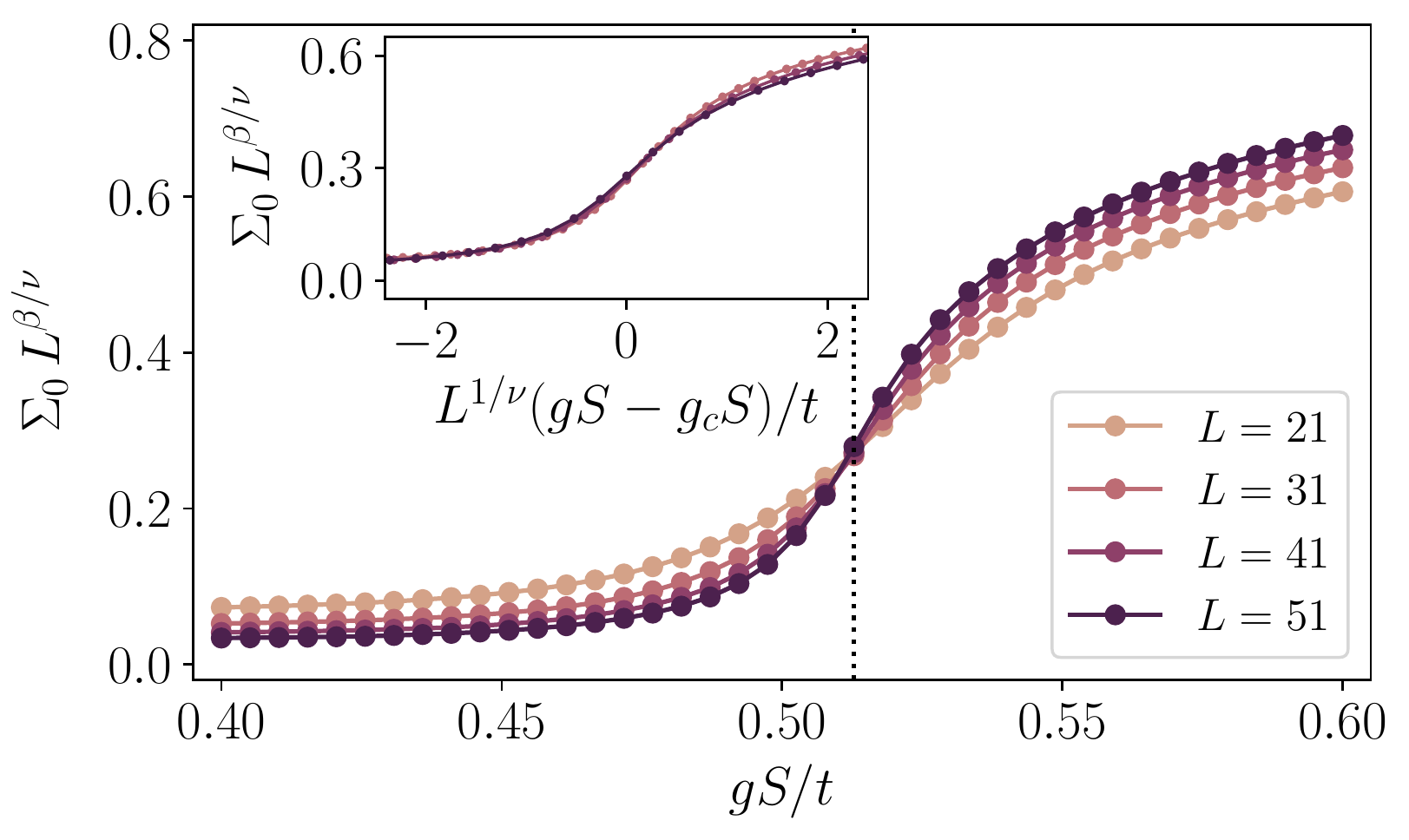}
\caption{\label{fig:scaling_g} {\textbf{Scalar condensate and chiral SSB: } We represent the  scalar  fermion condensate $\Sigma_0=\langle\overline{\Psi}\Psi\rangle$ in terms of $gS/t$ for the ground state of a half-filled chain of different lengths $N_s$, with $h_{\ell}S=0.3t$ and $h_{\mathsf{t}}S = 0.05t$. The results are obtained using DMRG for $S=1/2$. Using the  critical exponents  of the 2D Ising universality class ($\nu = 1$, $\beta = 1/8$), the  lines cross at the critical point obtained for the infinite system with iDMRG, and  collapse for an appropriate rescaling (inset).}}
\end{figure}

  In Figs.~\ref{fig:phase_diagram}{\bf (b)} and {\bf (c)}, we represent the phase diagram as the transverse field is switched on, which introduces quantum fluctuations on the spins. In this limit, we observe  how the long-range N\'eel phase shrinks as a result of the competing quantum fluctuations. We also observe that, as the value of $S$ increases,  a better agreement with the HF and QFT predictions is obtained, confirming the generic expectation. Note that this agreement is remarkable, given  that the considered values of the spins $S$ are still very far away from the large-$S$ limit. 
  
  So far, our numerical benchmark has focused on the SSB captured by the N\'eel field. Let us note, however, that the dynamical mass generation refers to the fermionic sector, and the gap opening is associated to an underlying non-zero scalar condensate $\Sigma_0=\langle\overline{\Psi}\Psi\rangle$. In order to extract the value of this condensate from the lattice simulations, we use $\Sigma_0=\frac{1}{N_{\rm s}}\sum_i(-1)^i\langle c_i^\dagger c^{\phantom{\dagger}}_i\rangle$, where the expectation value is calculated with the MPS groundstate obtained using a DMRG algorithm with bond dimension $D = 200$  for  finite chains  of  variable length $L=N_{\rm s}a$ and unit lattice spacing $a=1$. In Fig.~\ref{fig:scaling_g}, we represent the finite-size scaling  for the scalar condensate of the $S=1/2$ model. The crossing of the lines in the main panel serve to locate the critical point of the model $gS_{\rm c}/t$, which agrees with our previous iDMRG results based on the N\'eel order parameter. Hence, this shows that the chiral-SSB occurs via the simultaneous onset of a N\'eel antiferromagnet and a scalar fermion condensate, which corresponds to a charge density wave as seen from the lattice perspective. Moreover, these results allow us to identify the universality class of the corresponding chiral phase transition. As proved by the data collapse shown in the inset of Fig.~\ref{fig:scaling_g}, the critical exponents correspond to those of the (1+1) Ising universality class.


\vspace{0.5ex}
 \begin{figure}[t]
  \centering
  \includegraphics[width=0.8\linewidth]{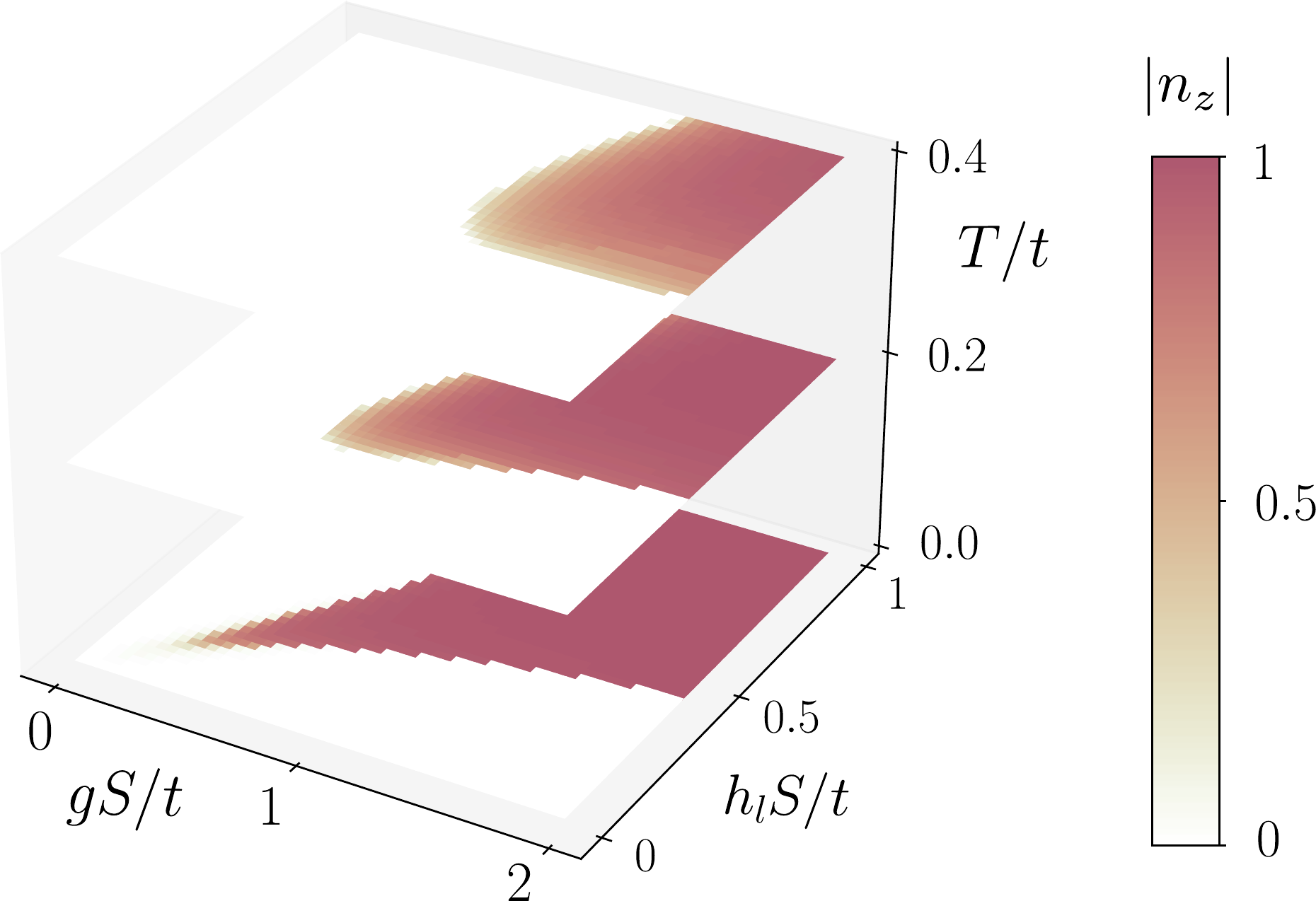}
\caption{\label{fig:finite_T} {\textbf{Chiral symmetry restoration at finite temperatures:} Phase diagram for different  $T/t$ and $h_{\mathsf{t}}S=0.05t$, where we represent the N\'eel order parameter $n_z$ calculated in the HF approximation. The AF phase, where chiral symmetry is spontaneously broken,  shrinks as the temperature increases. In the high-energy-physics lore, one says that chiral symmetry is restored at  high temperatures.}}
\end{figure}

{\bf Thermal fluctuations and chiral symmetry restoration.--} Let us now move on to the study of the corrections due to thermal fluctuations. In Fig.~\ref{fig:finite_T}, we represent sections of the phase diagram as a function of $(gS/t,h_{\ell}S/t)$ for several values of the temperature $T$ obtained with the HF method. One can observe how the area that encloses a dynamically-generated mass, characterised by the N\'eel order parameter $n_z$, shrinks with increasing $T$. Therefore, for sufficiently high temperatures, the chiral SSB phase   would eventually disappear in favour of the disordered paramagnet and the massless Dirac fermions, both of which  respect the chiral symmetry.

 As shown in Fig.~\ref{fig:finite_T}, the required temperature scale for a robust observation of the different phases may lie above that of the tunnel coupling $t$, which is a very promising feature of our QS scheme.  Using state-of-the-art cooling techniques, temperatures as low as $T/t=0.2$ have been reported for two-component Fermi gases~\cite{mazurenko_cold-atom_2017} and $T/U=0.05$ for bosonic atoms~\cite{yang_cooling_2020}. As displayed in Fig.~\ref{fig:finite_T}, the dynamical mass generation leading to the anti-ferromagnetic order and the scalar condensate can  be accessed at larger temperatures $T/t\sim 0.4$. The underlying reason is that the onset of antiferromagnetic correlations does not rest on the super-exchange mechanism, the scale of which is $t^2/U\ll U$. In our case, the mechanism is the dynamical breaking of chiral symmetry, the scale /gap of which is directly set by the interaction coupling $g\sim U_{bf}$, which can be on the order of the Rb-K interactions $U_{bf}$.

Let us now interpret these results in light of  \textit{thermal chiral symmetry restoration} in the standard model. In particular, for a large region of its phase diagram, the vacuum of QCD  is expected to break spontaneously a chiral symmetry associated to the quarks' flavour~\cite{Coleman_1980}. This mechanism is confirmed experimentally by the measured mass of light baryons, such as protons and neutrons, where chiral symmetry breaking yields the largest contribution to  their masses~\cite{wilczek_1999}, while only a small part comes from the masses of their constituent quarks.  On the other hand, at very high temperatures or densities, corresponding to the first instants after the Big Bang or to the  dense core of neutron stars, respectively, chiral symmetry is restored and quarks become massless, as shown in experiments involving  heavy-ion collisions~\cite{Rapp_2000}. This phenomenon is captured by several effective theories of nucleons, such as the Nambu-Jona-Lasinio quantum field theory~\cite{Nambu_1961} and, as discussed above, also occurs in our model. 

There are, however, several unsolved questions in QCD regarding the restoration of chiral symmetry. One is whether a phase transition  at finite $T_{\rm c}$ or a crossover exists for intermediate values of the baryon chemical potential $\mu_{\rm B}$~\cite{Brambilla_2014}. Another one and is the relation to the deconfinement of quarks where, instead of forming hadronic bound states, deconfinement gives rise to a so-called quark-gluon plasma~\cite{kogut_stephanov_2003}. For large values of $\mu_{\rm B}$, it is not known if both transitions occur simultaneously or, alternatively, intermediate phases exist~\cite{McLerran_2007}. Many effective theories, however, fail to address these questions, since they do not include any confinement mechanism even if they correctly capture the essence of dynamical mass generation. As we show in the next section, our model presents such confinement-deconfinement phase transition between fractionally-charged quasi-particles, allowing for the investigation of the interplay between the latter and chiral symmetry restoration in a simple setup.

\vspace{0.5ex}
{\bf Confinement of fractionally-charged quasi-particles.--} As mentioned before, the fundamental fields of the QCD sector of the standard model correspond to  fractionally-charged fermion fields, the so-called quarks, coupled to bosonic Yang-Mills fields, the so-called gluons~\cite{Peskin:1995ev}.  In contrast, the fundamental fields of our model~\eqref{eq:rotor_JR} are fermion fields with integer charges coupled to the constrained N\'eel and canting fields. As noted in the introduction, however, the renormalised quasi-particles of a strongly-coupled QFT may sometimes differ completely from its fundamental constituents. As shown below, our QFT~\eqref{eq:rotor_JR} displays a Jackiw-Rebbi-like mechanism of fractionalisation~\cite{PhysRevD.13.3398}, whereby soliton configurations of the N\'eel field host localised fermions with a fractional charge $q=\pm e/2$, which will play the role of quarks, allowing us to discuss various aspects of a confinement mechanism.

\begin{figure}[t]
  \centering
  \includegraphics[width=1.0\linewidth]{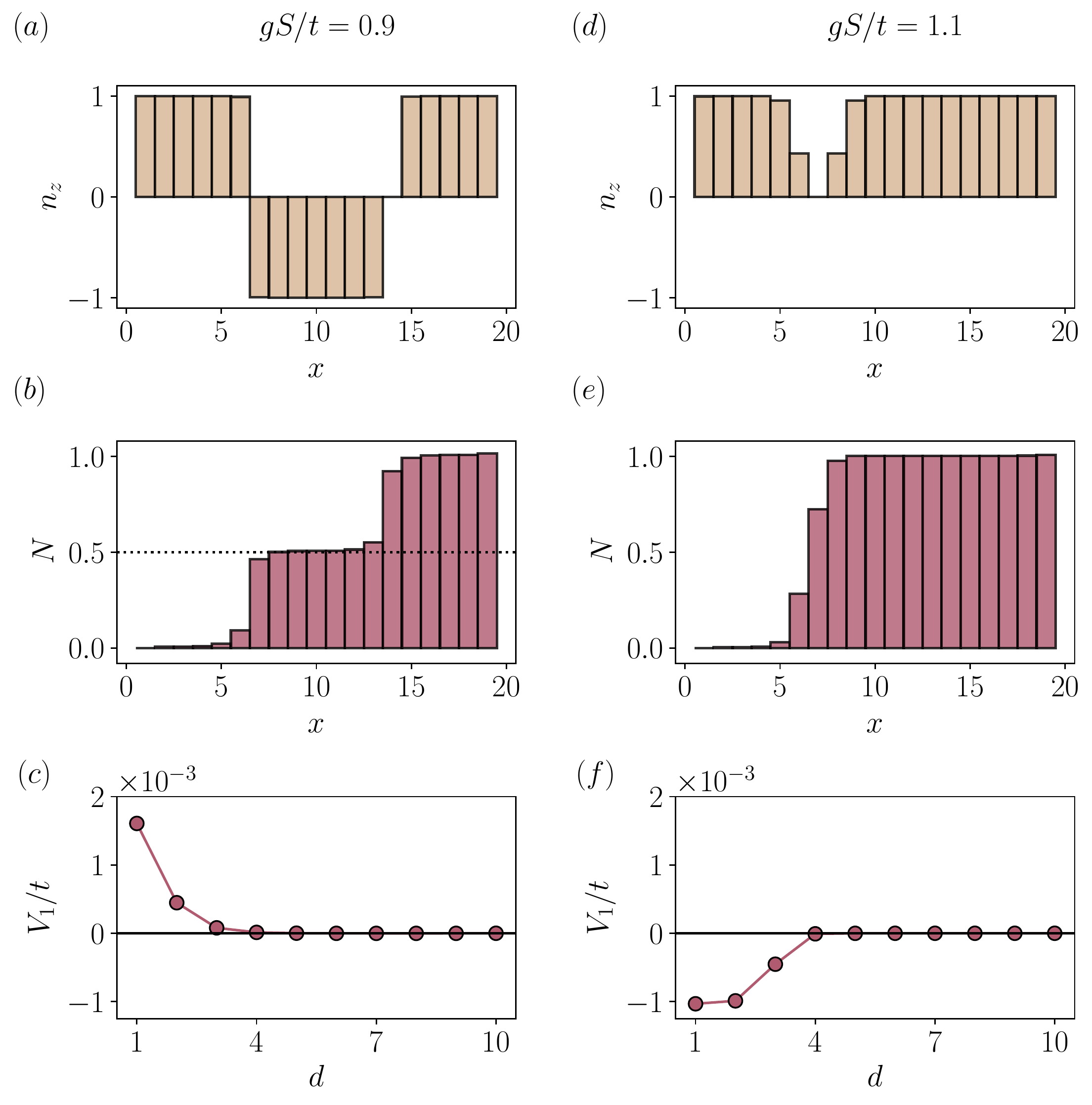}
\caption{\label{fig:real_space_deconfinement} {\bf Fractionalization and confinement:} {\bf (a)} Space dependence of $n_z$ in the ground state of a chain with $N_{\rm s}=41$ sites and $N_{\rm f}=22$ fermions, for $gS/t=0.9$, $h_{\ell}S/t = 0.4$ and $h_{\mathsf{t}}S/t = 0.01$. Two solitons appear on the otherwise homogeneous configuration found at half filling. {\bf (b)} The accumulated particle density shows how each defect carries a fermion with fractional charge $e/2$. {\bf (c)} The solitons repel each other at short distances, implying that the composite soliton-charge quasi-particles are deconfined. {\bf (d)} At larger values of $gS/t$, solitons merge forming a quark-bag quasi-particle with integer charge {\bf (e)}. {\bf (f)} In this case, the potential energy between the solitons is attractive, indicating a transition to a confined phase.}
\end{figure}

In Figs.~\ref{fig:real_space_deconfinement}{\bf (a)-(b)}, we present the MPS numerical results for the real-space configurations of the N\'eel lattice field $n_z(j)=\frac{1}{2S}(\langle {S}^z_{2j}\rangle-\langle {S}^z_{2j-1}\rangle)$, and the integrated fermion charge above the half-filled vacuum $N(j)=Q(j)/e=\sum_{i<j}(\langle c_i^\dagger c_i^{\phantom{\dagger}}\rangle-\half)$ when one extra fermion is introduced above half filling. These figures  show that the N\'eel field presents a kink-antikink pair that interpolates between the different SSB sectors, and that each of these topological solitons hosts a localised fermionic excitation with charge $q=e/2$, henceforth referred to as a `quark' by the analogy with the fractionally-charged fundamental fermion fields of QCD. We note that  similar quasi-particles with charges $q=-e/2$ would appear for dopings below half filling, playing the role of `anti-quarks', and that quark-antiquark pairs could appear in the vacuum due to thermal fluctuations, as they correspond to higher-energy states of the theory.

Contrary to the general situation in (1+1) lattice gauge theories, which can only host confining phases~\cite{greensite_2020}, we can find regimes where quarks/anti-quarks  can be confined/deconfined depending on the microscopic parameters. In order to explore this phenomenon, let us  remark that the distance of the pair of quarks of Fig.~\ref{fig:real_space_deconfinement}{\bf (b)} is determined by the external pinning of the N\'eel solitons of Fig.~\ref{fig:real_space_deconfinement}{\bf (a)}. As described in more detail in Appendix~\ref{app:sol_deconf}, we introduce an external potential that breaks explicitly the translational invariance and localises the solitons, which would otherwise travel freely through the chain, at the desired positions. This pinning  makes our fractionally-charged quasi-particles static, such that we can discuss the analogue of the static quark potential~\cite{BALI20011}. We note that it is a standard practice    in lattice gauge theories to use this terminology whether or not the charges  actually correspond to dynamical quarks~\cite{greensite_2020}. Therefore,   the static quark potential  quantifies the interaction energy between any pair of  static charges as their distance is modified, giving information about confinement not only in QCD, but in other effective models. 

In our model, the static quark potential $V_1(d)=E_1(d)-E_0$ can be obtained using the MPS numerics by calculating the energy $E_1(d)$ of the doped system with  an  extra  fermion that fractionalises into a pair of quarks pinned at a distance $d$, measured in unit cells ($2a$), with respect to the energy $E_0$ of a pair pinned far apart ($d\gg 1$). We note that the standard situation of (1+1) lattice gauge theories, such as the Schwinger model~\cite{PhysRev.128.2425}, is that the preservation of gauge symmetry requires that the static charges must be connected through an intermediate electric-field string, such that the energy increases with the separation $d$ and  leads to a linearly-increasing static quark potential~\cite{COLEMAN1975267,PhysRevX.6.041040}. In our case, the situation is completely different, as the energy of the deformed N\'eel field that connects the two quarks in Fig.~\ref{fig:real_space_deconfinement}{\bf (c)} is independent of the pinning distance, and confinement is thus not enforced a priori. As argued below, there exists a competing  mechanism that either favours confinement or deconfinement depending on the    microscopic parameters of the model.

In Fig.~\ref{fig:real_space_deconfinement}{\bf (c)}, we depict the distance-dependence of the static quark potential for the $S=1/2$ rotor Jackiw-Rebbi model for coupling $gS/t=0.9$, and setting the other parameters such that we lie in the chiral-SSB phase. As can be observed in this figure, the potential decreases with the distance for small soliton separations, which means that the quarks repel each other. Hence, in the absence of the external pinning, the fractionally-charged quasiparticles would move freely at large distances from each other, and thus appear as asymptotic excitations in the spectrum of the rJR QFT.  As we increase the coupling to $gS/t=1.1$, Figs.~\ref{fig:real_space_deconfinement}{\bf (d)-(e)} show that a completely-different quasi-particle emerges. In this case, the N\'eel field no longer interpolates between the two SSB sectors, but is instead  suppressed within a  small region of space where, as  shown in Fig.~\ref{fig:real_space_deconfinement}{\bf (e)}, an integer-charged fermion is localised. This situation is reminiscent of the so-called  quark bag models~\cite{PhysRevD.9.3471,PhysRevD.11.1094}, where  quarks  and gluons are locked within a finite region of space, in which a phenomenological term that compresses the bag compensates the outward pressure of the quarks that are held inside, and  confinement  results from the competition of these two terms. The present situation is closer in spirit to the soliton quark  model~\cite{PhysRevD.15.1694,PhysRevD.16.1096}, where the fermions deplete a SSB condensate in a finite region of space, gaining kinetic energy at the expense of the  cost of deforming  the condensate.  In our case, as the N\'eel field vanishes in the inner region, there scalar condensate will become zero $\Sigma_0(x)=0, \forall x\in[x_0-{\xi},x_0+\xi]$, such that the bound fermions have a vanishing dynamically-generated mass, increasing their kinetic energy and the outward pressure. As mentioned above, this is compensated by the energy cost due to the inhomogeneous layout of the N\'eel field and the accompanying scalar condensate. 

\begin{figure}[t]
  \centering
  \includegraphics[width=1.0\linewidth]{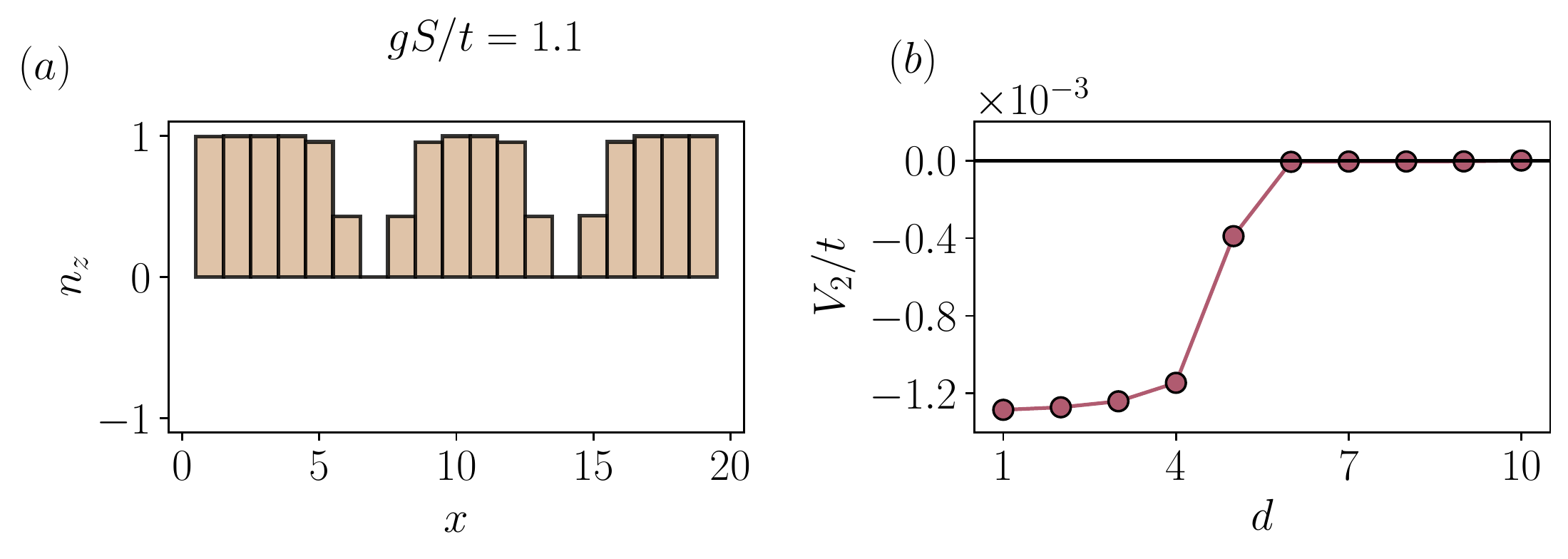}
\caption{\label{fig:real_space_instability} {\bf Static quark-bag  potential:} {\bf (a)} Real-space configuration of the N\'eel field $n_z(x)$ in the ground state of a chain with $N_{\rm s}=41$ sites and $N_{\rm f}=23$ fermions, this is, two fermions above half filling, for $gS/t=1.1$, $h_{\ell}S/t = 0.4$ and $h_{\mathsf{t}}S/t = 0.01$. The two quark-bag quasi-particles are pinned at a certain distance $d$ (see main text). {\bf (b)} The potential energy between them decreases with the distance, and the minimum is reached when they are located at neighbouring sites.}
\end{figure}

To substantiate this neat picture and connect it to the quark confinement,  we should provide evidence that the integer-valued charges shown in Fig.~\ref{fig:real_space_deconfinement}{\bf (e)} are the result of an attractive force between the fractionally-charged quasiparticles of Figs.~\ref{fig:real_space_deconfinement}{\bf (a)-(b)}. This evidence of confinement is supported by the numerical results presented in Fig.~\ref{fig:real_space_deconfinement}{\bf (f)}, where we show that the static quark potential increases with the inter-quark distance in this case. Hence, as advanced in the introduction, it is possible to understand the microscopic confinement mechanism  in our model.  In the regime $g<g_{\rm c}$, the outward pressure of the quarks overcomes the inner force that tends to re-establish the homogeneity of the condensate, such that the quarks get deconfined and can move synchronous with the kin/antikink. This changes for $g>g_{\rm c}$, where the inward  force to reestablish condensate homogeneity  prevails, and the quarks get confined within the so-called quark bag. For $h_{\ell}S/t = 0.4$ and $h_{\mathsf{t}}S/t = 0.01$, we estimate a critical value of  this confinement-deconfinement phase transition to be $g_{\rm c} S/t \approx 1.01$ (see App.~\ref{app:sol_deconf}).

 We note that these integer-charged excitations also occur in other QFT with  a non-classical scalar condensate due to chiral-SSB, such as the Gross-Neveu model~\cite{PhysRevD.9.3471}. However, to the best of our knowledge, there is no deconfinement transition where they become a pair of fractionally-charged fermions and, additionally, they require at least two fermion flavours to exist, which anyway masks the occurrence of fractionalization as happens for polyacetilene~\cite{PhysRevLett.42.1698,CAMPBELL1982297}.

\begin{figure}[t]
  \centering
  \includegraphics[width=1.0\linewidth]{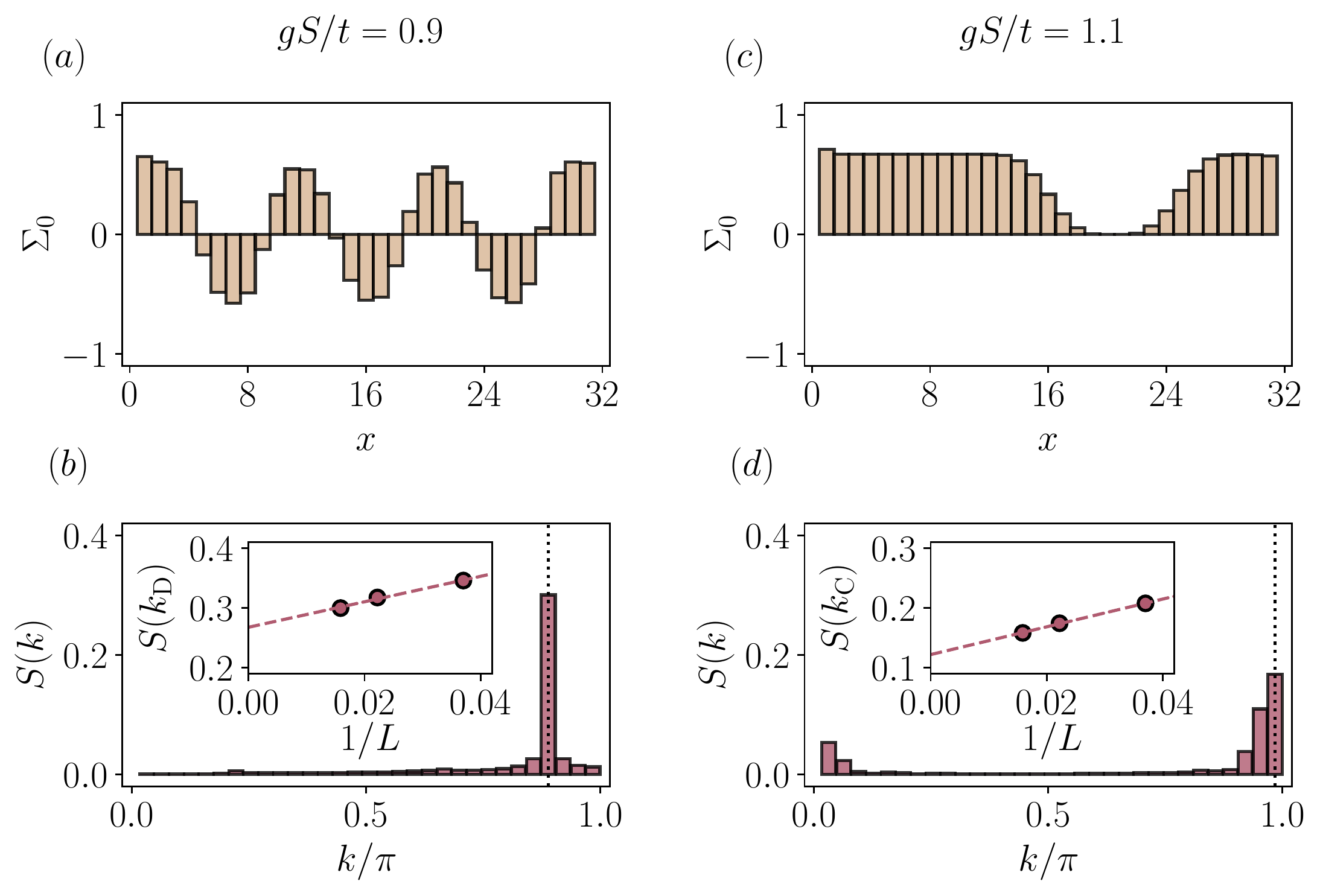}
\caption{\label{fig:real_space_finite_doping} {\bf Quark crystal and phase separation:} {\bf (a)} Space dependence of the scalar condensate  $\Sigma_0$ in the ground state of a chain with $N_{\rm s}=63$ sites and $N_{\rm f}=35$ fermions, for $gS/t=0.9$, $h_{\ell}S/t = 0.4$ and $h_{\mathsf{t}}S/t = 0.01$. For a finite density of particles above half filling, the deconfined quarks rearrange forming an ordered structure. {\bf (b)} Long-range order can be detected using the structure factor $S(k)$, which shows a peak at a momentum commensurate with the fermionic density, $k_{\rm D}/\pi = 2N_{\rm f} / N_{\rm s}$. {\bf (c)} In the confined phase, the attractive quark bags create an extensive region where the dynamical fermionic mass is screened to zero. {\bf (d)} In this case, the peak at $k_{\rm D}$ disappears, while $S(k)$ is non-zero around $k_{\rm C}=\pi$, signalling a reduced but non-vanishing AF order. The insets show the finite-size scaling of the peak in $S(k)$ for each case, where the density is kept fixed.}
\end{figure}

\vspace{0.5ex}
{\bf Quark crystals and chiral symmetry restoration.--}  Let us now move towards finite densities, and explore the rJR groundstate properties   in the confined and deconfined regimes. In the deconfined regime, we have shown that the isolated quarks repel each other, such that they will maximise the inter-quark distance and broaden the density profiles if left unpinned. Accordingly, for finite fermion densities above the half-filled vacuum, one would expect the formation of a crystalline structure of equidistant kinks and anti-kinks with the corresponding periodic distribution of fractionally-charged fermions, namely a {\it quark crystal}. 

On the other hand, in the confined regime, we have no a priori intuition  of the possible groundstate ordering. To gain such intuition, let us first  calculate the static potential between two distant quark-bag excitations, each of which contains a pair of confined quarks and thus an integer-charged fermion. The static bag potential  can be obtained using the MPS numerics by considering in this case the energy of the system doped with a pair of fermions that are held inside the bags $E_2(d)$, and  pinned at a distance $d$, $V_2(d)=E_2(d)-E_0$, where $E_0$ is again the energy of two quark bags pinned far apart. In order to fix the quark-bag distance as depicted in Fig.~\ref{fig:real_space_instability}{\bf (a)}, we impose again an external potential that localises them at two given locations (see App.~\ref{app:sol_deconf}). Our numerical results show that the static fermion-bag potential  of Fig.~\ref{fig:real_space_instability}{\bf (b)} increases with the inter-bag distance, proving that the quark-bags attract each other. Therefore, if left unpinned, we expect that the two bags will merge yielding a wider depletion  region of the condensate that can accommodate two integer-charged fermions, each of which can be thought of being composed of  two confined quarks.  This trend can be generalised to  finite density regimes, where we expect the appearance of an extensive quark bag that is sufficiently wide to host all  of the extra fermions. In the context of ultracold atoms, this phase can be understood as a phase separation phenomenon.

Let us now confirm this intuition by presenting the MPS numerics for the finite-density regime. In Fig.~\ref{fig:real_space_finite_doping}{\bf (a)} and {\bf (c)}, we display the real-space dependence of the scalar condensate for the deconfined and confined regimes without the pinning potentials. As can be clearly observed,  Fig.~\ref{fig:real_space_finite_doping}{\bf (a)} presents a periodic sequence of   kinks and antikinks, each of which hosts a single localised quark, giving rise to the aforementioned quark crystal. On the other hand, as we increase the coupling strength, Fig.~\ref{fig:real_space_finite_doping}{\bf (c)} displays an extensive region where the N\'eel field vanishes, and the dynamical fermionic mass is screened to zero. This wide bag accommodates for all the extra fermions, leading to a phase separation with respect to the region where the vacuum displays a large dynamically-generated mass inhibiting the penetration of the massless confined quarks. 

We note that the corresponding  phases can be quantitatively distinguished by means of the static spin structure factor $S(k)=\frac{1}{N^2_{\rm s}}\sum_{i,j}\langle S_i^zS^z_j\rangle\ee^{\ii k(i-j)}$, which will peak at different momenta  $k_{\rm D}$, $k_{\rm C}$ for the deconfined/confined phases. For the deconfined quark crystal, Fig.~\ref{fig:real_space_finite_doping}{\bf (b)} shows that the corresponding  peak of the structure factor occurs for  a momentum that is commensurate with the fermionic density modulation of the scalar condensate $k_{\rm D} = 2\pi N_{\rm f} / N_{\rm s}$. Conversely, for the confined phase-separated bag phase,  Fig.~\ref{fig:real_space_finite_doping}{\bf (d)} shows that the peak at $k_{\rm D}$ vanishes, and one gets instead a non-zero  structure factor around $k_{\rm C}=\pi$, signalling  N\'eel order, which is partially broadened by the condensate deformation due to the quark bag. The inset of both figures displays the finite-size scaling of each peak, where we increase both the size $N_{\rm s}$ and the number of fermions $N_{\rm f}$, such that the density $\rho=N_{\rm f}/N_{\rm s}$ remains fixed. The extrapolated non-zero values of the corresponding peaks   for $1/L\to0$ show that the quark-crystal and phase-separated bag phases are both stable in the thermodynamic limit.

The vanishing value of of the structure factor at momentum $k=\pi$ in the quark crystal suggests the possibility of a zero-temperature \textit{quantum chiral symmetry restoration} for finite dopings\textemdash since this quantity is equivalent to the N\'eel order parameter  $n_z$ used in our discussion of thermal chiral symmetry restoration for the rJR vacuum.  This is confirmed by calculating the average value of the fermionic condensate $\overline{\Sigma}_0 = \frac{1}{N_s}\sum_i \Sigma_0(i)$, where we get $\overline{\Sigma}_0=0.06$ and $\overline{\Sigma}_0=0.48$ for the parameters used in Fig.~\ref{fig:real_space_finite_doping}{\bf (a)} and Fig.~\ref{fig:real_space_finite_doping}{\bf (b)}, respectively. In this case, it is not the thermal fluctuations, but instead the finite density of topological solitons, which reduces the average value of the fermionic condensate to zero (up to finite-size effects), indicating that chiral symmetry is restored in the quark-crystal phase. In this phase, therefore, chiral symmetry coexists with deconfinement. The situation is analogous in QCD, where both properties appear in the quark-gluon plasma. The presence of a single transition from this phase to a confined symmetry-broken phase, or the possibility of intermediate phases with only one of these properties, is still an open question in particles physics~\cite{McLerran_2007}. The investigation of such interplay in simple models could help to gain a better understanding of it in more complicated theories, specially in regimes where the chemical potential is large and Monte Carlo simulations suffer from the sign problem~\cite{Brambilla_2014}.

\begin{figure}[t]
  \centering
\includegraphics[width=0.8\linewidth]{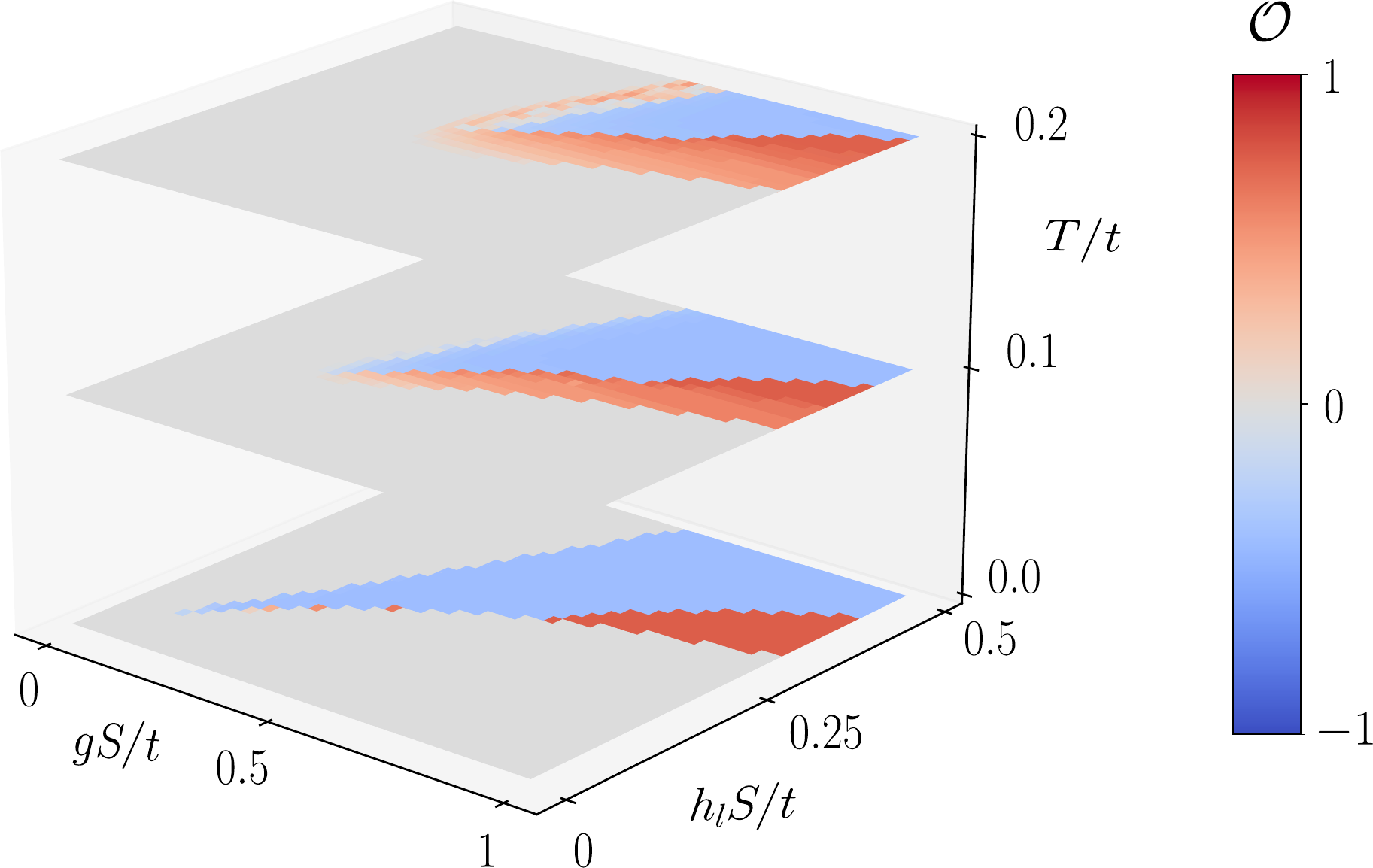}
\caption{\label{fig:finite_T_finite_doping} {\bf Deconfinement and chiral symmetry restoration:} Phase diagram for different values of $T/t$ obtained in the HF approximation for a periodic chain with $N_{\rm s} =64$ sites and $N_{\rm f} = 35$ fermions, with $h_{\mathsf{t}}S = 0.01t$. The order parameter $\mathcal{O}$ (see main text) allows us to distinguish between the three different phases that appear for a finite doping. As the temperature increases, the phase transition line between the quark-bag and the quark-crystal phases gets modified, until both of them disappear at a sufficiently high value of $T/t$.}
\end{figure}

Let us thus explore this interplay in the presence of thermal fluctuations. Figure~\ref{fig:finite_T_finite_doping} depicts the HF phase diagram for a finite density of fermions over the half-filled vacuum for different values of $T/t$. To distinguish the two phases, we use as an order parameter the difference between the structure factor at  the two different momenta of the confined and deconfined phases, $\mathcal{O} = S(k_{\rm C}) - S(k_{\rm D})$. This quantity is zero in the disordered paramagnetic phase, which corresponds to a Luttinger liquid, as in the case of half filling. Positive and negative finite values of $\mathcal{O}$ corresponds, on the other hand, to the quark-bag and quark-crystal phases, respectively. For $T/t = 0$, we can clearly distinguished these three phases. Note that, it is only in the ordered phases, surrounded by a disorder LL, where the notion of confinement and deconfinement of fractionally-charged quasi-particles is well defined. Within this region, we observe that both the quark-bag and the quark-crystal phases have a finite extension, with a phase transition line separating them. The ordered region shrinks as the temperature increases (Fig.~\ref{fig:finite_T_finite_doping}). It is interesting to notice that, in our model, the quark crystal disappears more rapidly than the quark-bag phase.

\begin{figure}[t]
  \centering
  \includegraphics[width=1.0\linewidth]{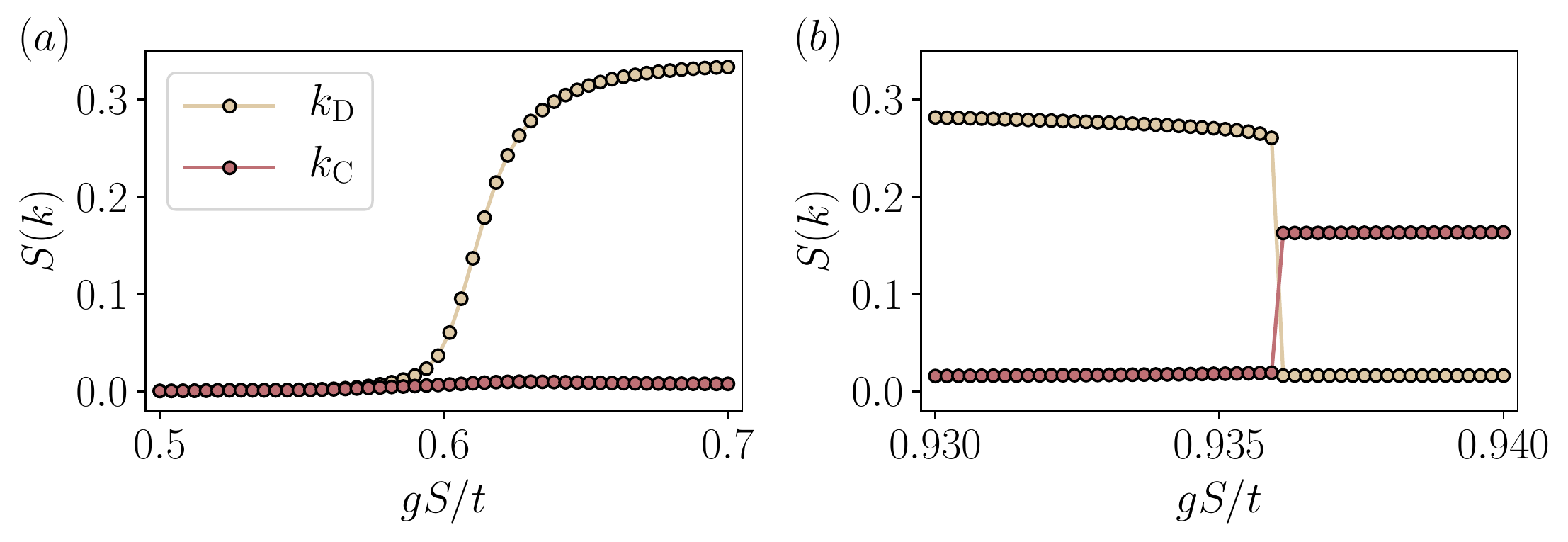}
\caption{\label{fig:phase_transitions_finite_doping} {\bf Deconfinement phase transitions:} Structure factor at $k_{\rm D}=2\pi N_{\rm f} / N_{\rm s}$ and $k_{\rm C}=\pi$ as a function of $gS/t$ for a chain with $N_{\rm s}=63$ sites and $N_{\rm f}=35$ particles, at $h_{\ell}S = 0.4t$ and $h_{\mathsf{t}}S = 0.01t$. {\bf (a)} In this case, $S(k_{\rm D})$ serves as an order parameter, signalling a direct transition between a disorder LL and an ordered quark crystal, while $S(k_{\rm C})$ remains zero all the way. {\bf (b)} At higher values of $gS/t$, we observe a deconfinement-confinement transition between the two ordered phases, the quark crystal and the quark bag, where both $S(k_{\rm D})$ and $S(k_{\rm C})$ are the respective order parameters.}
\end{figure}

{\bf Confinement-deconfinement phase transition.--} As we have shown in the previous sections, a characteristic feature of our (1+1)-dimensional QFT~\eqref{eq:rotor_JR} is the possibility to understand  the mechanism of confinement microscopically and, moreover, the existence of a deconfinement quantum phase transition as we vary the microscopic parameters. In order to study the latter in more detail, we make use again of the static structure factor peaks, which serve as order parameters to detect the corresponding phase transitions. In this section, we study two different types of phase transitions occurring at finite densities with DMRG, confirming the results obtained above using the HF method. The first one, corresponding to Fig.~\ref{fig:phase_transitions_finite_doping}{\bf (a)}, describes the transition between the quark crystal and a longitudinal paramagnetic phase. As shown in this figure, following the the spin structure factor at the two characteristic momenta $k_{\rm D}$ and $k_{\rm C}$, one can see that we start from a disordered phase at small interactions, where both $S(k_{\rm D})$ and $S(k_{\rm C})$ are zero. As $gS/t$ increases, the quark crystal order parameter reaches a non-zero value $S(k_{\rm D})>0$, while $S(k_{\rm C})$ remains zero. This phase transition is continuous, similarly to the direct order-disorder transition we found for half filling. For larger values of $gS/t$, we find another phase transition, again in correspondence with the HF results. Fig.~\ref{fig:phase_transitions_finite_doping}{\bf (b)} shows the transition between the deconfined quark crystal $S(k_{\rm D})>0,S(k_{\rm C})=0$ towards the confined  quark-bag phase  $S(k_{\rm D})=0,S(k_{\rm C})>0$. In this case, this deconfinement transition is  a first-order phase transition. This is believed to be the case also for the confinement-deconfinement transition in QCD at finite chemical potential. For $h_{\ell}S/t = 0.4$ and $h_{\mathsf{t}}S/t = 0.01$, this transition is located at $g_{\rm c} S / t = 0.94$, which  roughly agrees with the prediction we obtained using the static quark potential (i.e. $g_{\rm c} S / t = 1.01$). The difference shows the presence of many-body effects in the case of a quark crystal, where the interaction between two quarks is influenced by the presence of a finite density of them. This agreement supports our claim that the mechanism behind the transition between a quark crystal and a quark-bag phase is the confinement of quark-like fractional quasi-particles.

\vspace{0.5ex}
{\bf Conclusions and outlook.--} In this work, we studied several high-energy non-perturbative phenomena using a neat (1+1) quantum field theory, the rotor Jackiw-Rebbi model, and proposed a quantum simulation scheme using a Fermi-Bose mixture of ultracold atoms in an optical lattice.  Dirac fermions, whose interactions are mediated by spin-$S$ rotors in this model, acquire a dynamical mass through the spontaneous breaking of chiral symmetry. The generation of a mass term in the theory is accompanied here by antiferromagnetic order in the rotors, that we predict analytically in the large-$S$ limit of the continuum model. Using a lattice model that regularises the theory, we studied the phase diagram at half filling in the presence of quantum and thermal fluctuation, showing how dynamical mass generation, in particular, survives in the ultimate quantum limit of $S=1/2$. We also showed how, in this limit, the chiral symmetry breaking quantum phase transition lies in the Ising universality class and,  moreover, we observed  chiral  symmetry restoration at sufficiently high temperatures.

We then focused on the regime of finite chemical potentials, where we find a confinement-deconfinement phase transition between quark bags and a crystal of fractional quark-like quasi-particles. This transition is characterised using the spin structure factor, and its microscopic origin is uncovered by means of the static quark potential. We also showed how deconfinement coexists with a restoration of chiral symmetry, even in the absence of thermal fluctuations. In this case, the latter occurs due to a proliferation of topological solitons at finite densities, drawing an interesting analogy to the quark-gluon plasma of QCD. Our results indicate how confinement between fractional charges could be further investigated in atomic experiments, shedding light into one of the long-standing questions of particle physics.

 In the future, it would be interesting to extend the large-$S$ techniques to a different family of  lattice models that allow for symmetry-protected topological phases, and which have been studied  so far for bosonic matter coupled to link spins in the $S=1/2$ limit~\cite{PhysRevB.99.045139,Gonzalez-Cuadra2019,1908.02186,2003.10994}, or for fermionic matter coupled to  auxiliary scalar~\cite{BERMUDEZ2018149,PhysRevB.99.125106,PhysRevX.7.031057} or gauge~\cite{2002.06013,PhysRevD.99.014503,PhysRevB.100.115152} fields, which can be substituted by the N\'eel and canting fields hereby introduced. These extensions would allow to explore the interplay of confinement-deconiment transitions and topological phases of matter in a completely new framework. It would also be interesting to use matrix-product-state simulations to study real-time dynamics, serving as alternative benchmark for quantum simulations in addition to the finite-density regime hereby studied, where Monte Carlo simulations for a single flavour of fermion fields are expected to suffer from the sign problem. The model can also be easily extended to higher dimensions, where the simulation proposal can be generalized in a straightforward manner by using higher-dimensional optical lattices, which would also reach the limits of efficient tensor-network numerical techniques. It is precisely in the cases where numerical simulations show limitations where cold atoms represent an efficient alternative to provide a full solution of the quantum many-body problem.

\begin{acknowledgements}

This project has received funding from the European Union Horizon 2020 research and innovation programme under the Marie Sk\l{}odowska-Curie grant agreement No 665884, the Spanish Ministry MINECO and State Research Agency AEI (FIDEUA PID2019-106901GB-I00/10.13039 / 501100011033, SEVERO OCHOA No. SEV-2015-0522, FPI), European Social Fund, Fundaci\'o Cellex, Fundaci\'o Mir-Puig, Generalitat de Catalunya (AGAUR Grant No. 2017 SGR 1341, CERCA program, QuantumCAT $\_$U16-011424, co-funded by ERDF Operational Program of Catalonia 2014-2020), MINECO-EU QUANTERA MAQS (funded by The State Research Agency (AEI) PCI2019-111828-2 / 10.13039/501100011033), and the National Science Centre, Poland-Symfonia Grant No. 2016/20/W/ST4/00314.  A.D. akcnowledges the Juan de la Cierva program (IJCI-2017-33180) and the financial support from a fellowship granted by la Caixa Foundation (fellowship code LCF/BQ/PR20/11770012). A.B. acknowledges support from the Ram\'on y Cajal program RYC-2016-20066, and CAM/FEDER Project S2018/TCS- 4342 (QUITEMAD-CM) and the ''Plan Nacional Generaci\'on de Conocimiento PGC2018-095862-B-C22. M.A. was supported by the Deutsche Forschungsgemeinschaft (DFG, German Research Foundation) via Research Unit FOR 2414 under project number 277974659 and Germany's Excellence Strategy - EXC-2111 - 39081486 and by the European Union via the ERC Starting grant LaGaTYb (grant agreement number 803047).

\end{acknowledgements}

\appendix

\section{Implementation with cold atoms}
\label{app:implcoldatoms}
We propose to realize the $S=\half$ limit of the lattice Hamiltonian~\eqref{eq:spin_fermion_lattice} with a mixture of bosonic and spinfull fermionic atoms. Notice that the roles of bosonic and fermion degrees of freedom are interchanged here. In particular, we propose to use hard-core bosons to simulate the dynamical fermionic matter, while the fermionic ones will be used to implement the rotor fields. As will become clear below, this approach is motivated by the specific choice of  bosonic and  fermionic species, which presents a well-characterised  Feshbach resonance that will allow an accurate experimental control of the inter-species scattering, the crucial ingredient in our scheme.

 {\it Bose-Fermi Hamiltonian.--} We aim at realizing the following grand-canonical Hamiltonian for the Bose-Fermi mixture
\begin{equation}
{H}= {H}_b + {H}_f + {H}_{bf},
\end{equation}
where ${H}_b$ describes the dynamics of the bosonic atoms of mass $m_b$, ${H}_f$ the one of the fermionic atoms of mass $m_f$, and $ {H}_{bf}$ the interaction between the two species. 
The bosonic part of the Hamiltonian is defined as follows
\begin{align}
{H}_b = &- t \sum_i \left( {b}^\dagger_i {b}^{\phantom{\dagger}}_{i+1} + \text{h.c.} \right)\nonumber  + \frac{U}{2} \sum_i {b}^{\dagger}_i {b}_i^{\dagger}{b}_i{b}^{\phantom{\dagger}}_i - \sum_i\mu_{i} {b}_i^{\dagger}{b}^{\phantom{\dagger}}_i,
\end{align}
where ${b}_i$ and ${b}^{\dagger}_i$ are the bosonic creation and annihilation operators acting on lattice site $i$. This grand-canonical Hamiltonian describes the tunnelling of bosonic atoms in a 1D lattice with strength $t$, and the Hubbard interactions with energy $U$. Here, $\mu_i=\mu-V_{b,i}$ is expressed in terms of the chemical potential $\mu$ and the  on-site optical trapping potential $V_{b,i}$, and   controls the bosonic filling in the local density approximation. The fermionic contribution can be divided in two parts ${H}_f ={H}^m_f+{H}^i_f $, where the first term describes the external motional degrees of freedom
\begin{align}
\label{eq:H_mot_fermions}
{H}_f^m = -t_f& \sum_{i,\sigma} \left( {f}^\dagger_{i,\sigma} {f}_{i+1,\sigma}^{\phantom{\dagger}} + \text{H.c.} \right)+ U_{\uparrow\downarrow} \sum_i  {f}^{\dagger}_{i,\downarrow} {f}_{i,\uparrow}^{\dagger}{f}^{\phantom{\dagger}}_{i,\uparrow}{f}^{\phantom{\dagger}}_{i,\downarrow}\nonumber \\ &  -\sum_i \mu_{i,\uparrow} {f}_{i,\uparrow}^{\dagger}{f}_{i,\uparrow}^{\phantom{\dagger}}- \sum_i\mu_{i\downarrow} {f}_{i,\downarrow}^{\dagger}{f}^{\phantom{\dagger}}_{i,\downarrow}.
\end{align}
Here, ${f}^{\phantom{\dagger}}_{i\sigma}$ and ${f}^{\dagger}_{i\sigma}$ are the fermionic creation/annihilation operators acting on lattice site $i$ with internal state $\sigma=\{\uparrow,\downarrow\}$. Fermions are trapped in a very deep optical lattice, such that $t_{f}\ll t$, and we can neglect their  tunnelling along the 1D lattice during the timescale of interest.  They interact with Hubbard interaction $U_{\uparrow\downarrow}$, and their filling is controlled by the  local chemical potentials  $\mu_{i\sigma}=\mu_\sigma-V_{f,i}$, where $\mu_\sigma$ is the  chemical potential for the fermionic atoms in each internal state,  and $V_{f,i}$ is an optical trapping potential. In addition, the internal degrees of freedom shall be described by 
 \beq
{H}_{f}^i=\sum_{i,\sigma} \! \epsilon_{ \sigma}{f}_{i,\sigma}^{\dagger}{f}^{\phantom{\dagger}}_{i,\sigma}+\sum_i\big(\Omega_{\rm d}\cos{\omega_{\rm d}t}{f}_{i,\uparrow}^{\dagger}{f}^{\phantom{\dagger}}_{i,\downarrow}+{\rm H.c.}\big),
\eeq
where we have introduced the atomic energy levels $\epsilon_{ \uparrow},\epsilon_{ \downarrow}$ for the two states of the fermionic species, and a local driving of frequency $\omega_{\rm d}$ that induces local oscillations between these two states with a Rabi frequency $\Omega_{\rm d}$, where $\hbar=1$ henceforth. As will become clear below, this driving stems from  radio-frequency radiation, which has a negligible momentum, and one can thus neglect recoil effects that would couple the internal and external degrees of freedom. Finally, the interaction between the two species  is
\begin{equation}
\hat{H}_{bf}= \sum_i {b}^{\dagger}_i {b}^{\phantom{\dagger}}_i \left(U_{b\uparrow} {f}^{\dagger}_{i,\uparrow} {f}^{\phantom{\dagger}}_{i,\uparrow}+U_{b\downarrow} {f}^{\dagger}_{i,\downarrow} {f}^{\phantom{\dagger}}_{i,\downarrow}\right),
\end{equation}
which describes the on-site interaction between bosonic and fermionic atoms, and depends on the internal spin state of the fermions, i.e. in general $U_{b\uparrow} \neq U_{b\downarrow}$. As will be clear for the particular Bose-Fermi mixture discussed below, there are also  boson-fermion scattering processes where the internal states of the fermions is changed by populating other bosonic states such that the total angular momentum along the quantisation axis is conserved. Nonetheless, these so-called spin-flipping collisions of strength $U_{\rm sf}$  are negligible for a sufficiently-large  difference of the on-site energies $U_{\rm sf}\ll |\epsilon_{ \uparrow}-\epsilon_{ \downarrow}|$.

\begin{figure}[t]
  \centering
  \includegraphics[width=1.0\linewidth]{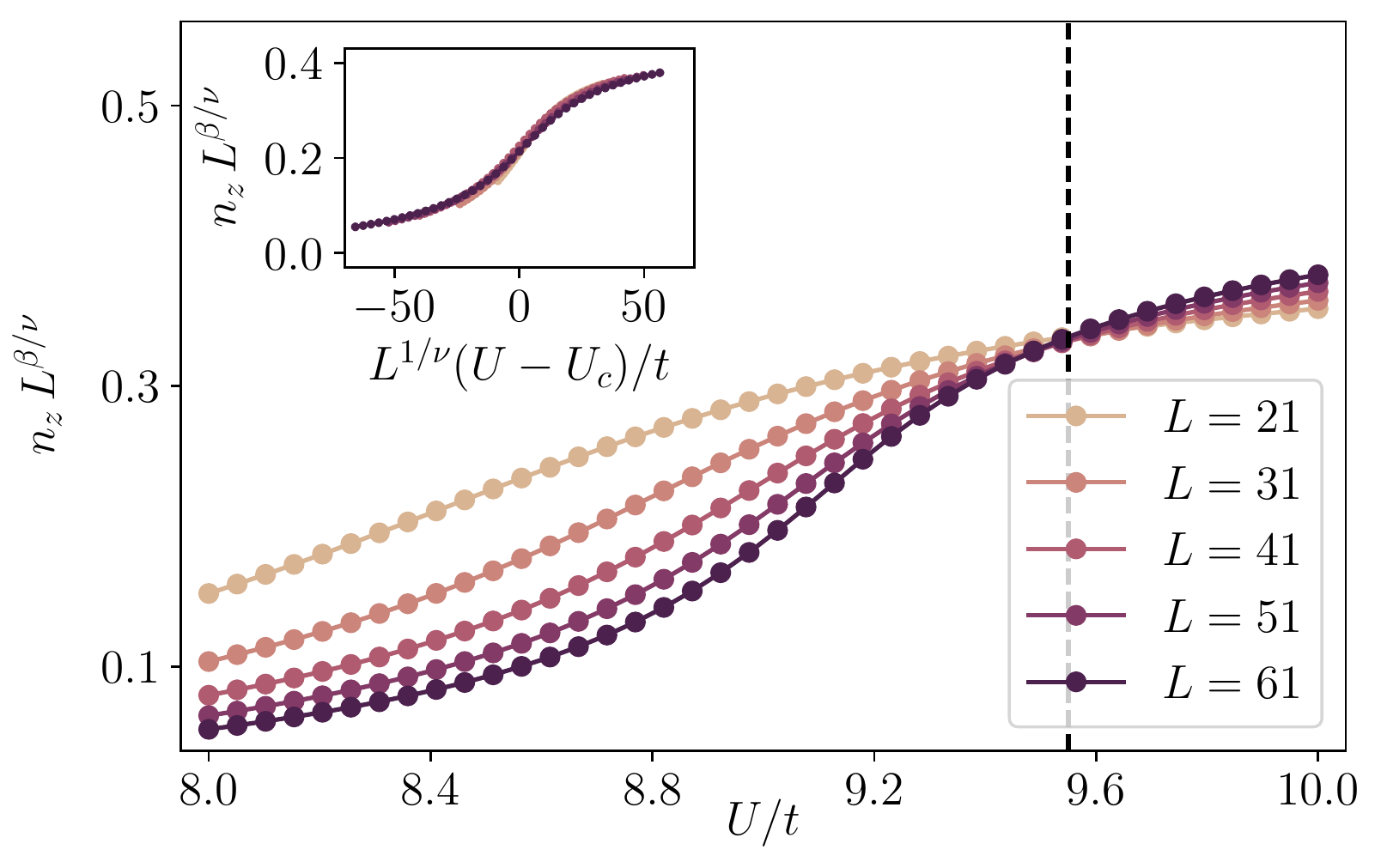}
\caption{\label{fig:scaling_U} {\textbf{Bosonic mass generation: } Order parameter $n_z$ in terms of $U/t$ for the ground state of a finite chain of different lengths $L$, with $gS/t = 1.0$, $h_{\ell}S=0.5t$ and $h_{\mathsf{t}}S/t = 0.05$. The different lines cross at the critical point, consistent with a phase transition in the Ising universality class ($\nu = 1$, $\beta = 1/8$). The inset shows how the rescaled lines collapse into a single one.}}
\end{figure}

 Let us now discuss the steps to arrive at the desired model~\eqref{eq:spin_fermion_lattice}, and find the relation between the model microscopic couplings and the cold-atom experimental parameters.  Using two internal states $\ket{\uparrow}$ and $\ket{\downarrow}$ with different magnetic moments, the energy splitting  corresponds to the Zeeman energy and can be tuned via an external magnetic. The driving frequency will be near-resonant and fulfils  $\omega_{\rm d}=(\epsilon_{\uparrow}-\epsilon_{\downarrow})-\Delta_{\rm d}$, where $\Delta_{\rm d} $ is the so-called detuning. Let us note at this point that the  optical trapping potential $V_{f,i}$ of Eq.~\eqref{eq:H_mot_fermions}  is assumed to be state-independent, which is generally the case, such that it does not modify the above resonance condition, and need not be included in the present discussion.   Moving to an interaction picture with respect to 
 \beq
 H_{f,0}^i=\sum_{i,\sigma} \! \epsilon_{ \sigma}{f}_{i,\sigma}^{\dagger}{f}^{\phantom{\dagger}}_{i,\sigma},
 \eeq
  the fermionic operators become  $f^{\phantom{\dagger}}_{i,\sigma}\to f^{\phantom{\dagger}}_{i,\sigma}(t)=f^{\phantom{\dagger}}_{i,\sigma}\ee^{-\ii\epsilon_\sigma t}$. Furthermore, assuming that $|\Omega_{\rm d}|,|\Delta_{\rm d}|\ll\omega_{\rm d}$, we can perform a rotating-wave approximation such that 
  \beq
{H}_{f}^i(t)\approx\sum_i\left(\frac{\Omega_{\rm d}}{2}\ee^{\ii \Delta_{\rm d}t}{f}_{i,\uparrow}^{\dagger}{f}^{\phantom{\dagger}}_{i,\downarrow}+{\rm H.c.}\right).
\eeq
Finally, by  moving to a  frame that rotates with the drive frequency, this Hamiltonian can be written as the following time-independent term 
\beq
\hat{H}_{f}^i=\sum_i\left(\frac{\Delta_{\rm d}}{2}\big(\hat{f}_{i,\uparrow}^{\dagger}\hat{f}^{\phantom{\dagger}}_{i,\uparrow}-\hat{f}_{i,\downarrow}^{\dagger}\hat{f}^{\phantom{\dagger}}_{i,\downarrow}\big)+\frac{\Omega_{\rm d}}{2}\big(\hat{f}_{i,\uparrow}^{\dagger}\hat{f}^{\phantom{\dagger}}_{i,\downarrow}+\hat{f}_{i,\downarrow}^{\dagger}\hat{f}^{\phantom{\dagger}}_{i,\uparrow}\big)\right).
\eeq
 where we define the Bloch sphere in a way that  $\Omega_{\rm d}\in\mathbb{R}$, i.e. we use the phase of the driving as a reference for subsequent measurements. Here, we  have introduced the following relation between the fermionic operators in the original Schrodinger picture $f^{\phantom{\dagger}}_{i,\sigma}$, and those $\hat{f}^{\phantom{\dagger}}_{i,\sigma}$ in the  rotating frame
 \beq
 \begin{split}
 \hat{f}^{\phantom{\dagger}}_{i,\uparrow}=f^{\phantom{\dagger}}_{i,\uparrow}\ee^{-\ii\left(\epsilon_\uparrow-\frac{\Delta_{\rm d}}{2}\right) t},\\
  \hat{f}^{\phantom{\dagger}}_{i,\downarrow}=f^{\phantom{\dagger}}_{i,\downarrow}\ee^{-\ii\left(\epsilon_\downarrow+\frac{\Delta_{\rm d}}{2}\right) t},
   \end{split}
 \eeq
 
We can now define the following spin-$\half$ operators in terms of these fermionic annihilation and creation operators
\begin{equation}
\begin{split}
S_i^z&=\half(\hat{f}_{i,\uparrow}^{\dagger}\hat{f}^{\phantom{\dagger}}_{i,\uparrow}-\hat{f}_{i\downarrow}^{\dagger}\hat{f}^{\phantom{\dagger}}_{i,\downarrow})=\half({f}_{i,\uparrow}^{\dagger}{f}^{\phantom{\dagger}}_{i,\uparrow}-{f}_{i,\downarrow}^{\dagger}{f}^{\phantom{\dagger}}_{i,\downarrow}), \\
 S_i^x&=\half(\hat{f}^{\dagger}_{i,\uparrow}\hat{f}^{\phantom{\dagger}}_{i,\downarrow}+\hat{f}^{\dagger}_{i,\downarrow}\hat{f}^{\phantom{\dagger}}_{i,\uparrow})=\half({f}^{\dagger}_{i,\uparrow}{f}^{\phantom{\dagger}}_{i,\downarrow}\ee^{\ii\omega_{\rm d}t}+{\rm H.c.}).
\end{split}
\end{equation}
Therefore, to measure the spin operators discussed in the main text, one would have to lock the phase evolution to the one set by the source that drives the transition.

 After these derivations, we should enforce that  only 1 fermion resides at each lattice site, which can be adjusted by the filling, and maintained by working in the regime where $t_{f}\ll t$, thus suppressing double occupancies. In this case, we obtain the desired $S = \langle S_i^z\rangle=\half$ limit. Moreover, since the fermion tunneling $t_{f}$ is negligible,   super-exchange processes stemming from virtual double occupancies occurring  at order $t_f^2/U_{\uparrow\downarrow}$ are also negligible, and we can finally arrive at an effective description according to the following  grand-canonical Hamiltonian
\begin{align}
\hat{H} =& -t \sum_i \left(\hat{b}^\dagger_i \hat{b}_{i+1} + \text{h.c.} \right) + \frac{U}{2}\sum_i \hat{n}_i (\hat{n}_i - 1)- \sum_i\mu_{i} \hat{b}_i^{\dagger}\hat{b}^{\phantom{\dagger}}_i, \nonumber \\
& + g \sum_i \hat{n}_i S_i^z - \sum_i (h_{\ell} S^z_i + h_{\mathsf{t}} S_i^x),
\label{ref:eqeffspin}
\end{align}
where the boson operators are not altered by moving to the rotating frame  $\hat{b}_i=b_i$. In the Hamiltonian above, the coupling constant is defined as
  \beq
  \label{eq:g_param}
 g=2(U_{b\uparrow} - U_{b\downarrow}),
 \eeq
  and the external field given by the driving term
  \beq
  h_\ell=-\Delta_{\rm d}, \hspace{2ex}h_{\mathsf{t}}=-\Omega_{\rm d},
  \eeq
  the specifics of which are discussed below. The local chemical potential $\mu_i$ will be adjusted to vary the filling of the bosonic atoms, which is homogeneous in the central region of the trap. The variation of the filling will allow to explore the different quasi-particle regimes discussed in the main text.

\paragraph*{Softcore bosons.--} We can now address the final step in the derivation. Dispensing with the chemical potential, and working in the central region of the trap to neglect the residual on-site potentials, the  Hamiltonian associated to Eq.~\eqref{ref:eqeffspin} coincides with the  lattice model~\eqref{eq:spin_fermion_lattice} in the hardcore limit, $U/t \rightarrow \infty$, where bosons can be mapped to fermions via the Jordan-Wigner transformation. Nevertheless, we emphasize that the phases investigated in the main text appear also away from this singular point, i.e. for finite values of $U/t$, provided that the Hubbard interactions are sufficiently strong. This is true in particular for the AF phase found in the fermionic case, which survives away from the hardcore limit, giving rise to the generation of a dynamical mass for strongly-correlated bosons. In Fig.~\ref{fig:scaling_U}, we represent the finite-size scaling for the N\'eel order parameter, and find that the disorder-order transitions occurs for interactions of the order $U_c \approx 10t$. Below this value the soft-core bosons are in a superfluid state, while the spins form a longitudinal paramagnet, similar to the disorder phase in the hardcore limit. Moreover, we find the the phase transition is also in the Ising universality class.

\begin{figure}[t]
  \centering
  \includegraphics{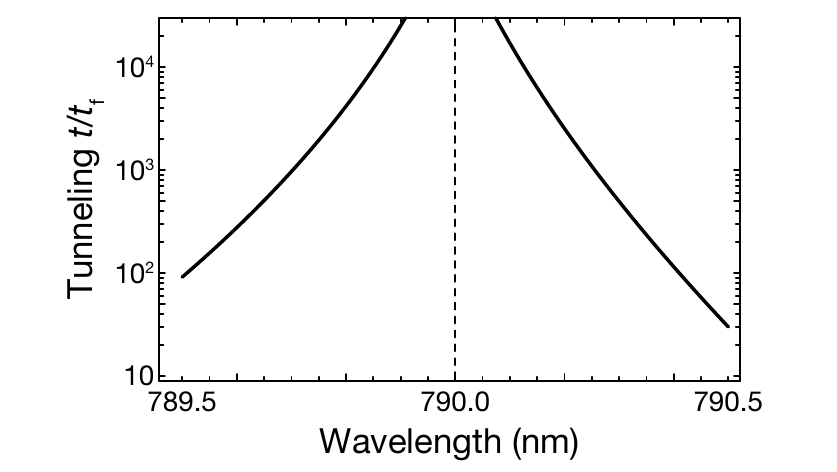}
\caption{\label{fig:Tratio} {\bf Tunnelling ratio $t/t_f$ as a function of the lattice wavelength}. The zero crossing of the polarizability of Rb at $\lambda_{bz}$ is indicated by the dashed vertical line. For the numerical calculation, we have chosen a fixed laser intensity for the lattice, and have tuned it to a regime where $t_f/h\simeq 5\,$mHz. Since $t_f$ is more or less constant throughout the whole regime, $t/t_f$ illustrates the range of available tunneling couplings for $t$. This can be optimized further by going to shallower lattice depths.}
\end{figure}

Once all the steps for the derivation of the target model~\eqref{ref:eqeffspin}  have been discussed for a generic cold-atom setting, let us estimate the specific parameters for a mixture of bosonic $^{87}$Rb and spinfull fermionic $^{40}$K atoms, and discuss the viability of the experimental realisation. The following main ingredients are relevant for the  implementation:

\paragraph*{Optical lattice potential.--}
The dominant transitions of the two alkali atoms, bosonic $^{87}$Rb and fermionic $^{40}$K are 
\begin{align}
\lambda_{\text{Rb,D2}}\simeq 780\,\textrm{nm}, \quad &\lambda_{\text{Rb,D1}} \simeq 795\,\textrm{nm} \\ \nonumber
\lambda_{\text{K,D2}}\simeq 767\,\textrm{nm}, \quad &\lambda_{\text{K,D1}} \simeq 770\,\textrm{nm}
\end{align}
which allows for widely tunable polarizabilities and optical-lattice potentials. In order to achieve a large separation of timescales, it is desirable to achieve a large ratio of the tunnelling couplings $t/t_f$, where $t$ is the strength of the tunnel coupling for the bosonic species, and $t_f$ denotes the spin-independent tunnelling amplitude of the fermionic species. In essence, this means that the lattice potential experienced by the bosonic species should be much weaker compared to the one seen by the fermionic atoms. In order to reduce off-resonant photon scattering, which would result in additional heating, at the same time the detuning from any internal transition has to be maximized. Due to the large fine-structure splitting of Rb, there is a convenient tuning range around the zero crossing of the polarizability at $\lambda_{bz}=790\,$nm. This range offers both a wide tunability of the tunnelling ratio $t/t_f$ (Fig.~\ref{fig:Tratio}) and a large detuning from all resonances to minimize heating.

\begin{figure}[t]
  \centering
  \includegraphics{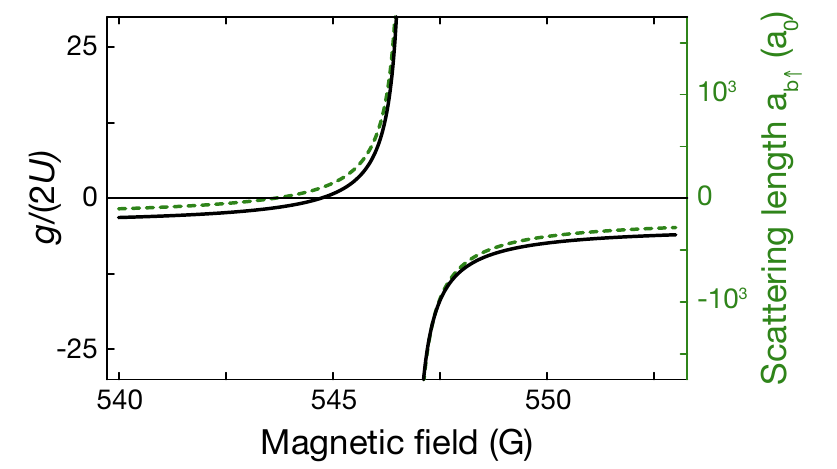}
\caption{\label{fig:gU} {\bf Ratio of interactions strengths $gS/U$ as a function of magnetic field}. The tunability here stems from the tunability of the interspecies scattering length $a_{b\uparrow}$, here  shown in green and defined in Eq.~(\ref{eq:FR}).}
\end{figure}

\paragraph*{Tunable interspecies interactions.--}
In order to provide good control over the parameter $g$~\eqref{eq:g_param} that appears in the effective Hamiltonian~(\ref{ref:eqeffspin}), we propose to make use of the well-calibrated and easily-accessible interspecies Feshbach resonance between the absolute ground states $\ket{F=1,m_F=1}$ of $^{87}$Rb and $\ket{\uparrow} \equiv \ket{F=9/2,m_F=-9/2}$ of $^{40}$K~\cite{simoni_nearthreshold_2008}. Although there does not seem to be a Feshbach resonance nearby to tune the scattering length between the ground state of Rubidium and $\ket{\downarrow}\equiv \ket{F=9/2,m_F=-7/2}$ of $^{40}$K, the interaction parameter $g$ can still be fully  tuned over a wide range of values as shown in Fig.~\ref{fig:gU}, as it depends on the difference of scattering lengths. In this figure,  we have plotted the scattering length $a_{b\uparrow}$ according to 
\begin{equation}
a_{b\uparrow}(B)=a_{bg}\left( 1- \frac{\Delta}{B-B_0} \right)
\label{eq:FR}
\end{equation}
based on the theoretical values reported in Ref.~\cite{simoni_nearthreshold_2008}, i.e. $B_0=546.75(6)\,$G, $\Delta=-3.1\,$G and $a_{bg}=-189\,a_0$, where $a_0$ is the Bohr radius.The on-site interspecies interaction energy is determined by
\begin{equation}
U_{b\sigma}=\frac{2\pi \hbar^2 a_{b\sigma}}{\mu_{b\sigma}} \int |w_b(\mathbf{r})|^2 |w_f(\mathbf{r})|^2 \text{d}^3\mathbf{r},
\end{equation}
where $\mu_{b\sigma} = m_{\text{Rb}} m_{\text{K}} /(m_{\text{Rb}} +m_{\text{K}} )$ is the reduced mass and $m_{\text{Rb}} =86.9\,u$ is the mass of one $^{87}$Rb atom and $m_{\text{K}} =39.96\,u$ the mass of one $^{40}$K atom; $u$ is the atomic mass unit. The functions $w_b(\mathbf{r})$ and $w_f(\mathbf{r})$ denote the Wannier functions of Rb and K respectively, which are  different because the two species see a lattice potential of different depth. Through these expressions, we can obtain the   parameter $g$, as discussed below.

\paragraph*{Interaction ratio $gS/U$.--}
Rubidium in its absolute ground state has a scattering length of $a_{b}\simeq 100.4\,a_0$, and the on-site Hubbard interaction is defined as 
\begin{equation}
U=\frac{4\pi \hbar^2 a_{b}}{m_{\text{Rb}}} \int |w_b(\mathbf{r})|^4 \text{d}^3\mathbf{r},
\end{equation}
Neglecting the contribution from the Wannier functions, which is dependent on the lattice depth, we can achieve a wide tunability as illustrated in Fig.~\ref{fig:gU}, where we plot $gS/U \approx \frac{ a_{b\uparrow}(m_{\text{Rb}} +m_{\text{K}} )}{2 a_{b} m_{\text{K}}}$ as a function of the external magnetic field. 

\paragraph*{Coupling between spin states.--} Ket us now discuss the specifics of the driving term discussed previously. 
The external fields $h_\ell,h_{\mathsf{t}}$ can be realized with a  radio-frequency or two-photon microwave transitions at frequency $\omega_{\rm d}$ almost resonant with the Zeeman energy difference $\Delta E^Z_{\uparrow,\downarrow}=\epsilon_{i,\uparrow}-\epsilon_{i,\downarrow}$ between $\ket{\uparrow}$ and $\ket{\downarrow}$ atoms. For the Feshbach resonance shown in Fig.~\ref{fig:gU}, the resonance occurs at $B_0$, which corresponds to $\Delta E^Z_{\uparrow,\downarrow}/h\approx 80\,$MHz, where $h$ denotes Planck's constant. The energy offset $h_{\ell}$ is then realized by detuning the coupling frequency from resonance, i.e. $h_{\ell}=\Delta E^Z_{\uparrow,\downarrow}/\hbar-\omega_{\rm d}$. With single-photon transitions, Rabi frequencies $\Omega_{\rm d}/2\pi$ of several $10\,{\rm kHz}$ can be easily achieved, which corresponds to the regime $|h_{\mathsf{t}}|= |\Omega_{\rm d}|\gg t$. Moreover, the pair of states $\ket{\uparrow}$ and $\ket{\downarrow}$ is well isolated from the other internal states, even in the presence of this coupling. The energetically closest transition is between the levels $\ket{\downarrow}$ and $\ket{F=9/2,m_F=-5/2}$, which is detuned by $\sim5\,$MHz$\,\gg \Omega_{\rm d}/h$.

\begin{figure}[t]
  \centering
  \includegraphics{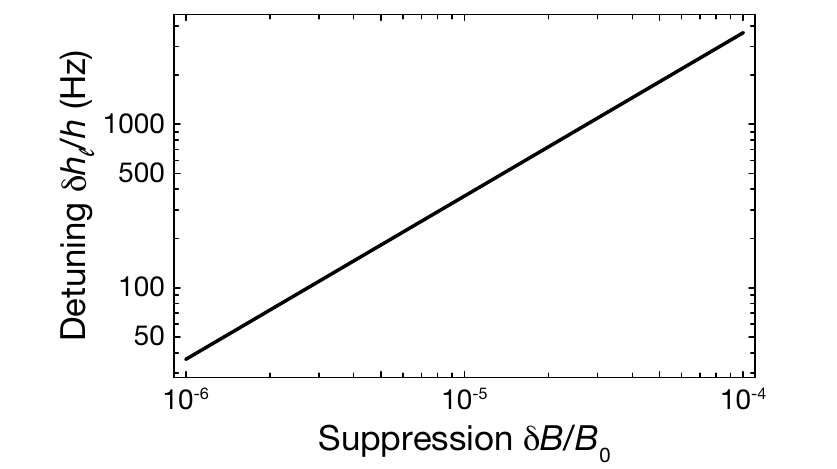}
\caption{\label{fig:dBfield} {\bf Detuning of the microwave transition $\delta h_{\ell}$.} The residual detuning is technically limited by magnetic field fluctuations $\delta B$. Estimating the residual detuning serves as a lower bound for the coupling strength $\omega$ and the control over the detuning $h_\ell$.}
\end{figure}

For some of the proposed phenomena, it is desirable to tune the coupling in order to enter the regime where $h_{\mathsf{t}}$ is on the order of the tunnel coupling $t$. The level of control one can achieve is limited by the stability of the magnetic field that defines the Zeeman shift $\Delta E^Z_{\uparrow,\downarrow}$. We have calculated the detuning from resonance $\delta h_{\ell}$ that occurs due to an imperfect control of the external magnetic field (Fig.~\ref{fig:dBfield}). It has been demonstrated that the fluctuations in the external field can be suppressed below $3\times 10^{-6}$, as reported in Ref.~\cite{johansen_testing_2017}. The most sensitive regime occurs for $h_{\ell}=0$, where the lower bound for $h_{\mathsf{t}}$ would be on the order of a few 100\,Hz, which coincides with typical experimental values for $t$. Note, that the typical timescale for these fluctuations is large compared to the duration of the experiment, hence, the detuning $\delta h_{\ell}$ can be assumed constant but will fluctuate between individual experimental realizations.

\section{Continuum limit of the Hamiltonian lattice theory}
\label{app:H_con}
In this Appendix, we present the details for the derivation of the continuum Hamiltonian field theory for the  rotor-Jackiw-Rebbi (rJR) model. Splitting the Hamiltonian in Eq.~\eqref{eq:spin_fermion_lattice} as $H=H_{\rm f}+H_{\rm sf}+H_{\rm s}$, the  bare fermion tunnelling $H_{\rm f}$ leads to a periodic band structure with a  pair of Fermi points at $\pm k_{ F}=\pm\pi/2a$. Making the long-wavelength approximation 
\beq
\label{eq:dirac_fields}
c_i\approx\ee^{-\ii k_{ F}x_i}\sqrt{a}\psi_+(x)+\ee^{+\ii k_{ F}x_i}\sqrt{a}\psi_-(x),
\eeq
yields a pair of slowly-varying fields with the correct canonical algebra in the continuum limit $a\to0$
\beq
\{\psi_\eta(x),\psi_{\eta'}(x')\}=\delta_{\eta,\eta'}\frac{\delta_{x,x'}}{a}\to\delta_{\eta,\eta'}\delta(x-x').
\eeq 
By making a gradient expansion on the slowly-varying fields $\psi_\eta(x+a)=\psi_\eta(x)+a\partial_x\psi_\eta(x)+\mathcal{O}(a^2)$, neglecting rapidly oscillating terms, and setting $a\sum_j\to\int{\rm d}x$ in the continuum limit,  the tunnelling term becomes
\beq
H_{\rm f}=\int\!{\rm d}x\!\sum_{\eta=\pm}\psi_\eta^{\dagger}(x)(-\ii\eta c\partial_x)\psi_{\eta}(x),\hspace{2ex}c=2ta,
\eeq
where $c=2ta$ plays the role of an effective speed of light. Defining the two-component spinors $\Psi(x)=(\psi_+(x),\psi_-(x))^{\rm t}$, $\overline{\Psi}(x)=(\psi^\dagger_-(x),\psi_+^\dagger(x))$, one readily sees that the Hamiltonian corresponds to a relativistic QFT of massless Dirac spinors 
\beq
H_{\rm f}=\int\!{\rm d}x\overline{\Psi}(x)\left(-\ii c\gamma^1\partial_1\right)\Psi(x)
\eeq
 in a (1+1) Minkowski spacetime $x^0=ct, x^1=x$ of metric $g_{\mu\nu}={\rm diag}(1,-1)$, where $\gamma^0=\sigma^x$, $\gamma^1=-\ii\sigma^y$  are the gamma matrices $\{\gamma_\mu,\gamma_\nu\}=2\delta_{\mu,\nu}$ in the so-called Weyl basis, and $\overline{\Psi}(x)={\Psi}^\dagger(x)\gamma^0$. In the context of lattice gauge theories, the tunnelling Hamiltonian corresponds  to the staggered-fermion discretization of the Dirac equation~\cite{PhysRevD.11.395} by  setting $t=1/2a$ and applying a Kawamoto-Smit rotation $c_j\to \ee^{\ii\pi j/2}c_j$~\cite{KAWAMOTO1981100}.
 
 Let us now turn our attention to the spin operators, which yield a $(2S+1)$-dimensional representation of the $\mathfrak{su}(2)$ algebra $\boldsymbol{S}_i\times\boldsymbol{S}_j=\ii\delta_{i,j}\boldsymbol{S}_i$. We introduce the so-called N\'eel $\boldsymbol{n}(x)$ and canting $\boldsymbol{\ell}(x)$ slowly-varying fields 
 \beq
 \label{eq:rotor_fields}
 \boldsymbol{S}_i\approx\cos( k_{ N}x_i)S\hspace{0.25ex}\boldsymbol{n}(x)+a\boldsymbol{\ell}(x),
 \eeq
 where the wave-vector $k_{N}=\pi/a$ captures the N\'eel  alternation of antiferromagnetic ordering. 
 These operators satisfy the following algebra in the continuum limit
 \beq
 \begin{split}
  \boldsymbol{\ell}(x)\times\boldsymbol{n}(x')&=\ii\delta(x-x')\boldsymbol{n}(x),\\
   \boldsymbol{\ell}(x)\times\boldsymbol{\ell}(x')&=\ii\delta(x-x')\boldsymbol{\ell}(x),\\
    \boldsymbol{n}(x)\times\boldsymbol{n}(x')&=\ii\delta(x-x')\boldsymbol{\ell}(x)(a/S)^2,\\
 \end{split}
 \eeq
 where one must consider that there is a two-site unit cell, such that the continuum limit  yields $\delta_{x,x'}/2a\to\delta(x-x')$ in this case.
In situations dominated by the N\'eel field, the contribution of the canting field will be negligible $\langle\boldsymbol{\ell}(a/S)^2\rangle\to \boldsymbol{0}$, and one obtains the algebra of position $\boldsymbol{n}$ and angular momentum $\boldsymbol{\ell}=\boldsymbol{n}\times(-\ii\boldsymbol{\nabla}_{\boldsymbol{n}})$ of a quantum-mechanical particle~\cite{HALDANE1983464,AFFLECK1985397}. Moreover, in this limit, one also finds that 
\beq
    \boldsymbol{n}(x)\cdot   \boldsymbol{n}(x)=1+1/S
\eeq
such that the particle will be confined to a unit sphere in the large-spin limit $S\gg1$. Therefore, in the large-$S$ limit, the N\'eel and canting fields represent a the orientation of a quantum rotor and its angular momentum, respectively.

Combining the expressions for the Dirac~\eqref{eq:dirac_fields} and rotor~\eqref{eq:rotor_fields} fields, and neglecting again  rapidly-oscillating terms, we find that the spin-fermion coupling can be expressed as
\beq
\label{eq:g_term}
H_{\rm sf}=\int\!{\rm d}x \overline{\Psi}(x)\boldsymbol{g}_s\cdot\left(\boldsymbol{n}(x)+\boldsymbol{\ell}(x)\gamma^0/S\right)\Psi(x),\hspace{2ex} \boldsymbol{g}_s=gS{\bf e}_z.
\eeq
The first term  can be understood as a Yukawa-like term that couples the  fermion bilinear $\overline{\Psi}(x)\Psi(x)$ to the rotor projection $n_z(x)$ instead of the standard scalar field in a Yukawa theory~\cite{Peskin:1995ev}. The second term couples the  time-like component of the fermion current $j^0(x)=\overline{\Psi}(x)\gamma^0\Psi(x)$, i.e. the charge density, to the projection of the rotor angular angular $\ell_z(x)$.

Finally, the continuum limit of the spin precession  yields  
\beq
\label{eq:h_term}
H_{\rm s}=-\int\!{\rm d}x\hspace{1ex}\boldsymbol{h}\cdot\boldsymbol{\ell}(x).
\eeq 
In analogy to the original situation~\eqref{eq:spin_fermion_lattice}, the rotor angular momentum is subjected to a magnetic field with both   longitudinal and transverse components, being the latter responsible for introducing  quantum fluctuations since $[\ell_x(x),n_z(x)]\neq 0$. Altogether, the continuum limit of the  lattice model~\eqref{eq:spin_fermion_lattice} corresponds to the quantum field theory of Eq.~\eqref{eq:rotor_JR}.

\section{Path integral formulation of the continuum QFT}
\label{app:S_con}
In this Appendix, we  provide a path-integral derivation of the continuum-limit rotor-fermion QFT~\eqref{eq:rotor_JR}. This derivation serves for two purposes: one the one hand,  it allows to clarify the absence of a topological $\theta$ term for any spin $S$,  differing markedly from $O(N)$ non-linear sigma models arising from Heisenberg models~\cite{HALDANE1983464}; on the other hand, it sets the stage for a large-$S$ limit approach to dynamical mass generation.

We are interested in the partition function  $\mathsf{Z}={\rm Tr}\{\ee^{-\beta H}\}$, where $H$ is  the spin-fermion lattice Hamiltonian~\eqref{eq:spin_fermion_lattice}. In the $\beta\to\infty$ limit of zero temperature, this partition function contains all the relevant information about quantum phase transitions related to the chiral SSB and dynamical mass generation we are seeking. Using fermionic and spin coherent states in Euclidean time $\tau=\ii t\in(0,\beta)$~\cite{fradkin_2013}, this partition function can be expressed as a functional integral in terms of anti-commuting Grassmann fields for the fermions $\psi(\tau,x_i),\psi^\star\!(\tau,x_i)$ and commuting vector fields for the spins $\boldsymbol{\Omega}(\tau,x_i)$ lying in a 2-sphere of radius $S$, which can be rescaled  in terms of unit vector fields $\boldsymbol{\Omega}(\tau,x_i)=S\boldsymbol{\omega}_i(\tau)$. Using the resolution of the identity in terms of both types of fields~\cite{fradkin_2013}, one finds that
\beq
\mathsf{Z}=\int{\rm D}[\psi^\star, \psi, \boldsymbol{\omega}]\ee^{-\mathsf{S}_{\rm E}[\psi^\star,\psi,S\boldsymbol{\omega}]}
\eeq 
where the Euclidean action   $\mathsf{S}_{\rm E}=\mathsf{S}_{\rm sf}-\ii S\mathsf{A}_{\rm WZ}$ can be expressed  as a functional over the fermion and spin fields
\beq
\mathsf{S}_{\rm E}\!=\!\int_0^\beta\!\!{\rm d}\tau\left(\sum_i\psi^\star\!(\tau,x_i)\partial_\tau\psi (\tau,x_i)+H(\psi^\star,\psi,S\boldsymbol{\omega})\!\right),
\eeq
where  $H(\psi^\star,\psi,S\boldsymbol{\omega})$ is obtained by substituting the fermion  and spin operators in the normal-ordered Hamiltonian~\eqref{eq:spin_fermion_lattice} by the Grassmann and vector fields. One also finds the so-called Wess-Zumino term, which corresponds to the area enclosed by the trajectory of the spin  field in the unit 2-sphere 
\beq
\mathsf{A}_{\rm WZ}=\!\int_{0}^\beta\!\!\!{\rm d}\tau\!\int_0^1\!\!{\rm d}s\sum_i\boldsymbol{\omega}_i(s,\tau)\cdot(\partial_s\boldsymbol{\omega}_i(s,\tau)\times\partial_\tau\boldsymbol{\omega}_i(s,\tau)).
\eeq
Here,  this area is parametrized by $s\in[0,1]$, such that $\boldsymbol{\omega}_i(s,0)=\boldsymbol{\omega}_i(s,\beta)$ correspond to the closed trajectories that cover a  spherical cap $\Sigma$ between  $\boldsymbol{\omega}_i(0,\tau)=\omega_i(\tau)$ and the north pole $\boldsymbol{\omega}_i(1,\tau)={\bf e}_z$, which is  used as the fiducial state in Bloch's sphere to define the spin coherent states (see Fig.~\ref{fig:scheme}).

In $O(3)$-symmetric situations, such as those arising in antiferromagnetic Heisenberg models, this Wess-Zumino term plays a crucial role as it is responsible for the mapping to a non-linear sigma model with an additional topological $\theta$ term that depends on the integer or half-integer nature of the spins~\cite{HALDANE1983464}. In the present case, however, there is no rotational symmetry in the classical Hamiltonian $H(\psi^\star,\psi,S\boldsymbol{\omega})$, and one can see from the particular form of the spin-fermion coupling and the external field that it will suffice to use coherent states pointing along the meridian at vanishing longitude 
\beq
\boldsymbol{\omega}_i(\tau)={\omega}^x_i(s,\tau){\bf e}_x+{\omega}^z_i(s,\tau){\bf e}_z,
\eeq
which correspond to the great circle in the $xz$ plane of  Fig.~\ref{fig:scheme}. Accordingly, $\boldsymbol{\omega}_i\cdot(\partial_s\boldsymbol{\omega}_i\times\partial_\tau\boldsymbol{\omega}_i)=0$ and there is no enclosed area by the precession of the spins, such that the Wess-Zumino term  vanishes $\mathsf{A}_{\rm WZ}=0$.

 We can now introduce the equivalent of the slowly-varying quantum fields in Eqs.~\eqref{eq:dirac_fields} and~\eqref{eq:rotor_fields} in terms of the Grasmmann fermion fields
\beq
\psi(\tau,x_i)\approx\ee^{-\ii k_{ F}x_i}\sqrt{a}\psi_+(\tau,x)+\ee^{+\ii k_{ F}x_i}\sqrt{a}\psi_-(\tau,x),
\eeq
and the spin vector fields 
\beq
{\omega}_i(\tau)\approx\cos( k_{ N}x_i)\hspace{0.25ex}\boldsymbol{n}(\tau,x)+\frac{a}{S}\boldsymbol{\ell}(\tau,x).
\eeq
Proceeding in analogy to Appendix~\ref{app:H_con}, we perform the gradient expansion in the continuum limit $a\to 0$,  neglect rapidly-oscillating terms, and find $\mathsf{S}_{\rm E}\approx\int{\rm d}^2x\hspace{0.5ex}\mathcal{L}(\boldsymbol{x})$ with the following  Lagrangian density
\beq
\label{eq:lag_app}
\mathcal{L}(\boldsymbol{x})=\overline{\Psi}(\boldsymbol{x})\left(\hat{\gamma}^\mu\partial_\mu+\boldsymbol{g}_s\cdot\boldsymbol{n}(\boldsymbol{x})\right){\Psi}(\boldsymbol{x})+(\boldsymbol{g}j_0(x)-\boldsymbol{h})\cdot\boldsymbol{\ell}(\boldsymbol{x}),
\eeq
with  coupling $\boldsymbol{g}_{\rm s}$ and external field $\boldsymbol{h}$ defined in Eqs.~\eqref{eq:g_term} and~\eqref{eq:h_term}, respectively.
Here, $\boldsymbol{x}=(c\tau,x)$ is a 2-dimensional Euclidean  space with  metric $g^{\rm E}_{\mu,\nu}={\rm diag}(1,1)$,  and the Euclidean gamma matrices are $\hat{\gamma}^0=\gamma^0, \hat{\gamma}^1=-\ii\gamma^1$. The Dirac spinor is composed of the right- and left-moving continuum Grassmann fields ${\Psi}(\boldsymbol{x})=(\psi_+(\tau,x),\psi_-(\tau,x))^{\rm t}$, such that the adjoint becomes $\overline{\Psi}(\boldsymbol{x})={\Psi}^\dagger(\boldsymbol{x})\hat{\gamma}^0=(\psi_-(\tau,x),\psi_+(\tau,x))$. Likewise,  the    N\'eel and canting   fields are the vector fields 
\beq 
\begin{split}
\boldsymbol{n}(\tau,x)&=\hspace{0.5ex}\frac{1}{2}\hspace{0.25ex}\left(\boldsymbol{\omega}_{2i}(\tau)-\boldsymbol{\omega}_{2i-1}(\tau)\right),\\
\boldsymbol{\ell}(\tau,x)&=\frac{S}{2a}\left(\boldsymbol{\omega}_{2i}(\tau)+\boldsymbol{\omega}_{2i-1}(\tau)\right).
\end{split}
\eeq
 In the large-$S$ limit, and in situations dominated by N\'eel correlations $|\boldsymbol{\ell}(a/S)^2|\to 0$, these fields  are additionally subjected to the rotor constraints
\beq
\boldsymbol{n}(\boldsymbol{x})\cdot\boldsymbol{n}(\boldsymbol{x})=1,\hspace{2ex}\boldsymbol{n}(\boldsymbol{x})\cdot\boldsymbol{\ell}(\boldsymbol{x})=0,
\eeq
which can be included  in the   partition function
\beq
Z=\int{\rm D}[\overline{\Psi}, \Psi,\boldsymbol{n},\boldsymbol{\ell}]\ee^{-\int{\rm d}^2x\mathcal{L}(\boldsymbol{x})}, 
\eeq
through the functional integral measure ${\rm D}[\overline{\Psi}, \Psi,\boldsymbol{n},\boldsymbol{\ell}]=\frac{(2S+1)}{4\pi}\prod_{\boldsymbol{x}}{\rm d}\overline{\Psi}(\boldsymbol{x}){\rm d}{\Psi}(\boldsymbol{x}){\rm d}^3{{n}}(\boldsymbol{x}){\rm d}^3{{\ell}}(\boldsymbol{x})\delta(\boldsymbol{n}^2-1)\delta(\boldsymbol{n}\cdot\boldsymbol{\ell})$.

\section{Effective rotor action and large-$S$ limit}
\label{app:S_mf}
In this Appendix, we give a  detailed  derivation of the effective rotor action~\eqref{eq:eff_action}. As outlined in the main text, one can  integrate out  the Grassmann fields from the partition function
\beq
\mathsf{Z}=\int{\rm D}[\boldsymbol{n},\boldsymbol{\ell}]\left(\int{\rm D}[\overline{\Psi}, \Psi]\ee^{-\int{\rm d}^2x\mathcal{L}(\boldsymbol{x})}\right), 
\eeq
since the corresponding functional integral reduces to a product of Gaussian integrals. These integrals are obtained after transforming the  fields  in terms of Matsubara frequencies $\omega_n=\frac{\pi}{\beta}(2n+1)$, and quasi-momentum $qa\in[-\pi,\pi)$, according to
\beq
\begin{split}
\Psi(\tau,x_i)&=\frac{1}{\sqrt{\beta L}}\sum_{n\in\mathbb{Z}}\sum_{q\in{\rm BZ}}\ee^{\ii(\omega_n\tau+qx_i)}\Psi(\omega_n,q),\\
\overline{\Psi}(\tau,x_i)&=\frac{1}{\sqrt{\beta L}}\sum_{n\in\mathbb{Z}}\sum_{q\in{\rm BZ}}\ee^{-\ii(\omega_n\tau+qx_i)}\overline{\Psi}(\omega_n,q).\\
\end{split}
\eeq
The corresponding integrals lead to 
\beq
\label{eq:int_fermions}
\int{\rm D}[\overline{\Psi}, \Psi]\ee^{-\int{\rm d}^2x\mathcal{L}}=\ee^{-\int{\rm d}^2x\mathcal{L}_{\rm r}}\prod_{n,q}{\rm det}\left(-\ii\omega_n+h_{\rm D}(q,\boldsymbol{g}_{s}\cdot\boldsymbol{n})\right),
\eeq
where we have introduced a renormalised rotor Lagrangian
\beq
\label{eq:bare_rotor}
\mathcal{L}_{\rm r}=\left(\frac{\boldsymbol{g}}{2}-\boldsymbol{h}\right)\cdot\boldsymbol{\ell},
\eeq
for a homogeneous canting field $\boldsymbol{\ell}(\boldsymbol{x})=\boldsymbol{\ell}$. This term~\eqref{eq:bare_rotor}, which contains the original precession under the external field $-\boldsymbol{h}$, gets a contribution from the back-action of the fermion current~\eqref{eq:lag_app}. For  a homogeneous canting field, and at half-filling conditions $\int{\rm dx}j_0(x)=N_{\rm s}/2$, this back-action  renormalizes  the  external field to $-(\boldsymbol{h}-\boldsymbol{g}/2)$.

 In  Eq.~\eqref{eq:int_fermions}, we have introduced
 the single-particle  Hamiltonian of a (1+1)-dimensional Dirac fermion   
 \beq
 h_{\rm D}(q,m)=cq{\gamma}^5+m{\gamma}^0,
 \eeq
   with a Dirac mass proportional to the  homogeneous N\'eel field $m=\boldsymbol{g}_{s}\cdot\boldsymbol{n}$. In the continuum  and $T=0$ limits, the product of determinants  involving these Dirac Hamiltonians can be expressed in terms of momentum integrals
 \beq
 \prod_{n,q}{\rm det}\left(-\ii\omega_n+h_{\rm D}(q,\boldsymbol{g}_{s}\cdot\boldsymbol{n})\right)=\ee^{\frac{1}{4\pi^2c}\int{{\rm d}^2k}\log\left(\boldsymbol{k}^2+(\boldsymbol{g}_{s}\cdot\boldsymbol{n})^2\right)}
 \eeq 
 where we have introduced  Euclidean momentum $\boldsymbol{k}=(\omega_n,cq)$. 
 
 This expression, together with Eq.~\eqref{eq:int_fermions}, leads to  an effective action which, up to an irrelevant term independent of the N\'eel and canting fields, reads as follows
\beq
\label{eq:s_eff_app}
\mathsf{S}_{\rm eff}=\!\!\int\!\!{\rm d}^2x\left(\!\left(\frac{\boldsymbol{g}}{2}-\boldsymbol{h}\right)\cdot\boldsymbol{\ell}-\frac{\left(\boldsymbol{g}_s\cdot\boldsymbol{n}\right)^2}{4\pi c}\left(\log\left(\frac{\Lambda_{\rm c}}{\boldsymbol{g}_s\cdot\boldsymbol{n}}\right)^{\!\!\!2}+1\right)\!\!\!\right),
\eeq
where the UV cutoff $\Lambda_{\rm c}=2t$ appears in the integrals over the Euclidean momentum $\int{\rm d}^2k=\frac{2\pi}{c}\int_0^{\Lambda_{\rm c}}\!k{\rm d}k$. Since the fields are homogeneous, this equation leads directly to the result~\eqref{eq:eff_action} used in the main text. This type of effective action, which contains a non-perturbative contribution $\sigma^2\log\sigma^2$, is a characteristic of the dynamical mass generation of the Gross-Neveu model~\cite{PhysRevD.10.3235}, where $\sigma$ is an auxiliary field. In that case, the above calculation is equivalent to a large-$N$ limit summation of the leading  Feynman diagrams, which contain  a single fermion loop and all possible  even numbers of legs for the $\sigma $ field~\cite{PhysRevD.10.3235}. This contributes to an effective potential that develops a double-well structure as soon as $g\neq 0$, and thus induces  the dynamical mass generation and chiral SSB. As noted in the main text, this SSB is dynamical as it requires quantum-mechanical effects via the fermion loops in order to take place, which makes it different from the classical SSB in the Jackiw-Rebbi model~\cite{PhysRevD.13.3398}. In the present case~\eqref{eq:s_eff_app}, the situation is different as the N\'eel and canting fields are not auxiliary, but have their own quantum dynamics that results from the non-commutativity of the rotor position and angular momentum. In particular, as a result of this competition, the dynamical SSB does not take place for an arbitrarily small $g$ regardless of the other microscopic parameters, but there will be critical lines that delimit the phase with a dynamically generated mass, as discussed in the main text.

  \begin{figure*}[t]
  \centering
  \includegraphics[width=0.85\linewidth]{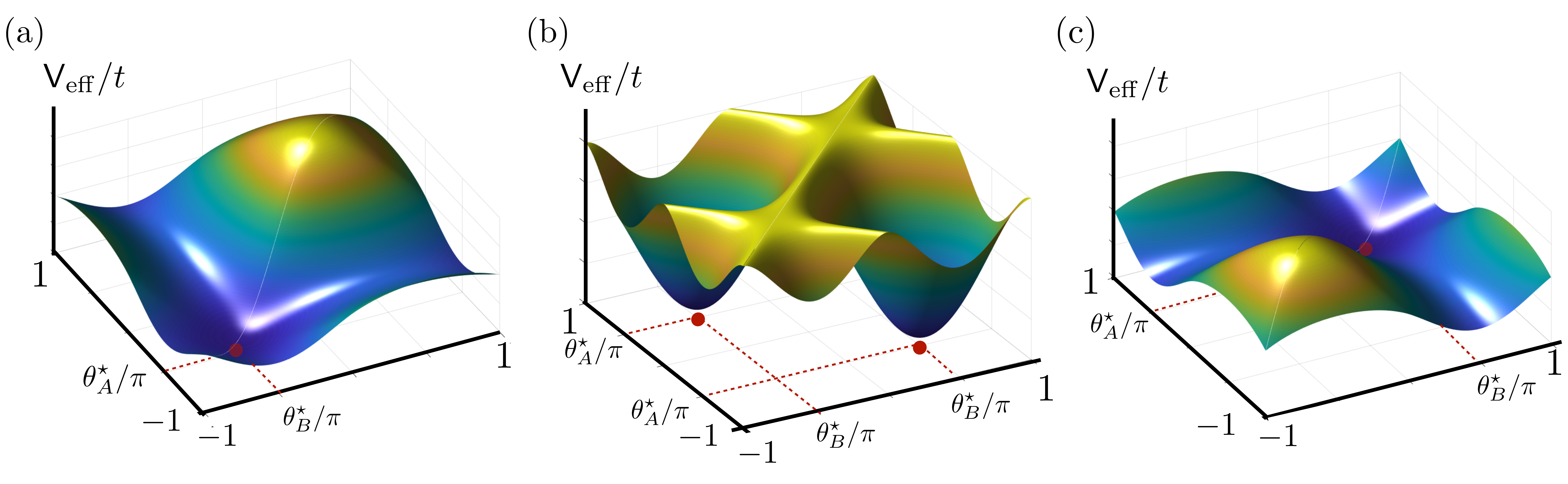}
\caption{\label{fig:eff_potential} {\bf Saddle points of the effective potential}  {\bf (a)} Effective potential for the south-pole paramagnetic regime $h_\ell=h_{\ell}^{-}-\delta/2$, where $\delta=h_{\ell}^{+}-h_{\ell}^{-}$. {\bf (b)} Effective potential for the N\'eel anti-ferromagnetic regime $h_\ell=h_{\ell}^{-}+\delta/2$. {\bf (c)} Effective potential for the north-pole paramagnetic regime $h_\ell=h_{\ell}^{-}+3\delta/2$.}
\end{figure*}

 Let us now describe the large-$S$ limit in more detail. Expressing the N\'eel and canting fields in the coherent-state basis~\eqref{eq:fields_angles}, one finds that the effective action~\eqref{eq:s_eff_app} can be expressed in terms of an effective potential $\mathsf{S}_{\rm eff}=  \beta LS\mathsf{V}_{\rm eff}(\theta_A,\theta_B)$, as
  there are no time derivatives of the fields. By letting $S\to\infty$, the solutions correspond to the saddle point equations $\left.\partial_{\theta_A}\mathsf{V}_{\rm eff}\right|_{\boldsymbol{\theta}^\star}=\left.\partial_{\theta_B}\mathsf{V}_{\rm eff}\right|_{\boldsymbol{\theta}^\star}=0$, which lead to the following system of non-linear equations
  \begin{widetext}
  \beq
  \label{eq:saddle_point_eqs}
  \begin{split}
 \left(\left( \frac{g}{2}-h_\ell\right)-\frac{t}{\pi S}\left(\frac{gS}{2t}\right)^2(\sin\theta^\star_A-\sin\theta^\star_B)\log\left(\frac{4t}{gS(\sin\theta^\star_A-\sin\theta^\star_B)}\right)^{\!\!2}\right)\cos\theta^\star_A&=-h_{\mathsf{t}}\sin\theta^\star_A,\\
\left(  \left( \frac{g}{2}-h_\ell\right)+\frac{t}{\pi S}\left(\frac{gS}{2t}\right)^2(\sin\theta^\star_A-\sin\theta^\star_B)\log\left(\frac{4t}{gS(\sin\theta^\star_A-\sin\theta^\star_B)}\right)^{\!\!2}\right)\cos\theta^\star_B&=-h_{\mathsf{t}}\sin\theta^\star_B,\\
  \end{split}
  \eeq
  \end{widetext}
  Despite the non-linearity, one can readily find an exact analytical solution to these saddle-point equations in the limit of vanishing transverse fields $h_{\mathsf{t}}=0$, where the only possible solutions correspond to $\cos\theta_A^\star=\cos\theta_B^\star=0$. There are four possible solutions within $\boldsymbol{\theta}^\star\in[-\pi,\pi)\times[-\pi,\pi)$, namely
  
  \vspace{0.5ex}
  {\it (i) North-pole paramagnet:} This solution is $\theta_A^\star=\theta_B^\star=\pi/2$, where the canting field points towards the north pole $\boldsymbol{\ell}(\boldsymbol{x})=(S/a)\boldsymbol{e}_z, \forall\boldsymbol{x}$. The corresponding spin coherent state  is
  \beq
   \ket{\mathsf{g}_{\ell\mathsf{P}}^+}=\ket{S,S}_{z,1}\otimes\ket{S,S}_{z,2}\otimes\cdots\otimes\ket{S,S}_{z,N_{\rm s}},
  \eeq
   where $\ket{S,m}_{\alpha,i}$ is the common eigenstate of $\boldsymbol{S}_i^2,S^\alpha_i$  with eigenvalues $S(S+1)$ and $m\in\{-S,-S+1,\cdots,S-1,S\}$. Hence, all spin-$S$ particles are aligned   towards the north pole. We note that this ordering is not caused by collective effects, but induced by the external longitudinal field, and that is why we refer to this state as a spin paramagnet.
   
    \vspace{0.5ex}
     {\it (ii) South-pole paramagnet:} This solution is $\theta_A^\star=\theta_B^\star=-\pi/2$, where the canting field points towards the south pole $\boldsymbol{\ell}(\boldsymbol{x})=-(S/a)\boldsymbol{e}_z, \forall\boldsymbol{x}$. The corresponding  coherent state  is
  \beq
   \ket{\mathsf{g}_{\ell\mathsf{P}}^-}=\ket{S,-S}_{z,1}\otimes\ket{S,-S}_{z,2}\otimes\cdots\otimes\ket{S,-S}_{z,N_{\rm s}}.
  \eeq
  Hence, all spin-$S$ particles are aligned   towards the south pole.

 \vspace{0.5ex}
 {\it (iii) N\'eel  anti-ferromagnets:} These solutions are $\theta_A^\star=-\theta_B^\star=\pi/2$,  or $\theta_A^\star=-\theta_B^\star=-\pi/2$. In this case, and  it is the N\'eel  field which points towards the north $\boldsymbol{n}(\boldsymbol{x})=\boldsymbol{e}_z, \forall\boldsymbol{x}$ and south pole $\boldsymbol{n}(\boldsymbol{x})=-\boldsymbol{e}_z, \forall\boldsymbol{x}$, respectively. The corresponding spin coherent states are  
  \beq
  \begin{split}
   \ket{\mathsf{g}_{\mathsf{N}}^+}=\ket{S,S}_{z,1}\otimes\ket{S,-S}_{z,2}\otimes\cdots\otimes\ket{S,S}_{z,N_{\rm s}-1}\otimes\ket{S,-S}_{z,N_{\rm s}},\\
   \ket{\mathsf{g}_{\mathsf{N}}^-}=\ket{S,-S}_{z,1}\otimes\ket{S,S}_{z,2}\otimes\cdots\otimes\ket{S,-S}_{z,N_{\rm s}-1}\otimes\ket{S,S}_{z,N_{\rm s}},
   \end{split}
  \eeq
 Hence, all neighbouring spin-$S$ particles are aligned  in an anti-parallel manner. We note that, in this case, this ordering is caused by collective effects, and the north- or south-pole solutions occur via the spontaneous symmetry breaking of the $\mathbb{Z}_2$ chiral symmetry $\boldsymbol{n}(\boldsymbol{x})\to-\boldsymbol{n}(\boldsymbol{x})$  mentioned in the main text. This is  why we refer to these states as anti-ferromagnets.
  
 In order to decide which of the above orderings occurs in the system, we can compare the corresponding free energies per unit length  $\mathsf{f}(\theta_A^\star,\theta_B^\star)=\frac{1}{L}\mathsf{F}(\theta_A^\star,\theta_B^\star)=-\frac{1}{L\beta }\log{\mathsf{Z}}(\theta_A^\star,\theta_B^\star)$, namely
 \beq
 \begin{split}
 \mathsf{f}\left(\pm\frac{\pi}{2},\pm\frac{\pi}{2}\right)&=\pm\left( \frac{g}{2}-h_\ell\right)\frac{S}{a},\\
   \mathsf{f}\left(\pm\frac{\pi}{2},\mp\frac{\pi}{2}\right)&=-\frac{t}{2\pi a }\left(\frac{gS}{2t}\right)^{\!\!\!2}\!\left(\log\left(\frac{2t}{gS}\right)^{\!\!\!2}+1\right).\\
  \end{split}
 \eeq
 Comparing these energies, we find two critical lines $h_{\ell}^{\pm}$ that separate the N\'eel anti-ferromagnet from the two longitudinal paramagnets. The first one
 \beq
  h_{\ell}^{-}=\frac{g}{2}-\frac{t}{\pi S}\left(\frac{gS}{2t}\right)^{\!\!2}\!\left(\log\left(\frac{2t}{gS}\right)+\frac{1}{2}\right),
 \eeq
 is obtained from $ \mathsf{f}\left(-{\pi/2},-{\pi/2}\right)=\mathsf{f}\left(+{\pi/2},-{\pi/2}\right)$. Accordingly, for $h_\ell<h_{\ell}^{-}$, the groundstate corresponds to the south-pole longitudinal paramagnet. Conversely, for $h_\ell>h_{\ell}^{-}$, we enter into the N\'eel anti-ferromagnet. The second  line is 
  \beq
  h_{\ell}^{+}=\frac{g}{2}+\frac{t}{\pi S}\left(\frac{gS}{2t}\right)^{\!\!2}\!\left(\log\left(\frac{2t}{gS}\right)+\frac{1}{2}\right),
 \eeq
  and is obtained from $ \mathsf{f}\left(+{\pi/2},-{\pi/2}\right)=\mathsf{f}\left(+{\pi/2},+{\pi/2}\right)$. In this case, for $h_\ell<h_{\ell}^{+}$, the groundstate corresponds to the N\'eel anti-ferromagnet, while or $h_\ell>h_{\ell}^{+}$, we enter into the north-pole longitudinal paramagnet. Both critical lines correspond to a situation where the effective potential changes from a single minimum into a double well, as represented in Fig.~\ref{fig:eff_potential}.

     \vspace{0.5ex}
     {\it (iv) Equator paramagnet:} 
 Finally, before concluding this Appendix, let us discuss the other limit where one can find a different saddle-point solution.  This corresponds to the $h_{\ell}=g=0$ and $h_{\mathsf{t}}>0$ limit, where both equations~\eqref{eq:saddle_point_eqs}  are solved when $\sin\theta_A^\star=\sin\theta_B^\star=0$. In this case, by comparing the corresponding free energies, one readily sees that there is a unique groundstate corresponding to   $\theta_A^\star=\theta_B^\star=\pi$, where the canting field points along the equator  $\boldsymbol{\ell}(\boldsymbol{x})=(S/a)\boldsymbol{e}_x, \forall\boldsymbol{x}$. In terms of the spins, this phase corresponds to  
   \beq
   \ket{\mathsf{g}_{\mathsf{t}\mathsf{P}}}=\ket{S,S}_{x,1}\otimes\ket{S,S}_{x,2}\otimes\cdots\otimes\ket{S,S}_{x,N_{\rm s}},
  \eeq
  where all spins align parallel to the external transverse field. Since the ordering is not due to collective effects, we refer to this phase as a transverse paramagnet.

\section{Self-consistent Hartree-Fock mean-field theory}
\label{app:MF}

In this Appendix, we discuss more in depth the details of the derivation of the self-consistent mean-field theory used in the main text. We start from the Hamiltonian of Eq.~\eqref{eq:spin_fermion_lattice}, and perform a Hartree-Fock decoupling of the spin-fermion coupling 
\begin{equation}
\label{eq:MF_decoupling}
\boldsymbol{g}\cdot\boldsymbol{S}^{\phantom{\dagger}}_i n_i\simeq \boldsymbol{g}\cdot \langle\boldsymbol{S}^{\phantom{\dagger}}_i\rangle n_i + \boldsymbol{g}\cdot \boldsymbol{S}^{\phantom{\dagger}}_i \langle n_i\rangle - \boldsymbol{g}\cdot \langle\boldsymbol{S}^{\phantom{\dagger}}_i\rangle \langle n_i\rangle,
\end{equation}
where we have introduced the  number operator $n_i=c_i^\dagger c_i^{\phantom{\dagger}}$.
After the Hartree-Fock decoupling, the Hamiltonian can be expressed as the sum of non-interacting  Hamiltonians for the fermion/spin sectors
\begin{align}
\label{eq:MF_hamiltonians_fermions}
H_{\rm f}&=-\sum_i tc_i^{\dagger}c_{i+1}^{\phantom{\dagger}}+\sum_i\epsilon_{{\rm f},i} n_i, \hspace{2ex} \epsilon_{{\rm f},i}=\boldsymbol{g}\cdot \langle\boldsymbol{S}_i\rangle^{\phantom{\dagger}}\\
\label{eq:MF_hamiltonians_spins}
H_{\rm s}&=-\sum_i\boldsymbol{h}_{{\rm s},i}\cdot \boldsymbol{S}_i^{\phantom{\dagger}},\hspace{11ex} \boldsymbol{h}_{{\rm s},i}=\boldsymbol{h}-\boldsymbol{g} \langle n_i\rangle
\end{align}
where ${\rm f}$ (${\rm s}$) stands for fermions (spins). Accordingly, the fermion tunnel in a potential-energy landscape  $\epsilon_{{\rm f},i}$ set by the expectation value of the spins (see Fig.~\ref{fig:scheme}), while the spins precess in an effective external field, which becomes inhomogeneous depending on the average distribution of fermions.


This  mean-field approximation  requires the obsrevables $\{\langle \boldsymbol{S}^{\phantom{\dagger}}_i\rangle, \langle n_i \rangle\}_{i=1}^{N_{\rm s}}$  to be determined self-consistently,   and we must deal with a number of self-consistency equations that grows linearly with the number of sites.   The self-consistent loop consists in the following steps: We start by setting an initial spin configuration $\langle \boldsymbol{S}_i \rangle$, and compute the expectations values $\langle n_i \rangle$ by solving the fermionic tight-biding model~\eqref{eq:MF_hamiltonians_fermions}  for a given temperature $T=\beta^{-1}$. These mean-field parameters are then used as input to determine the effective external field in the spin Hamiltonian~\eqref{eq:MF_hamiltonians_spins}, which is subsequently diagonalised, such that we can calculate the corresponding spin expectation values $\langle \boldsymbol{S}_i \rangle$  for a given temperature $T$.

  \begin{figure}[t]
  \centering
  \includegraphics[width=0.95\linewidth]{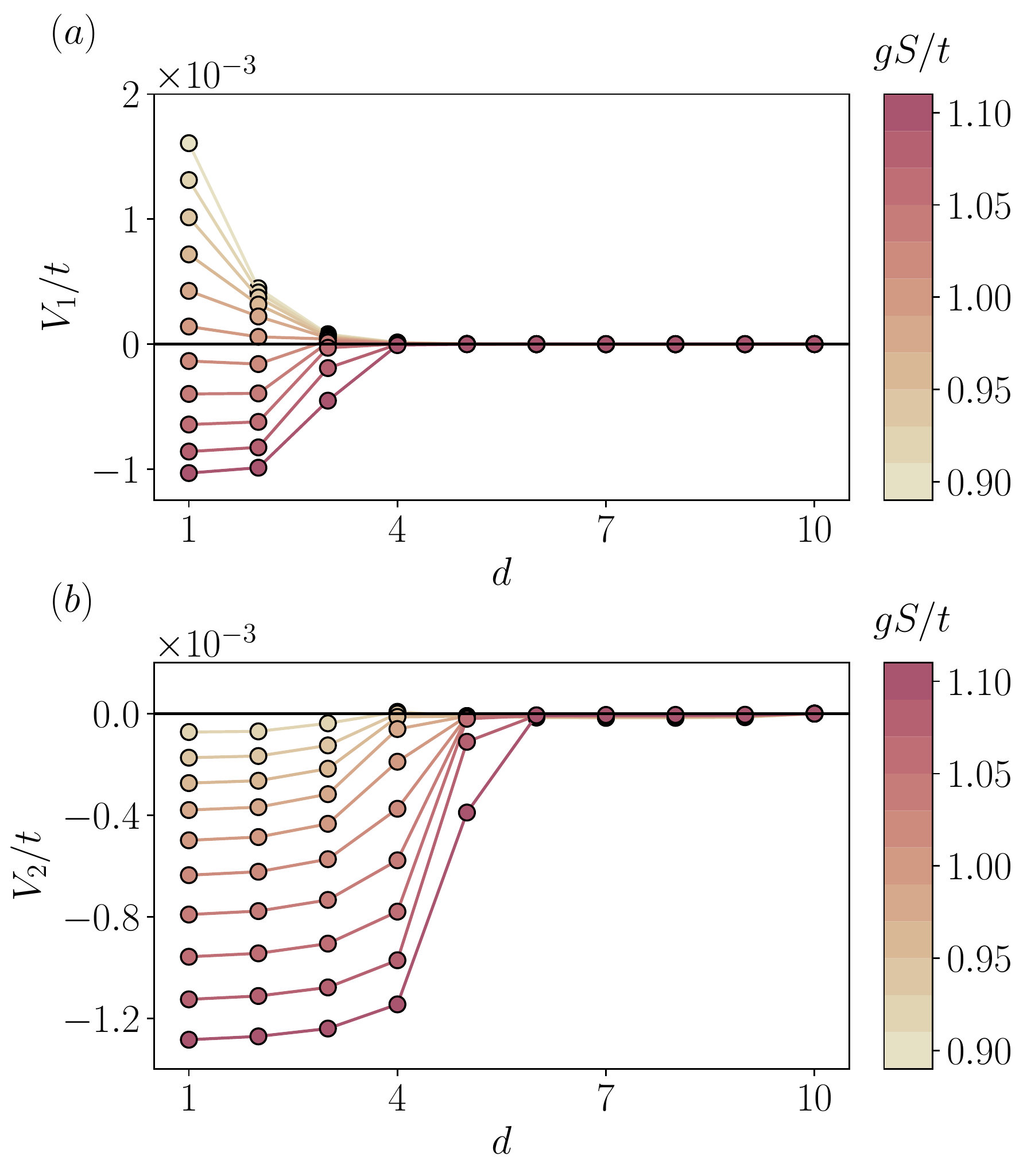}
\caption{\label{fig:deconfinement_transition} {\bf Deconfinement transition and quark-bag instability:} {\bf (a)} Potential energy between two defects pinned at a distance $d$ (see main text) for a chain with $N_{\rm s} = 41$ sites and $N_{\rm f} = 22$, at $h_{\ell}S = 0.4t$, $h_{\mathsf{t}}S = 0.01t$ and $\epsilon S = 0.02t$. As $gS/t$ is increased, the repulsive potential between two solitons turns into an atractive one, signalling a deconfinement-confinement transition. {\bf (b)} Attractive potential energy between two pinned soliton bags, for $N_{\rm s}=41$ and $N_{\rm f} = 23$.}
\end{figure}

This process must be   iterated  until reaching convergence for the free energy, which can be expressed as $\mathsf{F}_\text{MF}=\mathsf{F}_{\rm f}+\mathsf{F}_{\rm s}+C$, where $\mathsf{F}_{\rm f}$ ($\mathsf{F}_{\rm s}$) is the free energy of the fermions (spins) and $C$ is the constant term appearing in the Hartree-Fock decoupling~\eqref{eq:MF_decoupling}. The free energy for the system of  fermions can be expressed in terms of the grand partition function
\beq
\mathsf{Z}_{\rm f}={\rm Tr}\left\{\ee^{-\beta (H_{\rm f}-\mu N_{\rm f})}\right\},
\eeq
where we have introduced the chemical potential $\mu$, and the total fermion number $N_{\rm f}=\sum_i n_i$. For the mean-field decoupled model, this partition function can be readily expressed as
\beq
\mathsf{Z}_{\rm f}=\prod_{n=1}^{N_{\rm s}}\mathsf{Z}_{{\rm f},n},\hspace{2ex} \mathsf{Z}_{{\rm f},n}=1+\ee^{-\beta(E_{{\rm f},n}-\mu)},
\eeq
where $E_{{\rm f},n}$ are the eigenvalues of the quadratic fermionic Hamiltonian~\eqref{eq:MF_hamiltonians_fermions}. The free energy, which can be expressed as $\mathsf{F}_{\rm f}=-\frac{1}{\beta}\log\mathsf{Z}_{\rm f}+\frac{\mu}{\beta}\frac{\partial}{\partial \mu}\log\mathsf{Z}_{\rm f}$ , thus becomes
\begin{equation}
\mathsf{F}_{\rm f}=  -\frac{1}{\beta} \sum_n\log\mathsf{Z}_{{\rm f},n}+\mu \sum_nn_{\rm FD}(\mu,\beta),
\end{equation}
where $n_{\rm FD}(\mu,\beta)=\frac{\mu \partial_\mu\mathsf{Z}_{{\rm f},n}}{\beta\mathsf{Z}_{{\rm f},n}}=1/(\ee^{\beta(E_{{\rm f},n}-\mu)}+1)$ is the so-called Fermi-Dirac distribution. For the spins, the number of which is conserved,  the free energy  can be written  in terms of the canonical partition function
\beq
\mathsf{Z}_{\rm s}={\rm Tr}\left\{\ee^{-\beta H_{\rm s}}\right\},
\eeq
For the mean-field decoupled system, the  Hamiltonian~\eqref{eq:MF_hamiltonians_spins} can be diagonalised for each spin independently
\beq
\mathsf{Z}_{\rm s}=\prod_{i=1}^{N_{\rm s}}\mathsf{Z}_{{\rm s},i},\hspace{2ex} \mathsf{Z}_{{\rm s},i}=\sum_{n=1}^{2S+1}\ee^{-\beta E_{{\rm s},n}},
\eeq
where  $E_{{\rm s},n}$ are the the $(2S+1)$ energies of the spins, which are identical for all sites of the chain. The corresponding  free energy is
\begin{equation}
\mathsf{F}_s = -\frac{1}{\beta} \sum_i\log  \mathsf{Z}_{{\rm s},i}.
\end{equation}

\section{Static quasi-particle potentials}
\label{app:sol_deconf}

In this Appendix, we give further details on the calculation of the static potentials between fractional quark-like quasi-particles and between quark bags. In both cases, this is achieved by first pinning two of these quasi-particles to certain locations in the chain, separated by a distance $d$ measured in unit cells ($2a$). In the first case, since each of them is associated to a soliton in the N\'eel ordered rotor field, we introduce a small parallel field
\begin{equation}
\label{eq:parallel_field}
H_\epsilon = -\sum_i \epsilon_i(d) S^z_i,
\end{equation}
with
\begin{equation}
\label{eq:pinning_1}
\epsilon_i(d) = \begin{cases} 
      \epsilon(-1)^{i} & 1\leq i < i_0 \\
      \epsilon(-1)^{i+1} & i_0 \leq i < i_0 + d\\
      \epsilon(-1)^{i} & i_0 + d \leq i \leq N_s.
   \end{cases}
\end{equation}
This field breaks translational invariance and pines two solitons, and the associated fractional fermionic charges, to positions $i_0$ and $i_0 + d$. In Figure~\ref{fig:deconfinement_transition}{\bf(a)}, we choose $i_0 = (N_s + 1) / 3$ and $\epsilon = 0.02t$, and represent the static quark potential $V_1(d) = E_1(d) - E_0$, where $E_1(d)$ and $E_0$ are the energies of a chain with $N_s$ sites and $N_{\rm f} = (N_{\rm s} + 1) / 2 + 1$ fermions under the pinning potential~\eqref{eq:pinning_1} for a given $d$ and $i_0=10$, respectively. The potential changes from repulsive to attractive as we increased $gS/t$. For the parameters $h_{\ell}S = 0.4t$, $h_{\mathsf{t}}S = 0.01t$, the transition is located at $g_cS/t \approx 1.01$.

The static potential between two quark-bag quasi-particles is obtained by first pinning them at a distance $d$, in a chain with $N_{\rm s}$ sites and $N_{\rm f} = (N_{\rm s} + 1) / 2 + 2$ fermions, using now a parallel field \eqref{eq:parallel_field} with
\begin{equation}
\epsilon_i = \epsilon \left[(-1)^i - 2\delta_{i,i_0} - 2\delta_{i,i_0+d}\right].
\end{equation}
The potential $V_2(d) = E_2(d) - E_0$, where $E_2(d)$ is now the corresponding energy when two quark bags are  pinned at a distance $d$, is shown in Figure~\ref{fig:deconfinement_transition}{(b)} for different values of $gS/t$. In this case, the potential is always attractive.

\bibliographystyle{apsrev4-1}
\bibliography{bibliography}

\begin{thebibliography}{121}%
\makeatletter
\providecommand \@ifxundefined [1]{%
 \@ifx{#1\undefined}
}%
\providecommand \@ifnum [1]{%
 \ifnum #1\expandafter \@firstoftwo
 \else \expandafter \@secondoftwo
 \fi
}%
\providecommand \@ifx [1]{%
 \ifx #1\expandafter \@firstoftwo
 \else \expandafter \@secondoftwo
 \fi
}%
\providecommand \natexlab [1]{#1}%
\providecommand \enquote  [1]{``#1''}%
\providecommand \bibnamefont  [1]{#1}%
\providecommand \bibfnamefont [1]{#1}%
\providecommand \citenamefont [1]{#1}%
\providecommand \href@noop [0]{\@secondoftwo}%
\providecommand \href [0]{\begingroup \@sanitize@url \@href}%
\providecommand \@href[1]{\@@startlink{#1}\@@href}%
\providecommand \@@href[1]{\endgroup#1\@@endlink}%
\providecommand \@sanitize@url [0]{\catcode `\\12\catcode `\$12\catcode
  `\&12\catcode `\#12\catcode `\^12\catcode `\_12\catcode `\%12\relax}%
\providecommand \@@startlink[1]{}%
\providecommand \@@endlink[0]{}%
\providecommand \url  [0]{\begingroup\@sanitize@url \@url }%
\providecommand \@url [1]{\endgroup\@href {#1}{\urlprefix }}%
\providecommand \urlprefix  [0]{URL }%
\providecommand \Eprint [0]{\href }%
\providecommand \doibase [0]{http://dx.doi.org/}%
\providecommand \selectlanguage [0]{\@gobble}%
\providecommand \bibinfo  [0]{\@secondoftwo}%
\providecommand \bibfield  [0]{\@secondoftwo}%
\providecommand \translation [1]{[#1]}%
\providecommand \BibitemOpen [0]{}%
\providecommand \bibitemStop [0]{}%
\providecommand \bibitemNoStop [0]{.\EOS\space}%
\providecommand \EOS [0]{\spacefactor3000\relax}%
\providecommand \BibitemShut  [1]{\csname bibitem#1\endcsname}%
\let\auto@bib@innerbib\@empty
\bibitem [{\citenamefont {Peskin}\ and\ \citenamefont
  {Schroeder}(1995)}]{Peskin:1995ev}%
  \BibitemOpen
  \bibfield  {author} {\bibinfo {author} {\bibfnamefont {M.~E.}\ \bibnamefont
  {Peskin}}\ and\ \bibinfo {author} {\bibfnamefont {D.~V.}\ \bibnamefont
  {Schroeder}},\ }\href {https://www.taylorfrancis.com/books/9780429503559}
  {\emph {\bibinfo {title} {{An Introduction to quantum field theory}}}}\
  (\bibinfo  {publisher} {Addison-Wesley},\ \bibinfo {address} {Reading, USA},\
  \bibinfo {year} {1995})\BibitemShut {NoStop}%
\bibitem [{\citenamefont {Wen}(2007)}]{Xiao:803748}%
  \BibitemOpen
  \bibfield  {author} {\bibinfo {author} {\bibfnamefont {X.-G.}\ \bibnamefont
  {Wen}},\ }\href {\doibase 10.1093/acprof:oso/9780199227259.001.0001} {\emph
  {\bibinfo {title} {{Quantum field theory of many-body systems: from the
  origin of sound to an origin of light and electrons}}}}\ (\bibinfo
  {publisher} {Oxford University Press},\ \bibinfo {address} {Oxford},\
  \bibinfo {year} {2007})\BibitemShut {NoStop}%
\bibitem [{\citenamefont {Landau}(1937)}]{landau}%
  \BibitemOpen
  \bibfield  {author} {\bibinfo {author} {\bibfnamefont {L.}~\bibnamefont
  {Landau}},\ }\href@noop {} {\bibfield  {journal} {\bibinfo  {journal} {Zh.
  Eksp. Teor. Fiz}\ }\textbf {\bibinfo {volume} {11}},\ \bibinfo {pages} {26}
  (\bibinfo {year} {1937})}\BibitemShut {NoStop}%
\bibitem [{\citenamefont {Landau}(1956)}]{landau_fl}%
  \BibitemOpen
  \bibfield  {author} {\bibinfo {author} {\bibfnamefont {L.}~\bibnamefont
  {Landau}},\ }\href@noop {} {\bibfield  {journal} {\bibinfo  {journal} {Sov.
  Phys. JETP}\ }\textbf {\bibinfo {volume} {30}},\ \bibinfo {pages} {1058}
  (\bibinfo {year} {1956})}\BibitemShut {NoStop}%
\bibitem [{\citenamefont {Anderson}(1972)}]{Anderson393}%
  \BibitemOpen
  \bibfield  {author} {\bibinfo {author} {\bibfnamefont {P.~W.}\ \bibnamefont
  {Anderson}},\ }\href {\doibase 10.1126/science.177.4047.393} {\bibfield
  {journal} {\bibinfo  {journal} {Science}\ }\textbf {\bibinfo {volume}
  {177}},\ \bibinfo {pages} {393} (\bibinfo {year} {1972})}\BibitemShut
  {NoStop}%
\bibitem [{\citenamefont {Wilson}\ and\ \citenamefont
  {Kogut}(1974)}]{WILSON197475}%
  \BibitemOpen
  \bibfield  {author} {\bibinfo {author} {\bibfnamefont {K.~G.}\ \bibnamefont
  {Wilson}}\ and\ \bibinfo {author} {\bibfnamefont {J.}~\bibnamefont {Kogut}},\
  }\href {\doibase https://doi.org/10.1016/0370-1573(74)90023-4} {\bibfield
  {journal} {\bibinfo  {journal} {Physics Reports}\ }\textbf {\bibinfo {volume}
  {12}},\ \bibinfo {pages} {75 } (\bibinfo {year} {1974})}\BibitemShut
  {NoStop}%
\bibitem [{\citenamefont {Wilson}(1975)}]{RevModPhys.47.773}%
  \BibitemOpen
  \bibfield  {author} {\bibinfo {author} {\bibfnamefont {K.~G.}\ \bibnamefont
  {Wilson}},\ }\href {\doibase 10.1103/RevModPhys.47.773} {\bibfield  {journal}
  {\bibinfo  {journal} {Rev. Mod. Phys.}\ }\textbf {\bibinfo {volume} {47}},\
  \bibinfo {pages} {773} (\bibinfo {year} {1975})}\BibitemShut {NoStop}%
\bibitem [{\citenamefont {Hollowood}(2013)}]{hollowood_2013}%
  \BibitemOpen
  \bibfield  {author} {\bibinfo {author} {\bibfnamefont {T.~J.}\ \bibnamefont
  {Hollowood}},\ }\href {https://www.springer.com/gp/book/9783642363115} {\emph
  {\bibinfo {title} {Renormalization Group and Fixed Points in Quantum Field
  Theory}}}\ (\bibinfo  {publisher} {Springer},\ \bibinfo {year}
  {2013})\BibitemShut {NoStop}%
\bibitem [{\citenamefont {Wilson}(1974)}]{PhysRevD.10.2445}%
  \BibitemOpen
  \bibfield  {author} {\bibinfo {author} {\bibfnamefont {K.~G.}\ \bibnamefont
  {Wilson}},\ }\href {\doibase 10.1103/PhysRevD.10.2445} {\bibfield  {journal}
  {\bibinfo  {journal} {Phys. Rev. D}\ }\textbf {\bibinfo {volume} {10}},\
  \bibinfo {pages} {2445} (\bibinfo {year} {1974})}\BibitemShut {NoStop}%
\bibitem [{\citenamefont {Kogut}(1979)}]{RevModPhys.51.659}%
  \BibitemOpen
  \bibfield  {author} {\bibinfo {author} {\bibfnamefont {J.~B.}\ \bibnamefont
  {Kogut}},\ }\href {\doibase 10.1103/RevModPhys.51.659} {\bibfield  {journal}
  {\bibinfo  {journal} {Rev. Mod. Phys.}\ }\textbf {\bibinfo {volume} {51}},\
  \bibinfo {pages} {659} (\bibinfo {year} {1979})}\BibitemShut {NoStop}%
\bibitem [{\citenamefont {Tarruell}\ \emph {et~al.}(2012)\citenamefont
  {Tarruell}, \citenamefont {Greif}, \citenamefont {Uehlinger}, \citenamefont
  {Jotzu},\ and\ \citenamefont {Esslinger}}]{Tarruell2012}%
  \BibitemOpen
  \bibfield  {author} {\bibinfo {author} {\bibfnamefont {L.}~\bibnamefont
  {Tarruell}}, \bibinfo {author} {\bibfnamefont {D.}~\bibnamefont {Greif}},
  \bibinfo {author} {\bibfnamefont {T.}~\bibnamefont {Uehlinger}}, \bibinfo
  {author} {\bibfnamefont {G.}~\bibnamefont {Jotzu}}, \ and\ \bibinfo {author}
  {\bibfnamefont {T.}~\bibnamefont {Esslinger}},\ }\href {\doibase
  10.1038/nature10871} {\bibfield  {journal} {\bibinfo  {journal} {Nature}\
  }\textbf {\bibinfo {volume} {483}},\ \bibinfo {pages} {302} (\bibinfo {year}
  {2012})}\BibitemShut {NoStop}%
\bibitem [{\citenamefont {Duca}\ \emph {et~al.}(2015)\citenamefont {Duca},
  \citenamefont {Li}, \citenamefont {Reitter}, \citenamefont {Bloch},
  \citenamefont {Schleier-Smith},\ and\ \citenamefont {Schneider}}]{Duca288}%
  \BibitemOpen
  \bibfield  {author} {\bibinfo {author} {\bibfnamefont {L.}~\bibnamefont
  {Duca}}, \bibinfo {author} {\bibfnamefont {T.}~\bibnamefont {Li}}, \bibinfo
  {author} {\bibfnamefont {M.}~\bibnamefont {Reitter}}, \bibinfo {author}
  {\bibfnamefont {I.}~\bibnamefont {Bloch}}, \bibinfo {author} {\bibfnamefont
  {M.}~\bibnamefont {Schleier-Smith}}, \ and\ \bibinfo {author} {\bibfnamefont
  {U.}~\bibnamefont {Schneider}},\ }\href {\doibase 10.1126/science.1259052}
  {\bibfield  {journal} {\bibinfo  {journal} {Science}\ }\textbf {\bibinfo
  {volume} {347}},\ \bibinfo {pages} {288} (\bibinfo {year}
  {2015})}\BibitemShut {NoStop}%
\bibitem [{\citenamefont {Fl{\"a}schner}\ \emph {et~al.}(2016)\citenamefont
  {Fl{\"a}schner}, \citenamefont {Rem}, \citenamefont {Tarnowski},
  \citenamefont {Vogel}, \citenamefont {L{\"u}hmann}, \citenamefont
  {Sengstock},\ and\ \citenamefont {Weitenberg}}]{flaschner_experimental_2016}%
  \BibitemOpen
  \bibfield  {author} {\bibinfo {author} {\bibfnamefont {N.}~\bibnamefont
  {Fl{\"a}schner}}, \bibinfo {author} {\bibfnamefont {B.~S.}\ \bibnamefont
  {Rem}}, \bibinfo {author} {\bibfnamefont {M.}~\bibnamefont {Tarnowski}},
  \bibinfo {author} {\bibfnamefont {D.}~\bibnamefont {Vogel}}, \bibinfo
  {author} {\bibfnamefont {D.-S.}\ \bibnamefont {L{\"u}hmann}}, \bibinfo
  {author} {\bibfnamefont {K.}~\bibnamefont {Sengstock}}, \ and\ \bibinfo
  {author} {\bibfnamefont {C.}~\bibnamefont {Weitenberg}},\ }\href
  {https://science.sciencemag.org/content/352/6289/1091.abstract} {\bibfield
  {journal} {\bibinfo  {journal} {Science}\ }\textbf {\bibinfo {volume}
  {352}},\ \bibinfo {pages} {1091} (\bibinfo {year} {2016})}\BibitemShut
  {NoStop}%
\bibitem [{\citenamefont {Schweizer}\ \emph {et~al.}(2019)\citenamefont
  {Schweizer}, \citenamefont {Grusdt}, \citenamefont {Berngruber},
  \citenamefont {Barbiero}, \citenamefont {Demler}, \citenamefont {Goldman},
  \citenamefont {Bloch},\ and\ \citenamefont {Aidelsburger}}]{Schweizer2019}%
  \BibitemOpen
  \bibfield  {author} {\bibinfo {author} {\bibfnamefont {C.}~\bibnamefont
  {Schweizer}}, \bibinfo {author} {\bibfnamefont {F.}~\bibnamefont {Grusdt}},
  \bibinfo {author} {\bibfnamefont {M.}~\bibnamefont {Berngruber}}, \bibinfo
  {author} {\bibfnamefont {L.}~\bibnamefont {Barbiero}}, \bibinfo {author}
  {\bibfnamefont {E.}~\bibnamefont {Demler}}, \bibinfo {author} {\bibfnamefont
  {N.}~\bibnamefont {Goldman}}, \bibinfo {author} {\bibfnamefont
  {I.}~\bibnamefont {Bloch}}, \ and\ \bibinfo {author} {\bibfnamefont
  {M.}~\bibnamefont {Aidelsburger}},\ }\href {\doibase
  10.1038/s41567-019-0649-7} {\bibfield  {journal} {\bibinfo  {journal} {Nature
  Physics}\ }\textbf {\bibinfo {volume} {15}},\ \bibinfo {pages} {1168}
  (\bibinfo {year} {2019})}\BibitemShut {NoStop}%
\bibitem [{\citenamefont {Mil}\ \emph {et~al.}(2020)\citenamefont {Mil},
  \citenamefont {Zache}, \citenamefont {Hegde}, \citenamefont {Xia},
  \citenamefont {Bhatt}, \citenamefont {Oberthaler}, \citenamefont {Hauke},
  \citenamefont {Berges},\ and\ \citenamefont {Jendrzejewski}}]{Mil1128}%
  \BibitemOpen
  \bibfield  {author} {\bibinfo {author} {\bibfnamefont {A.}~\bibnamefont
  {Mil}}, \bibinfo {author} {\bibfnamefont {T.~V.}\ \bibnamefont {Zache}},
  \bibinfo {author} {\bibfnamefont {A.}~\bibnamefont {Hegde}}, \bibinfo
  {author} {\bibfnamefont {A.}~\bibnamefont {Xia}}, \bibinfo {author}
  {\bibfnamefont {R.~P.}\ \bibnamefont {Bhatt}}, \bibinfo {author}
  {\bibfnamefont {M.~K.}\ \bibnamefont {Oberthaler}}, \bibinfo {author}
  {\bibfnamefont {P.}~\bibnamefont {Hauke}}, \bibinfo {author} {\bibfnamefont
  {J.}~\bibnamefont {Berges}}, \ and\ \bibinfo {author} {\bibfnamefont
  {F.}~\bibnamefont {Jendrzejewski}},\ }\href {\doibase
  10.1126/science.aaz5312} {\bibfield  {journal} {\bibinfo  {journal}
  {Science}\ }\textbf {\bibinfo {volume} {367}},\ \bibinfo {pages} {1128}
  (\bibinfo {year} {2020})}\BibitemShut {NoStop}%
\bibitem [{\citenamefont {Yang}\ \emph
  {et~al.}(2020{\natexlab{a}})\citenamefont {Yang}, \citenamefont {Sun},
  \citenamefont {Ott}, \citenamefont {Wang}, \citenamefont {Zache},
  \citenamefont {Halimeh}, \citenamefont {Yuan}, \citenamefont {Hauke},\ and\
  \citenamefont {Pan}}]{2003.08945}%
  \BibitemOpen
  \bibfield  {author} {\bibinfo {author} {\bibfnamefont {B.}~\bibnamefont
  {Yang}}, \bibinfo {author} {\bibfnamefont {H.}~\bibnamefont {Sun}}, \bibinfo
  {author} {\bibfnamefont {R.}~\bibnamefont {Ott}}, \bibinfo {author}
  {\bibfnamefont {H.-Y.}\ \bibnamefont {Wang}}, \bibinfo {author}
  {\bibfnamefont {T.~V.}\ \bibnamefont {Zache}}, \bibinfo {author}
  {\bibfnamefont {J.~C.}\ \bibnamefont {Halimeh}}, \bibinfo {author}
  {\bibfnamefont {Z.-S.}\ \bibnamefont {Yuan}}, \bibinfo {author}
  {\bibfnamefont {P.}~\bibnamefont {Hauke}}, \ and\ \bibinfo {author}
  {\bibfnamefont {J.-W.}\ \bibnamefont {Pan}},\ }\href@noop {} {\enquote
  {\bibinfo {title} {Observation of gauge invariance in a 71-site quantum
  simulator},}\ } (\bibinfo {year} {2020}{\natexlab{a}}),\ \Eprint
  {http://arxiv.org/abs/arXiv:2003.08945} {arXiv:2003.08945} \BibitemShut
  {NoStop}%
\bibitem [{\citenamefont {Gerritsma}\ \emph {et~al.}(2010)\citenamefont
  {Gerritsma}, \citenamefont {Kirchmair}, \citenamefont {Z{\"a}hringer},
  \citenamefont {Solano}, \citenamefont {Blatt},\ and\ \citenamefont
  {Roos}}]{Gerritsma2010}%
  \BibitemOpen
  \bibfield  {author} {\bibinfo {author} {\bibfnamefont {R.}~\bibnamefont
  {Gerritsma}}, \bibinfo {author} {\bibfnamefont {G.}~\bibnamefont
  {Kirchmair}}, \bibinfo {author} {\bibfnamefont {F.}~\bibnamefont
  {Z{\"a}hringer}}, \bibinfo {author} {\bibfnamefont {E.}~\bibnamefont
  {Solano}}, \bibinfo {author} {\bibfnamefont {R.}~\bibnamefont {Blatt}}, \
  and\ \bibinfo {author} {\bibfnamefont {C.~F.}\ \bibnamefont {Roos}},\ }\href
  {\doibase 10.1038/nature08688} {\bibfield  {journal} {\bibinfo  {journal}
  {Nature}\ }\textbf {\bibinfo {volume} {463}},\ \bibinfo {pages} {68}
  (\bibinfo {year} {2010})}\BibitemShut {NoStop}%
\bibitem [{\citenamefont {Gerritsma}\ \emph {et~al.}(2011)\citenamefont
  {Gerritsma}, \citenamefont {Lanyon}, \citenamefont {Kirchmair}, \citenamefont
  {Z\"ahringer}, \citenamefont {Hempel}, \citenamefont {Casanova},
  \citenamefont {Garc\'{\i}a-Ripoll}, \citenamefont {Solano}, \citenamefont
  {Blatt},\ and\ \citenamefont {Roos}}]{PhysRevLett.106.060503}%
  \BibitemOpen
  \bibfield  {author} {\bibinfo {author} {\bibfnamefont {R.}~\bibnamefont
  {Gerritsma}}, \bibinfo {author} {\bibfnamefont {B.~P.}\ \bibnamefont
  {Lanyon}}, \bibinfo {author} {\bibfnamefont {G.}~\bibnamefont {Kirchmair}},
  \bibinfo {author} {\bibfnamefont {F.}~\bibnamefont {Z\"ahringer}}, \bibinfo
  {author} {\bibfnamefont {C.}~\bibnamefont {Hempel}}, \bibinfo {author}
  {\bibfnamefont {J.}~\bibnamefont {Casanova}}, \bibinfo {author}
  {\bibfnamefont {J.~J.}\ \bibnamefont {Garc\'{\i}a-Ripoll}}, \bibinfo {author}
  {\bibfnamefont {E.}~\bibnamefont {Solano}}, \bibinfo {author} {\bibfnamefont
  {R.}~\bibnamefont {Blatt}}, \ and\ \bibinfo {author} {\bibfnamefont {C.~F.}\
  \bibnamefont {Roos}},\ }\href {\doibase 10.1103/PhysRevLett.106.060503}
  {\bibfield  {journal} {\bibinfo  {journal} {Phys. Rev. Lett.}\ }\textbf
  {\bibinfo {volume} {106}},\ \bibinfo {pages} {060503} (\bibinfo {year}
  {2011})}\BibitemShut {NoStop}%
\bibitem [{\citenamefont {Martinez}\ \emph {et~al.}(2016)\citenamefont
  {Martinez}, \citenamefont {Muschik}, \citenamefont {Schindler}, \citenamefont
  {Nigg}, \citenamefont {Erhard}, \citenamefont {Heyl}, \citenamefont {Hauke},
  \citenamefont {Dalmonte}, \citenamefont {Monz}, \citenamefont {Zoller},\ and\
  \citenamefont {Blatt}}]{Martinez2016}%
  \BibitemOpen
  \bibfield  {author} {\bibinfo {author} {\bibfnamefont {E.~A.}\ \bibnamefont
  {Martinez}}, \bibinfo {author} {\bibfnamefont {C.~A.}\ \bibnamefont
  {Muschik}}, \bibinfo {author} {\bibfnamefont {P.}~\bibnamefont {Schindler}},
  \bibinfo {author} {\bibfnamefont {D.}~\bibnamefont {Nigg}}, \bibinfo {author}
  {\bibfnamefont {A.}~\bibnamefont {Erhard}}, \bibinfo {author} {\bibfnamefont
  {M.}~\bibnamefont {Heyl}}, \bibinfo {author} {\bibfnamefont {P.}~\bibnamefont
  {Hauke}}, \bibinfo {author} {\bibfnamefont {M.}~\bibnamefont {Dalmonte}},
  \bibinfo {author} {\bibfnamefont {T.}~\bibnamefont {Monz}}, \bibinfo {author}
  {\bibfnamefont {P.}~\bibnamefont {Zoller}}, \ and\ \bibinfo {author}
  {\bibfnamefont {R.}~\bibnamefont {Blatt}},\ }\href {\doibase
  10.1038/nature18318} {\bibfield  {journal} {\bibinfo  {journal} {Nature}\
  }\textbf {\bibinfo {volume} {534}},\ \bibinfo {pages} {516} (\bibinfo {year}
  {2016})}\BibitemShut {NoStop}%
\bibitem [{\citenamefont {Feynman}(1982)}]{Feynman_1982}%
  \BibitemOpen
  \bibfield  {author} {\bibinfo {author} {\bibfnamefont {R.~P.}\ \bibnamefont
  {Feynman}},\ }\href {\doibase 10.1007/bf02650179} {\bibfield  {journal}
  {\bibinfo  {journal} {Int. J. Theor. Phys.}\ }\textbf {\bibinfo {volume}
  {21}},\ \bibinfo {pages} {467} (\bibinfo {year} {1982})}\BibitemShut
  {NoStop}%
\bibitem [{\citenamefont {Lewenstein}\ \emph {et~al.}(2007)\citenamefont
  {Lewenstein}, \citenamefont {Sanpera}, \citenamefont {Ahufinger},
  \citenamefont {Damski}, \citenamefont {Sen(De)},\ and\ \citenamefont
  {Sen}}]{doi:10.1080/00018730701223200}%
  \BibitemOpen
  \bibfield  {author} {\bibinfo {author} {\bibfnamefont {M.}~\bibnamefont
  {Lewenstein}}, \bibinfo {author} {\bibfnamefont {A.}~\bibnamefont {Sanpera}},
  \bibinfo {author} {\bibfnamefont {V.}~\bibnamefont {Ahufinger}}, \bibinfo
  {author} {\bibfnamefont {B.}~\bibnamefont {Damski}}, \bibinfo {author}
  {\bibfnamefont {A.}~\bibnamefont {Sen(De)}}, \ and\ \bibinfo {author}
  {\bibfnamefont {U.}~\bibnamefont {Sen}},\ }\href {\doibase
  10.1080/00018730701223200} {\bibfield  {journal} {\bibinfo  {journal}
  {Advances in Physics}\ }\textbf {\bibinfo {volume} {56}},\ \bibinfo {pages}
  {243} (\bibinfo {year} {2007})}\BibitemShut {NoStop}%
\bibitem [{\citenamefont {Bloch}\ \emph {et~al.}(2012)\citenamefont {Bloch},
  \citenamefont {Dalibard},\ and\ \citenamefont {Nascimb{\`e}ne}}]{Bloch2012}%
  \BibitemOpen
  \bibfield  {author} {\bibinfo {author} {\bibfnamefont {I.}~\bibnamefont
  {Bloch}}, \bibinfo {author} {\bibfnamefont {J.}~\bibnamefont {Dalibard}}, \
  and\ \bibinfo {author} {\bibfnamefont {S.}~\bibnamefont {Nascimb{\`e}ne}},\
  }\href {\doibase 10.1038/nphys2259} {\bibfield  {journal} {\bibinfo
  {journal} {Nature Physics}\ }\textbf {\bibinfo {volume} {8}},\ \bibinfo
  {pages} {267} (\bibinfo {year} {2012})}\BibitemShut {NoStop}%
\bibitem [{\citenamefont {Blatt}\ and\ \citenamefont {Roos}(2012)}]{Blatt2012}%
  \BibitemOpen
  \bibfield  {author} {\bibinfo {author} {\bibfnamefont {R.}~\bibnamefont
  {Blatt}}\ and\ \bibinfo {author} {\bibfnamefont {C.~F.}\ \bibnamefont
  {Roos}},\ }\href {\doibase 10.1038/nphys2252} {\bibfield  {journal} {\bibinfo
   {journal} {Nature Physics}\ }\textbf {\bibinfo {volume} {8}},\ \bibinfo
  {pages} {277} (\bibinfo {year} {2012})}\BibitemShut {NoStop}%
\bibitem [{\citenamefont {Schwinger}(1962)}]{PhysRev.128.2425}%
  \BibitemOpen
  \bibfield  {author} {\bibinfo {author} {\bibfnamefont {J.}~\bibnamefont
  {Schwinger}},\ }\href {\doibase 10.1103/PhysRev.128.2425} {\bibfield
  {journal} {\bibinfo  {journal} {Phys. Rev.}\ }\textbf {\bibinfo {volume}
  {128}},\ \bibinfo {pages} {2425} (\bibinfo {year} {1962})}\BibitemShut
  {NoStop}%
\bibitem [{\citenamefont {Manton}(1985)}]{MANTON1985220}%
  \BibitemOpen
  \bibfield  {author} {\bibinfo {author} {\bibfnamefont {N.}~\bibnamefont
  {Manton}},\ }\href {\doibase https://doi.org/10.1016/0003-4916(85)90199-X}
  {\bibfield  {journal} {\bibinfo  {journal} {Annals of Physics}\ }\textbf
  {\bibinfo {volume} {159}},\ \bibinfo {pages} {220 } (\bibinfo {year}
  {1985})}\BibitemShut {NoStop}%
\bibitem [{\citenamefont {Thirring}(1958)}]{THIRRING195891}%
  \BibitemOpen
  \bibfield  {author} {\bibinfo {author} {\bibfnamefont {W.~E.}\ \bibnamefont
  {Thirring}},\ }\href {\doibase https://doi.org/10.1016/0003-4916(58)90015-0}
  {\bibfield  {journal} {\bibinfo  {journal} {Annals of Physics}\ }\textbf
  {\bibinfo {volume} {3}},\ \bibinfo {pages} {91 } (\bibinfo {year}
  {1958})}\BibitemShut {NoStop}%
\bibitem [{\citenamefont {Coleman}(1975)}]{PhysRevD.11.2088}%
  \BibitemOpen
  \bibfield  {author} {\bibinfo {author} {\bibfnamefont {S.}~\bibnamefont
  {Coleman}},\ }\href {\doibase 10.1103/PhysRevD.11.2088} {\bibfield  {journal}
  {\bibinfo  {journal} {Phys. Rev. D}\ }\textbf {\bibinfo {volume} {11}},\
  \bibinfo {pages} {2088} (\bibinfo {year} {1975})}\BibitemShut {NoStop}%
\bibitem [{\citenamefont {Gross}\ and\ \citenamefont
  {Neveu}(1974)}]{PhysRevD.10.3235}%
  \BibitemOpen
  \bibfield  {author} {\bibinfo {author} {\bibfnamefont {D.~J.}\ \bibnamefont
  {Gross}}\ and\ \bibinfo {author} {\bibfnamefont {A.}~\bibnamefont {Neveu}},\
  }\href {\doibase 10.1103/PhysRevD.10.3235} {\bibfield  {journal} {\bibinfo
  {journal} {Phys. Rev. D}\ }\textbf {\bibinfo {volume} {10}},\ \bibinfo
  {pages} {3235} (\bibinfo {year} {1974})}\BibitemShut {NoStop}%
\bibitem [{\citenamefont {Jackiw}\ and\ \citenamefont
  {Rebbi}(1976)}]{PhysRevD.13.3398}%
  \BibitemOpen
  \bibfield  {author} {\bibinfo {author} {\bibfnamefont {R.}~\bibnamefont
  {Jackiw}}\ and\ \bibinfo {author} {\bibfnamefont {C.}~\bibnamefont {Rebbi}},\
  }\href {\doibase 10.1103/PhysRevD.13.3398} {\bibfield  {journal} {\bibinfo
  {journal} {Phys. Rev. D}\ }\textbf {\bibinfo {volume} {13}},\ \bibinfo
  {pages} {3398} (\bibinfo {year} {1976})}\BibitemShut {NoStop}%
\bibitem [{\citenamefont {Ba{\~{n}}uls}\ \emph {et~al.}(2019)\citenamefont
  {Ba{\~{n}}uls}, \citenamefont {Blatt}, \citenamefont {Catani}, \citenamefont
  {Celi}, \citenamefont {Cirac}, \citenamefont {Dalmonte}, \citenamefont
  {Fallani}, \citenamefont {Jansen}, \citenamefont {Lewenstein}, \citenamefont
  {Montangero}, \citenamefont {Muschik}, \citenamefont {Reznik}, \citenamefont
  {Rico}, \citenamefont {Tagliacozzo}, \citenamefont {Acoleyen}, \citenamefont
  {Verstraete}, \citenamefont {Wiese}, \citenamefont {Wingate}, \citenamefont
  {Zakrzewski},\ and\ \citenamefont {Zoller}}]{1911.00003}%
  \BibitemOpen
  \bibfield  {author} {\bibinfo {author} {\bibfnamefont {M.~C.}\ \bibnamefont
  {Ba{\~{n}}uls}}, \bibinfo {author} {\bibfnamefont {R.}~\bibnamefont {Blatt}},
  \bibinfo {author} {\bibfnamefont {J.}~\bibnamefont {Catani}}, \bibinfo
  {author} {\bibfnamefont {A.}~\bibnamefont {Celi}}, \bibinfo {author}
  {\bibfnamefont {J.~I.}\ \bibnamefont {Cirac}}, \bibinfo {author}
  {\bibfnamefont {M.}~\bibnamefont {Dalmonte}}, \bibinfo {author}
  {\bibfnamefont {L.}~\bibnamefont {Fallani}}, \bibinfo {author} {\bibfnamefont
  {K.}~\bibnamefont {Jansen}}, \bibinfo {author} {\bibfnamefont
  {M.}~\bibnamefont {Lewenstein}}, \bibinfo {author} {\bibfnamefont
  {S.}~\bibnamefont {Montangero}}, \bibinfo {author} {\bibfnamefont {C.~A.}\
  \bibnamefont {Muschik}}, \bibinfo {author} {\bibfnamefont {B.}~\bibnamefont
  {Reznik}}, \bibinfo {author} {\bibfnamefont {E.}~\bibnamefont {Rico}},
  \bibinfo {author} {\bibfnamefont {L.}~\bibnamefont {Tagliacozzo}}, \bibinfo
  {author} {\bibfnamefont {K.~V.}\ \bibnamefont {Acoleyen}}, \bibinfo {author}
  {\bibfnamefont {F.}~\bibnamefont {Verstraete}}, \bibinfo {author}
  {\bibfnamefont {U.~J.}\ \bibnamefont {Wiese}}, \bibinfo {author}
  {\bibfnamefont {M.}~\bibnamefont {Wingate}}, \bibinfo {author} {\bibfnamefont
  {J.}~\bibnamefont {Zakrzewski}}, \ and\ \bibinfo {author} {\bibfnamefont
  {P.}~\bibnamefont {Zoller}},\ }\href@noop {} {\enquote {\bibinfo {title}
  {Simulating lattice gauge theories within quantum technologies},}\ }
  (\bibinfo {year} {2019}),\ \Eprint {http://arxiv.org/abs/arXiv:1911.00003}
  {arXiv:1911.00003} \BibitemShut {NoStop}%
\bibitem [{\citenamefont {Zohar}\ \emph {et~al.}(2015)\citenamefont {Zohar},
  \citenamefont {Cirac},\ and\ \citenamefont {Reznik}}]{Zohar_2015}%
  \BibitemOpen
  \bibfield  {author} {\bibinfo {author} {\bibfnamefont {E.}~\bibnamefont
  {Zohar}}, \bibinfo {author} {\bibfnamefont {J.~I.}\ \bibnamefont {Cirac}}, \
  and\ \bibinfo {author} {\bibfnamefont {B.}~\bibnamefont {Reznik}},\ }\href
  {\doibase 10.1088/0034-4885/79/1/014401} {\bibfield  {journal} {\bibinfo
  {journal} {Reports on Progress in Physics}\ }\textbf {\bibinfo {volume}
  {79}},\ \bibinfo {pages} {014401} (\bibinfo {year} {2015})}\BibitemShut
  {NoStop}%
\bibitem [{\citenamefont {Dalmonte}\ and\ \citenamefont
  {Montangero}(2016)}]{doi:10.1080/00107514.2016.1151199}%
  \BibitemOpen
  \bibfield  {author} {\bibinfo {author} {\bibfnamefont {M.}~\bibnamefont
  {Dalmonte}}\ and\ \bibinfo {author} {\bibfnamefont {S.}~\bibnamefont
  {Montangero}},\ }\href {\doibase 10.1080/00107514.2016.1151199} {\bibfield
  {journal} {\bibinfo  {journal} {Contemporary Physics}\ }\textbf {\bibinfo
  {volume} {57}},\ \bibinfo {pages} {388} (\bibinfo {year} {2016})}\BibitemShut
  {NoStop}%
\bibitem [{\citenamefont {Wiese}(2013)}]{doi:10.1002/andp.201300104}%
  \BibitemOpen
  \bibfield  {author} {\bibinfo {author} {\bibfnamefont {U.-J.}\ \bibnamefont
  {Wiese}},\ }\href {\doibase 10.1002/andp.201300104} {\bibfield  {journal}
  {\bibinfo  {journal} {Annalen der Physik}\ }\textbf {\bibinfo {volume}
  {525}},\ \bibinfo {pages} {777} (\bibinfo {year} {2013})}\BibitemShut
  {NoStop}%
\bibitem [{\citenamefont {Ba{\~{n}}uls}\ and\ \citenamefont
  {Cichy}(2020)}]{Carmen_Ba_uls_2020}%
  \BibitemOpen
  \bibfield  {author} {\bibinfo {author} {\bibfnamefont {M.~C.}\ \bibnamefont
  {Ba{\~{n}}uls}}\ and\ \bibinfo {author} {\bibfnamefont {K.}~\bibnamefont
  {Cichy}},\ }\href {\doibase 10.1088/1361-6633/ab6311} {\bibfield  {journal}
  {\bibinfo  {journal} {Reports on Progress in Physics}\ }\textbf {\bibinfo
  {volume} {83}},\ \bibinfo {pages} {024401} (\bibinfo {year}
  {2020})}\BibitemShut {NoStop}%
\bibitem [{\citenamefont {Kasper}\ \emph {et~al.}(2020)\citenamefont {Kasper},
  \citenamefont {Juzeliunas}, \citenamefont {Lewenstein}, \citenamefont
  {Jendrzejewski},\ and\ \citenamefont {Zohar}}]{2006.01258}%
  \BibitemOpen
  \bibfield  {author} {\bibinfo {author} {\bibfnamefont {V.}~\bibnamefont
  {Kasper}}, \bibinfo {author} {\bibfnamefont {G.}~\bibnamefont {Juzeliunas}},
  \bibinfo {author} {\bibfnamefont {M.}~\bibnamefont {Lewenstein}}, \bibinfo
  {author} {\bibfnamefont {F.}~\bibnamefont {Jendrzejewski}}, \ and\ \bibinfo
  {author} {\bibfnamefont {E.}~\bibnamefont {Zohar}},\ }\href@noop {} {\enquote
  {\bibinfo {title} {From the jaynes-cummings model to non-abelian gauge
  theories: a guided tour for the quantum engineer},}\ } (\bibinfo {year}
  {2020}),\ \Eprint {http://arxiv.org/abs/arXiv:2006.01258} {arXiv:2006.01258}
  \BibitemShut {NoStop}%
\bibitem [{\citenamefont {Greensite}(2020)}]{greensite_2020}%
  \BibitemOpen
  \bibfield  {author} {\bibinfo {author} {\bibfnamefont {J.}~\bibnamefont
  {Greensite}},\ }\href {https://www.springer.com/gp/book/9783642143816} {\emph
  {\bibinfo {title} {Introduction to the confinement problem}}}\ (\bibinfo
  {publisher} {Springer Nature},\ \bibinfo {year} {2020})\BibitemShut {NoStop}%
\bibitem [{\citenamefont {Kogut}\ and\ \citenamefont
  {Stephanov}(2003)}]{kogut_stephanov_2003}%
  \BibitemOpen
  \bibfield  {author} {\bibinfo {author} {\bibfnamefont {J.~B.}\ \bibnamefont
  {Kogut}}\ and\ \bibinfo {author} {\bibfnamefont {M.~A.}\ \bibnamefont
  {Stephanov}},\ }\href {\doibase 10.1017/CBO9780511534980} {\emph {\bibinfo
  {title} {The Phases of Quantum Chromodynamics: From Confinement to Extreme
  Environments}}},\ Cambridge Monographs on Particle Physics, Nuclear Physics
  and Cosmology\ (\bibinfo  {publisher} {Cambridge University Press},\ \bibinfo
  {year} {2003})\BibitemShut {NoStop}%
\bibitem [{\citenamefont {Bazavov}\ \emph {et~al.}(2019)\citenamefont
  {Bazavov}, \citenamefont {Ding}, \citenamefont {Hegde}, \citenamefont
  {Kaczmarek}, \citenamefont {Karsch}, \citenamefont {Karthik}, \citenamefont
  {Laermann}, \citenamefont {Lahiri}, \citenamefont {Larsen}, \citenamefont
  {Li}, \citenamefont {Mukherjee}, \citenamefont {Ohno}, \citenamefont
  {Petreczky}, \citenamefont {Sandmeyer}, \citenamefont {Schmidt},
  \citenamefont {Sharma},\ and\ \citenamefont {Steinbrecher}}]{Bazanov_2019}%
  \BibitemOpen
  \bibfield  {author} {\bibinfo {author} {\bibfnamefont {A.}~\bibnamefont
  {Bazavov}}, \bibinfo {author} {\bibfnamefont {H.-T.}\ \bibnamefont {Ding}},
  \bibinfo {author} {\bibfnamefont {P.}~\bibnamefont {Hegde}}, \bibinfo
  {author} {\bibfnamefont {O.}~\bibnamefont {Kaczmarek}}, \bibinfo {author}
  {\bibfnamefont {F.}~\bibnamefont {Karsch}}, \bibinfo {author} {\bibfnamefont
  {N.}~\bibnamefont {Karthik}}, \bibinfo {author} {\bibfnamefont
  {E.}~\bibnamefont {Laermann}}, \bibinfo {author} {\bibfnamefont
  {A.}~\bibnamefont {Lahiri}}, \bibinfo {author} {\bibfnamefont
  {R.}~\bibnamefont {Larsen}}, \bibinfo {author} {\bibfnamefont {S.-T.}\
  \bibnamefont {Li}}, \bibinfo {author} {\bibfnamefont {S.}~\bibnamefont
  {Mukherjee}}, \bibinfo {author} {\bibfnamefont {H.}~\bibnamefont {Ohno}},
  \bibinfo {author} {\bibfnamefont {P.}~\bibnamefont {Petreczky}}, \bibinfo
  {author} {\bibfnamefont {H.}~\bibnamefont {Sandmeyer}}, \bibinfo {author}
  {\bibfnamefont {C.}~\bibnamefont {Schmidt}}, \bibinfo {author} {\bibfnamefont
  {S.}~\bibnamefont {Sharma}}, \ and\ \bibinfo {author} {\bibfnamefont
  {P.}~\bibnamefont {Steinbrecher}},\ }\href {\doibase
  https://doi.org/10.1016/j.physletb.2019.05.013} {\bibfield  {journal}
  {\bibinfo  {journal} {Physics Letters B}\ }\textbf {\bibinfo {volume}
  {795}},\ \bibinfo {pages} {15 } (\bibinfo {year} {2019})}\BibitemShut
  {NoStop}%
\bibitem [{\citenamefont {Dai}\ \emph {et~al.}(2017)\citenamefont {Dai},
  \citenamefont {Yang}, \citenamefont {Reingruber}, \citenamefont {Sun},
  \citenamefont {Xu}, \citenamefont {Chen}, \citenamefont {Yuan},\ and\
  \citenamefont {Pan}}]{dai_four-body_2017}%
  \BibitemOpen
  \bibfield  {author} {\bibinfo {author} {\bibfnamefont {H.-N.}\ \bibnamefont
  {Dai}}, \bibinfo {author} {\bibfnamefont {B.}~\bibnamefont {Yang}}, \bibinfo
  {author} {\bibfnamefont {A.}~\bibnamefont {Reingruber}}, \bibinfo {author}
  {\bibfnamefont {H.}~\bibnamefont {Sun}}, \bibinfo {author} {\bibfnamefont
  {X.-F.}\ \bibnamefont {Xu}}, \bibinfo {author} {\bibfnamefont {Y.-A.}\
  \bibnamefont {Chen}}, \bibinfo {author} {\bibfnamefont {Z.-S.}\ \bibnamefont
  {Yuan}}, \ and\ \bibinfo {author} {\bibfnamefont {J.-W.}\ \bibnamefont
  {Pan}},\ }\href {\doibase 10.1038/nphys4243} {\bibfield  {journal} {\bibinfo
  {journal} {Nature Physics}\ }\textbf {\bibinfo {volume} {13}},\ \bibinfo
  {pages} {1195} (\bibinfo {year} {2017})}\BibitemShut {NoStop}%
\bibitem [{\citenamefont {Zohar}\ and\ \citenamefont
  {Reznik}(2011)}]{PhysRevLett.107.275301}%
  \BibitemOpen
  \bibfield  {author} {\bibinfo {author} {\bibfnamefont {E.}~\bibnamefont
  {Zohar}}\ and\ \bibinfo {author} {\bibfnamefont {B.}~\bibnamefont {Reznik}},\
  }\href {\doibase 10.1103/PhysRevLett.107.275301} {\bibfield  {journal}
  {\bibinfo  {journal} {Phys. Rev. Lett.}\ }\textbf {\bibinfo {volume} {107}},\
  \bibinfo {pages} {275301} (\bibinfo {year} {2011})}\BibitemShut {NoStop}%
\bibitem [{\citenamefont {Zohar}\ \emph {et~al.}(2012)\citenamefont {Zohar},
  \citenamefont {Cirac},\ and\ \citenamefont
  {Reznik}}]{PhysRevLett.109.125302}%
  \BibitemOpen
  \bibfield  {author} {\bibinfo {author} {\bibfnamefont {E.}~\bibnamefont
  {Zohar}}, \bibinfo {author} {\bibfnamefont {J.~I.}\ \bibnamefont {Cirac}}, \
  and\ \bibinfo {author} {\bibfnamefont {B.}~\bibnamefont {Reznik}},\ }\href
  {\doibase 10.1103/PhysRevLett.109.125302} {\bibfield  {journal} {\bibinfo
  {journal} {Phys. Rev. Lett.}\ }\textbf {\bibinfo {volume} {109}},\ \bibinfo
  {pages} {125302} (\bibinfo {year} {2012})}\BibitemShut {NoStop}%
\bibitem [{\citenamefont {Tagliacozzo}\ \emph
  {et~al.}(2013{\natexlab{a}})\citenamefont {Tagliacozzo}, \citenamefont
  {Celi}, \citenamefont {Zamora},\ and\ \citenamefont
  {Lewenstein}}]{TAGLIACOZZO2013160}%
  \BibitemOpen
  \bibfield  {author} {\bibinfo {author} {\bibfnamefont {L.}~\bibnamefont
  {Tagliacozzo}}, \bibinfo {author} {\bibfnamefont {A.}~\bibnamefont {Celi}},
  \bibinfo {author} {\bibfnamefont {A.}~\bibnamefont {Zamora}}, \ and\ \bibinfo
  {author} {\bibfnamefont {M.}~\bibnamefont {Lewenstein}},\ }\href {\doibase
  https://doi.org/10.1016/j.aop.2012.11.009} {\bibfield  {journal} {\bibinfo
  {journal} {Annals of Physics}\ }\textbf {\bibinfo {volume} {330}},\ \bibinfo
  {pages} {160 } (\bibinfo {year} {2013}{\natexlab{a}})}\BibitemShut {NoStop}%
\bibitem [{\citenamefont {Tagliacozzo}\ \emph
  {et~al.}(2013{\natexlab{b}})\citenamefont {Tagliacozzo}, \citenamefont
  {Celi}, \citenamefont {Orland}, \citenamefont {Mitchell},\ and\ \citenamefont
  {Lewenstein}}]{Tagliacozzo2013}%
  \BibitemOpen
  \bibfield  {author} {\bibinfo {author} {\bibfnamefont {L.}~\bibnamefont
  {Tagliacozzo}}, \bibinfo {author} {\bibfnamefont {A.}~\bibnamefont {Celi}},
  \bibinfo {author} {\bibfnamefont {P.}~\bibnamefont {Orland}}, \bibinfo
  {author} {\bibfnamefont {M.~W.}\ \bibnamefont {Mitchell}}, \ and\ \bibinfo
  {author} {\bibfnamefont {M.}~\bibnamefont {Lewenstein}},\ }\href {\doibase
  10.1038/ncomms3615} {\bibfield  {journal} {\bibinfo  {journal} {Nature
  Communications}\ }\textbf {\bibinfo {volume} {4}},\ \bibinfo {pages} {2615}
  (\bibinfo {year} {2013}{\natexlab{b}})}\BibitemShut {NoStop}%
\bibitem [{\citenamefont {Brennen}\ \emph {et~al.}(2016)\citenamefont
  {Brennen}, \citenamefont {Pupillo}, \citenamefont {Rico}, \citenamefont
  {Stace},\ and\ \citenamefont {Vodola}}]{PhysRevLett.117.240504}%
  \BibitemOpen
  \bibfield  {author} {\bibinfo {author} {\bibfnamefont {G.~K.}\ \bibnamefont
  {Brennen}}, \bibinfo {author} {\bibfnamefont {G.}~\bibnamefont {Pupillo}},
  \bibinfo {author} {\bibfnamefont {E.}~\bibnamefont {Rico}}, \bibinfo {author}
  {\bibfnamefont {T.~M.}\ \bibnamefont {Stace}}, \ and\ \bibinfo {author}
  {\bibfnamefont {D.}~\bibnamefont {Vodola}},\ }\href {\doibase
  10.1103/PhysRevLett.117.240504} {\bibfield  {journal} {\bibinfo  {journal}
  {Phys. Rev. Lett.}\ }\textbf {\bibinfo {volume} {117}},\ \bibinfo {pages}
  {240504} (\bibinfo {year} {2016})}\BibitemShut {NoStop}%
\bibitem [{\citenamefont {Rico}\ \emph {et~al.}(2018)\citenamefont {Rico},
  \citenamefont {Dalmonte}, \citenamefont {Zoller}, \citenamefont {Banerjee},
  \citenamefont {Bögli}, \citenamefont {Stebler},\ and\ \citenamefont
  {Wiese}}]{RICO2018466}%
  \BibitemOpen
  \bibfield  {author} {\bibinfo {author} {\bibfnamefont {E.}~\bibnamefont
  {Rico}}, \bibinfo {author} {\bibfnamefont {M.}~\bibnamefont {Dalmonte}},
  \bibinfo {author} {\bibfnamefont {P.}~\bibnamefont {Zoller}}, \bibinfo
  {author} {\bibfnamefont {D.}~\bibnamefont {Banerjee}}, \bibinfo {author}
  {\bibfnamefont {M.}~\bibnamefont {Bögli}}, \bibinfo {author} {\bibfnamefont
  {P.}~\bibnamefont {Stebler}}, \ and\ \bibinfo {author} {\bibfnamefont
  {U.-J.}\ \bibnamefont {Wiese}},\ }\href {\doibase
  https://doi.org/10.1016/j.aop.2018.03.020} {\bibfield  {journal} {\bibinfo
  {journal} {Annals of Physics}\ }\textbf {\bibinfo {volume} {393}},\ \bibinfo
  {pages} {466 } (\bibinfo {year} {2018})}\BibitemShut {NoStop}%
\bibitem [{\citenamefont {Park}\ \emph {et~al.}(2019)\citenamefont {Park},
  \citenamefont {Kuno},\ and\ \citenamefont {Ichinose}}]{PhysRevA.100.013629}%
  \BibitemOpen
  \bibfield  {author} {\bibinfo {author} {\bibfnamefont {J.}~\bibnamefont
  {Park}}, \bibinfo {author} {\bibfnamefont {Y.}~\bibnamefont {Kuno}}, \ and\
  \bibinfo {author} {\bibfnamefont {I.}~\bibnamefont {Ichinose}},\ }\href
  {\doibase 10.1103/PhysRevA.100.013629} {\bibfield  {journal} {\bibinfo
  {journal} {Phys. Rev. A}\ }\textbf {\bibinfo {volume} {100}},\ \bibinfo
  {pages} {013629} (\bibinfo {year} {2019})}\BibitemShut {NoStop}%
\bibitem [{\citenamefont {Liu}\ \emph {et~al.}(2019)\citenamefont {Liu},
  \citenamefont {Lundgren}, \citenamefont {Titum}, \citenamefont {Pagano},
  \citenamefont {Zhang}, \citenamefont {Monroe},\ and\ \citenamefont
  {Gorshkov}}]{PhysRevLett.122.150601}%
  \BibitemOpen
  \bibfield  {author} {\bibinfo {author} {\bibfnamefont {F.}~\bibnamefont
  {Liu}}, \bibinfo {author} {\bibfnamefont {R.}~\bibnamefont {Lundgren}},
  \bibinfo {author} {\bibfnamefont {P.}~\bibnamefont {Titum}}, \bibinfo
  {author} {\bibfnamefont {G.}~\bibnamefont {Pagano}}, \bibinfo {author}
  {\bibfnamefont {J.}~\bibnamefont {Zhang}}, \bibinfo {author} {\bibfnamefont
  {C.}~\bibnamefont {Monroe}}, \ and\ \bibinfo {author} {\bibfnamefont {A.~V.}\
  \bibnamefont {Gorshkov}},\ }\href {\doibase 10.1103/PhysRevLett.122.150601}
  {\bibfield  {journal} {\bibinfo  {journal} {Phys. Rev. Lett.}\ }\textbf
  {\bibinfo {volume} {122}},\ \bibinfo {pages} {150601} (\bibinfo {year}
  {2019})}\BibitemShut {NoStop}%
\bibitem [{\citenamefont {Tan}\ \emph {et~al.}(2019)\citenamefont {Tan},
  \citenamefont {Becker}, \citenamefont {Liu}, \citenamefont {Pagano},
  \citenamefont {Collins}, \citenamefont {De}, \citenamefont {Feng},
  \citenamefont {Kaplan}, \citenamefont {Kyprianidis}, \citenamefont
  {Lundgren}, \citenamefont {Morong}, \citenamefont {Whitsitt}, \citenamefont
  {Gorshkov},\ and\ \citenamefont {Monroe}}]{1912.11117}%
  \BibitemOpen
  \bibfield  {author} {\bibinfo {author} {\bibfnamefont {W.~L.}\ \bibnamefont
  {Tan}}, \bibinfo {author} {\bibfnamefont {P.}~\bibnamefont {Becker}},
  \bibinfo {author} {\bibfnamefont {F.}~\bibnamefont {Liu}}, \bibinfo {author}
  {\bibfnamefont {G.}~\bibnamefont {Pagano}}, \bibinfo {author} {\bibfnamefont
  {K.~S.}\ \bibnamefont {Collins}}, \bibinfo {author} {\bibfnamefont
  {A.}~\bibnamefont {De}}, \bibinfo {author} {\bibfnamefont {L.}~\bibnamefont
  {Feng}}, \bibinfo {author} {\bibfnamefont {H.~B.}\ \bibnamefont {Kaplan}},
  \bibinfo {author} {\bibfnamefont {A.}~\bibnamefont {Kyprianidis}}, \bibinfo
  {author} {\bibfnamefont {R.}~\bibnamefont {Lundgren}}, \bibinfo {author}
  {\bibfnamefont {W.}~\bibnamefont {Morong}}, \bibinfo {author} {\bibfnamefont
  {S.}~\bibnamefont {Whitsitt}}, \bibinfo {author} {\bibfnamefont {A.~V.}\
  \bibnamefont {Gorshkov}}, \ and\ \bibinfo {author} {\bibfnamefont
  {C.}~\bibnamefont {Monroe}},\ }\href@noop {} {\enquote {\bibinfo {title}
  {Observation of domain wall confinement and dynamics in a quantum
  simulator},}\ } (\bibinfo {year} {2019}),\ \Eprint
  {http://arxiv.org/abs/arXiv:1912.11117} {arXiv:1912.11117} \BibitemShut
  {NoStop}%
\bibitem [{\citenamefont {Borla}\ \emph {et~al.}(2020)\citenamefont {Borla},
  \citenamefont {Verresen}, \citenamefont {Grusdt},\ and\ \citenamefont
  {Moroz}}]{PhysRevLett.124.120503}%
  \BibitemOpen
  \bibfield  {author} {\bibinfo {author} {\bibfnamefont {U.}~\bibnamefont
  {Borla}}, \bibinfo {author} {\bibfnamefont {R.}~\bibnamefont {Verresen}},
  \bibinfo {author} {\bibfnamefont {F.}~\bibnamefont {Grusdt}}, \ and\ \bibinfo
  {author} {\bibfnamefont {S.}~\bibnamefont {Moroz}},\ }\href {\doibase
  10.1103/PhysRevLett.124.120503} {\bibfield  {journal} {\bibinfo  {journal}
  {Phys. Rev. Lett.}\ }\textbf {\bibinfo {volume} {124}},\ \bibinfo {pages}
  {120503} (\bibinfo {year} {2020})}\BibitemShut {NoStop}%
\bibitem [{\citenamefont {Notarnicola}\ \emph {et~al.}(2020)\citenamefont
  {Notarnicola}, \citenamefont {Collura},\ and\ \citenamefont
  {Montangero}}]{PhysRevResearch.2.013288}%
  \BibitemOpen
  \bibfield  {author} {\bibinfo {author} {\bibfnamefont {S.}~\bibnamefont
  {Notarnicola}}, \bibinfo {author} {\bibfnamefont {M.}~\bibnamefont
  {Collura}}, \ and\ \bibinfo {author} {\bibfnamefont {S.}~\bibnamefont
  {Montangero}},\ }\href {\doibase 10.1103/PhysRevResearch.2.013288} {\bibfield
   {journal} {\bibinfo  {journal} {Phys. Rev. Research}\ }\textbf {\bibinfo
  {volume} {2}},\ \bibinfo {pages} {013288} (\bibinfo {year}
  {2020})}\BibitemShut {NoStop}%
\bibitem [{\citenamefont {Surace}\ \emph {et~al.}(2020)\citenamefont {Surace},
  \citenamefont {Mazza}, \citenamefont {Giudici}, \citenamefont {Lerose},
  \citenamefont {Gambassi},\ and\ \citenamefont
  {Dalmonte}}]{PhysRevX.10.021041}%
  \BibitemOpen
  \bibfield  {author} {\bibinfo {author} {\bibfnamefont {F.~M.}\ \bibnamefont
  {Surace}}, \bibinfo {author} {\bibfnamefont {P.~P.}\ \bibnamefont {Mazza}},
  \bibinfo {author} {\bibfnamefont {G.}~\bibnamefont {Giudici}}, \bibinfo
  {author} {\bibfnamefont {A.}~\bibnamefont {Lerose}}, \bibinfo {author}
  {\bibfnamefont {A.}~\bibnamefont {Gambassi}}, \ and\ \bibinfo {author}
  {\bibfnamefont {M.}~\bibnamefont {Dalmonte}},\ }\href {\doibase
  10.1103/PhysRevX.10.021041} {\bibfield  {journal} {\bibinfo  {journal} {Phys.
  Rev. X}\ }\textbf {\bibinfo {volume} {10}},\ \bibinfo {pages} {021041}
  (\bibinfo {year} {2020})}\BibitemShut {NoStop}%
\bibitem [{\citenamefont {Magnifico}\ \emph {et~al.}(2020)\citenamefont
  {Magnifico}, \citenamefont {Dalmonte}, \citenamefont {Facchi}, \citenamefont
  {Pascazio}, \citenamefont {Pepe},\ and\ \citenamefont
  {Ercolessi}}]{Magnifico2020realtimedynamics}%
  \BibitemOpen
  \bibfield  {author} {\bibinfo {author} {\bibfnamefont {G.}~\bibnamefont
  {Magnifico}}, \bibinfo {author} {\bibfnamefont {M.}~\bibnamefont {Dalmonte}},
  \bibinfo {author} {\bibfnamefont {P.}~\bibnamefont {Facchi}}, \bibinfo
  {author} {\bibfnamefont {S.}~\bibnamefont {Pascazio}}, \bibinfo {author}
  {\bibfnamefont {F.~V.}\ \bibnamefont {Pepe}}, \ and\ \bibinfo {author}
  {\bibfnamefont {E.}~\bibnamefont {Ercolessi}},\ }\href {\doibase
  10.22331/q-2020-06-15-281} {\bibfield  {journal} {\bibinfo  {journal}
  {{Quantum}}\ }\textbf {\bibinfo {volume} {4}},\ \bibinfo {pages} {281}
  (\bibinfo {year} {2020})}\BibitemShut {NoStop}%
\bibitem [{\citenamefont {Gonz\'alez-Cuadra}\ \emph
  {et~al.}(2020{\natexlab{a}})\citenamefont {Gonz\'alez-Cuadra}, \citenamefont
  {Tagliacozzo}, \citenamefont {Lewenstein},\ and\ \citenamefont
  {Bermudez}}]{2002.06013}%
  \BibitemOpen
  \bibfield  {author} {\bibinfo {author} {\bibfnamefont {D.}~\bibnamefont
  {Gonz\'alez-Cuadra}}, \bibinfo {author} {\bibfnamefont {L.}~\bibnamefont
  {Tagliacozzo}}, \bibinfo {author} {\bibfnamefont {M.}~\bibnamefont
  {Lewenstein}}, \ and\ \bibinfo {author} {\bibfnamefont {A.}~\bibnamefont
  {Bermudez}},\ }\href@noop {} {\enquote {\bibinfo {title} {Aharonov-bohm
  instability in fermionic $\mathbb{Z}_2$ gauge theories: topological order and
  soliton-induced deconfinement},}\ } (\bibinfo {year} {2020}{\natexlab{a}}),\
  \Eprint {http://arxiv.org/abs/arXiv:2002.06013} {arXiv:2002.06013}
  \BibitemShut {NoStop}%
\bibitem [{\citenamefont {Chanda}\ \emph {et~al.}(2020)\citenamefont {Chanda},
  \citenamefont {Zakrzewski}, \citenamefont {Lewenstein},\ and\ \citenamefont
  {Tagliacozzo}}]{PhysRevLett.124.180602}%
  \BibitemOpen
  \bibfield  {author} {\bibinfo {author} {\bibfnamefont {T.}~\bibnamefont
  {Chanda}}, \bibinfo {author} {\bibfnamefont {J.}~\bibnamefont {Zakrzewski}},
  \bibinfo {author} {\bibfnamefont {M.}~\bibnamefont {Lewenstein}}, \ and\
  \bibinfo {author} {\bibfnamefont {L.}~\bibnamefont {Tagliacozzo}},\ }\href
  {\doibase 10.1103/PhysRevLett.124.180602} {\bibfield  {journal} {\bibinfo
  {journal} {Phys. Rev. Lett.}\ }\textbf {\bibinfo {volume} {124}},\ \bibinfo
  {pages} {180602} (\bibinfo {year} {2020})}\BibitemShut {NoStop}%
\bibitem [{\citenamefont {Liu}\ \emph {et~al.}(2020)\citenamefont {Liu},
  \citenamefont {Whitsitt}, \citenamefont {Bienias}, \citenamefont {Lundgren},\
  and\ \citenamefont {Gorshkov}}]{2007.07258}%
  \BibitemOpen
  \bibfield  {author} {\bibinfo {author} {\bibfnamefont {F.}~\bibnamefont
  {Liu}}, \bibinfo {author} {\bibfnamefont {S.}~\bibnamefont {Whitsitt}},
  \bibinfo {author} {\bibfnamefont {P.}~\bibnamefont {Bienias}}, \bibinfo
  {author} {\bibfnamefont {R.}~\bibnamefont {Lundgren}}, \ and\ \bibinfo
  {author} {\bibfnamefont {A.~V.}\ \bibnamefont {Gorshkov}},\ }\href@noop {}
  {\enquote {\bibinfo {title} {Realizing and probing baryonic excitations in
  rydberg atom arrays},}\ } (\bibinfo {year} {2020}),\ \Eprint
  {http://arxiv.org/abs/arXiv:2007.07258} {arXiv:2007.07258} \BibitemShut
  {NoStop}%
\bibitem [{\citenamefont {Polyakov}(1975)}]{POLYAKOV197579}%
  \BibitemOpen
  \bibfield  {author} {\bibinfo {author} {\bibfnamefont {A.}~\bibnamefont
  {Polyakov}},\ }\href {\doibase https://doi.org/10.1016/0370-2693(75)90161-6}
  {\bibfield  {journal} {\bibinfo  {journal} {Physics Letters B}\ }\textbf
  {\bibinfo {volume} {59}},\ \bibinfo {pages} {79 } (\bibinfo {year}
  {1975})}\BibitemShut {NoStop}%
\bibitem [{\citenamefont {Coleman}(1973)}]{Coleman1973}%
  \BibitemOpen
  \bibfield  {author} {\bibinfo {author} {\bibfnamefont {S.}~\bibnamefont
  {Coleman}},\ }\href {\doibase 10.1007/BF01646487} {\bibfield  {journal}
  {\bibinfo  {journal} {Communications in Mathematical Physics}\ }\textbf
  {\bibinfo {volume} {31}},\ \bibinfo {pages} {259} (\bibinfo {year}
  {1973})}\BibitemShut {NoStop}%
\bibitem [{\citenamefont {Haldane}(1983)}]{HALDANE1983464}%
  \BibitemOpen
  \bibfield  {author} {\bibinfo {author} {\bibfnamefont {F.}~\bibnamefont
  {Haldane}},\ }\href {\doibase https://doi.org/10.1016/0375-9601(83)90631-X}
  {\bibfield  {journal} {\bibinfo  {journal} {Physics Letters A}\ }\textbf
  {\bibinfo {volume} {93}},\ \bibinfo {pages} {464 } (\bibinfo {year}
  {1983})}\BibitemShut {NoStop}%
\bibitem [{\citenamefont {Affleck}(1985)}]{AFFLECK1985397}%
  \BibitemOpen
  \bibfield  {author} {\bibinfo {author} {\bibfnamefont {I.}~\bibnamefont
  {Affleck}},\ }\href {\doibase https://doi.org/10.1016/0550-3213(85)90353-0}
  {\bibfield  {journal} {\bibinfo  {journal} {Nuclear Physics B}\ }\textbf
  {\bibinfo {volume} {257}},\ \bibinfo {pages} {397 } (\bibinfo {year}
  {1985})}\BibitemShut {NoStop}%
\bibitem [{\citenamefont {Sachdev}(2011)}]{sachdev_2011}%
  \BibitemOpen
  \bibfield  {author} {\bibinfo {author} {\bibfnamefont {S.}~\bibnamefont
  {Sachdev}},\ }\href {\doibase 10.1017/CBO9780511973765} {\emph {\bibinfo
  {title} {Quantum Phase Transitions}}},\ \bibinfo {edition} {2nd}\ ed.\
  (\bibinfo  {publisher} {Cambridge University Press},\ \bibinfo {year}
  {2011})\BibitemShut {NoStop}%
\bibitem [{\citenamefont {Kondo}(1964)}]{10.1143/PTP.32.37}%
  \BibitemOpen
  \bibfield  {author} {\bibinfo {author} {\bibfnamefont {J.}~\bibnamefont
  {Kondo}},\ }\href {\doibase 10.1143/PTP.32.37} {\bibfield  {journal}
  {\bibinfo  {journal} {Progress of Theoretical Physics}\ }\textbf {\bibinfo
  {volume} {32}},\ \bibinfo {pages} {37} (\bibinfo {year} {1964})}\BibitemShut
  {NoStop}%
\bibitem [{\citenamefont {Jaksch}\ \emph {et~al.}(1998)\citenamefont {Jaksch},
  \citenamefont {Bruder}, \citenamefont {Cirac}, \citenamefont {Gardiner},\
  and\ \citenamefont {Zoller}}]{PhysRevLett.81.3108}%
  \BibitemOpen
  \bibfield  {author} {\bibinfo {author} {\bibfnamefont {D.}~\bibnamefont
  {Jaksch}}, \bibinfo {author} {\bibfnamefont {C.}~\bibnamefont {Bruder}},
  \bibinfo {author} {\bibfnamefont {J.~I.}\ \bibnamefont {Cirac}}, \bibinfo
  {author} {\bibfnamefont {C.~W.}\ \bibnamefont {Gardiner}}, \ and\ \bibinfo
  {author} {\bibfnamefont {P.}~\bibnamefont {Zoller}},\ }\href {\doibase
  10.1103/PhysRevLett.81.3108} {\bibfield  {journal} {\bibinfo  {journal}
  {Phys. Rev. Lett.}\ }\textbf {\bibinfo {volume} {81}},\ \bibinfo {pages}
  {3108} (\bibinfo {year} {1998})}\BibitemShut {NoStop}%
\bibitem [{\citenamefont {Hofstetter}\ \emph {et~al.}(2002)\citenamefont
  {Hofstetter}, \citenamefont {Cirac}, \citenamefont {Zoller}, \citenamefont
  {Demler},\ and\ \citenamefont {Lukin}}]{PhysRevLett.89.220407}%
  \BibitemOpen
  \bibfield  {author} {\bibinfo {author} {\bibfnamefont {W.}~\bibnamefont
  {Hofstetter}}, \bibinfo {author} {\bibfnamefont {J.~I.}\ \bibnamefont
  {Cirac}}, \bibinfo {author} {\bibfnamefont {P.}~\bibnamefont {Zoller}},
  \bibinfo {author} {\bibfnamefont {E.}~\bibnamefont {Demler}}, \ and\ \bibinfo
  {author} {\bibfnamefont {M.~D.}\ \bibnamefont {Lukin}},\ }\href {\doibase
  10.1103/PhysRevLett.89.220407} {\bibfield  {journal} {\bibinfo  {journal}
  {Phys. Rev. Lett.}\ }\textbf {\bibinfo {volume} {89}},\ \bibinfo {pages}
  {220407} (\bibinfo {year} {2002})}\BibitemShut {NoStop}%
\bibitem [{\citenamefont {Albus}\ \emph {et~al.}(2003)\citenamefont {Albus},
  \citenamefont {Illuminati},\ and\ \citenamefont
  {Eisert}}]{PhysRevA.68.023606}%
  \BibitemOpen
  \bibfield  {author} {\bibinfo {author} {\bibfnamefont {A.}~\bibnamefont
  {Albus}}, \bibinfo {author} {\bibfnamefont {F.}~\bibnamefont {Illuminati}}, \
  and\ \bibinfo {author} {\bibfnamefont {J.}~\bibnamefont {Eisert}},\ }\href
  {\doibase 10.1103/PhysRevA.68.023606} {\bibfield  {journal} {\bibinfo
  {journal} {Phys. Rev. A}\ }\textbf {\bibinfo {volume} {68}},\ \bibinfo
  {pages} {023606} (\bibinfo {year} {2003})}\BibitemShut {NoStop}%
\bibitem [{\citenamefont {Jordan}\ and\ \citenamefont
  {Wigner}(1928)}]{Jordan1928}%
  \BibitemOpen
  \bibfield  {author} {\bibinfo {author} {\bibfnamefont {P.}~\bibnamefont
  {Jordan}}\ and\ \bibinfo {author} {\bibfnamefont {E.}~\bibnamefont
  {Wigner}},\ }\href {\doibase 10.1007/BF01331938} {\bibfield  {journal}
  {\bibinfo  {journal} {Zeitschrift f{\"u}r Physik}\ }\textbf {\bibinfo
  {volume} {47}},\ \bibinfo {pages} {631} (\bibinfo {year} {1928})}\BibitemShut
  {NoStop}%
\bibitem [{\citenamefont {Haldane}(1981)}]{Haldane_1981}%
  \BibitemOpen
  \bibfield  {author} {\bibinfo {author} {\bibfnamefont {F.~D.~M.}\
  \bibnamefont {Haldane}},\ }\href {\doibase 10.1088/0022-3719/14/19/010}
  {\bibfield  {journal} {\bibinfo  {journal} {Journal of Physics C: Solid State
  Physics}\ }\textbf {\bibinfo {volume} {14}},\ \bibinfo {pages} {2585}
  (\bibinfo {year} {1981})}\BibitemShut {NoStop}%
\bibitem [{\citenamefont {Tomonaga}(1950)}]{10.1143/ptp/5.4.544}%
  \BibitemOpen
  \bibfield  {author} {\bibinfo {author} {\bibfnamefont {S.}~\bibnamefont
  {Tomonaga}},\ }\href {\doibase 10.1143/ptp/5.4.544} {\bibfield  {journal}
  {\bibinfo  {journal} {Progress of Theoretical Physics}\ }\textbf {\bibinfo
  {volume} {5}},\ \bibinfo {pages} {544} (\bibinfo {year} {1950})}\BibitemShut
  {NoStop}%
\bibitem [{\citenamefont {Luttinger}(1963)}]{doi:10.1063/1.1704046}%
  \BibitemOpen
  \bibfield  {author} {\bibinfo {author} {\bibfnamefont {J.~M.}\ \bibnamefont
  {Luttinger}},\ }\href {\doibase 10.1063/1.1704046} {\bibfield  {journal}
  {\bibinfo  {journal} {Journal of Mathematical Physics}\ }\textbf {\bibinfo
  {volume} {4}},\ \bibinfo {pages} {1154} (\bibinfo {year} {1963})}\BibitemShut
  {NoStop}%
\bibitem [{\citenamefont {Kogut}\ and\ \citenamefont
  {Susskind}(1975)}]{PhysRevD.11.395}%
  \BibitemOpen
  \bibfield  {author} {\bibinfo {author} {\bibfnamefont {J.}~\bibnamefont
  {Kogut}}\ and\ \bibinfo {author} {\bibfnamefont {L.}~\bibnamefont
  {Susskind}},\ }\href {\doibase 10.1103/PhysRevD.11.395} {\bibfield  {journal}
  {\bibinfo  {journal} {Phys. Rev. D}\ }\textbf {\bibinfo {volume} {11}},\
  \bibinfo {pages} {395} (\bibinfo {year} {1975})}\BibitemShut {NoStop}%
\bibitem [{\citenamefont {Coleman}(1985)}]{coleman_1985}%
  \BibitemOpen
  \bibfield  {author} {\bibinfo {author} {\bibfnamefont {S.}~\bibnamefont
  {Coleman}},\ }\href {\doibase 10.1017/CBO9780511565045} {\emph {\bibinfo
  {title} {Aspects of Symmetry: Selected Erice Lectures}}}\ (\bibinfo
  {publisher} {Cambridge University Press},\ \bibinfo {year}
  {1985})\BibitemShut {NoStop}%
\bibitem [{\citenamefont {Radcliffe}(1971)}]{Radcliffe_1971}%
  \BibitemOpen
  \bibfield  {author} {\bibinfo {author} {\bibfnamefont {J.~M.}\ \bibnamefont
  {Radcliffe}},\ }\href {\doibase 10.1088/0305-4470/4/3/009} {\bibfield
  {journal} {\bibinfo  {journal} {Journal of Physics A: General Physics}\
  }\textbf {\bibinfo {volume} {4}},\ \bibinfo {pages} {313} (\bibinfo {year}
  {1971})}\BibitemShut {NoStop}%
\bibitem [{\citenamefont {Fradkin}(2013)}]{fradkin_2013}%
  \BibitemOpen
  \bibfield  {author} {\bibinfo {author} {\bibfnamefont {E.}~\bibnamefont
  {Fradkin}},\ }\href {\doibase 10.1017/CBO9781139015509} {\emph {\bibinfo
  {title} {Field Theories of Condensed Matter Physics}}},\ \bibinfo {edition}
  {2nd}\ ed.\ (\bibinfo  {publisher} {Cambridge University Press},\ \bibinfo
  {year} {2013})\BibitemShut {NoStop}%
\bibitem [{\citenamefont {Peierls}(1955)}]{peierls}%
  \BibitemOpen
  \bibfield  {author} {\bibinfo {author} {\bibfnamefont {R.}~\bibnamefont
  {Peierls}},\ }\href {\doibase 10.1093/acprof:oso/9780198507819.001.0001}
  {\emph {\bibinfo {title} {Quantum Theory of Solids}}},\ International series
  of monographs on physics\ (\bibinfo  {publisher} {Clarendon Press},\ \bibinfo
  {year} {1955})\BibitemShut {NoStop}%
\bibitem [{\citenamefont {Giuliani}\ and\ \citenamefont
  {Vignale}(2008)}]{giuliani_vignale_2008}%
  \BibitemOpen
  \bibfield  {author} {\bibinfo {author} {\bibfnamefont {G.}~\bibnamefont
  {Giuliani}}\ and\ \bibinfo {author} {\bibfnamefont {G.}~\bibnamefont
  {Vignale}},\ }\href
  {https://www.cambridge.org/core/books/quantum-theory-of-the-electron-liquid/EA75F41350A1C41D5E1BD202D539BB9E}
  {\emph {\bibinfo {title} {Quantum theory of the electron liquid}}}\ (\bibinfo
   {publisher} {Cambridge University Press},\ \bibinfo {year}
  {2008})\BibitemShut {NoStop}%
\bibitem [{\citenamefont {Schollwöck}(2011)}]{SCHOLLWOCK201196}%
  \BibitemOpen
  \bibfield  {author} {\bibinfo {author} {\bibfnamefont {U.}~\bibnamefont
  {Schollwöck}},\ }\href {\doibase https://doi.org/10.1016/j.aop.2010.09.012}
  {\bibfield  {journal} {\bibinfo  {journal} {Annals of Physics}\ }\textbf
  {\bibinfo {volume} {326}},\ \bibinfo {pages} {96 } (\bibinfo {year}
  {2011})},\ \bibinfo {note} {january 2011 Special Issue}\BibitemShut {NoStop}%
\bibitem [{\citenamefont {Hauschild}\ and\ \citenamefont
  {Pollmann}(2018)}]{10.21468/SciPostPhysLectNotes.5}%
  \BibitemOpen
  \bibfield  {author} {\bibinfo {author} {\bibfnamefont {J.}~\bibnamefont
  {Hauschild}}\ and\ \bibinfo {author} {\bibfnamefont {F.}~\bibnamefont
  {Pollmann}},\ }\href {\doibase 10.21468/SciPostPhysLectNotes.5} {\bibfield
  {journal} {\bibinfo  {journal} {SciPost Phys. Lect. Notes}\ ,\ \bibinfo
  {pages} {5}} (\bibinfo {year} {2018})}\BibitemShut {NoStop}%
\bibitem [{\citenamefont {White}(1992)}]{PhysRevLett.69.2863}%
  \BibitemOpen
  \bibfield  {author} {\bibinfo {author} {\bibfnamefont {S.~R.}\ \bibnamefont
  {White}},\ }\href {\doibase 10.1103/PhysRevLett.69.2863} {\bibfield
  {journal} {\bibinfo  {journal} {Phys. Rev. Lett.}\ }\textbf {\bibinfo
  {volume} {69}},\ \bibinfo {pages} {2863} (\bibinfo {year}
  {1992})}\BibitemShut {NoStop}%
\bibitem [{\citenamefont {Duan}\ \emph {et~al.}(2003)\citenamefont {Duan},
  \citenamefont {Demler},\ and\ \citenamefont {Lukin}}]{PhysRevLett.91.090402}%
  \BibitemOpen
  \bibfield  {author} {\bibinfo {author} {\bibfnamefont {L.-M.}\ \bibnamefont
  {Duan}}, \bibinfo {author} {\bibfnamefont {E.}~\bibnamefont {Demler}}, \ and\
  \bibinfo {author} {\bibfnamefont {M.~D.}\ \bibnamefont {Lukin}},\ }\href
  {\doibase 10.1103/PhysRevLett.91.090402} {\bibfield  {journal} {\bibinfo
  {journal} {Phys. Rev. Lett.}\ }\textbf {\bibinfo {volume} {91}},\ \bibinfo
  {pages} {090402} (\bibinfo {year} {2003})}\BibitemShut {NoStop}%
\bibitem [{\citenamefont {Trotzky}\ \emph {et~al.}(2008)\citenamefont
  {Trotzky}, \citenamefont {Cheinet}, \citenamefont {F{\"o}lling},
  \citenamefont {Feld}, \citenamefont {Schnorrberger}, \citenamefont {Rey},
  \citenamefont {Polkovnikov}, \citenamefont {Demler}, \citenamefont {Lukin},\
  and\ \citenamefont {Bloch}}]{Trotzky295}%
  \BibitemOpen
  \bibfield  {author} {\bibinfo {author} {\bibfnamefont {S.}~\bibnamefont
  {Trotzky}}, \bibinfo {author} {\bibfnamefont {P.}~\bibnamefont {Cheinet}},
  \bibinfo {author} {\bibfnamefont {S.}~\bibnamefont {F{\"o}lling}}, \bibinfo
  {author} {\bibfnamefont {M.}~\bibnamefont {Feld}}, \bibinfo {author}
  {\bibfnamefont {U.}~\bibnamefont {Schnorrberger}}, \bibinfo {author}
  {\bibfnamefont {A.~M.}\ \bibnamefont {Rey}}, \bibinfo {author} {\bibfnamefont
  {A.}~\bibnamefont {Polkovnikov}}, \bibinfo {author} {\bibfnamefont {E.~A.}\
  \bibnamefont {Demler}}, \bibinfo {author} {\bibfnamefont {M.~D.}\
  \bibnamefont {Lukin}}, \ and\ \bibinfo {author} {\bibfnamefont
  {I.}~\bibnamefont {Bloch}},\ }\href {\doibase 10.1126/science.1150841}
  {\bibfield  {journal} {\bibinfo  {journal} {Science}\ }\textbf {\bibinfo
  {volume} {319}},\ \bibinfo {pages} {295} (\bibinfo {year}
  {2008})}\BibitemShut {NoStop}%
\bibitem [{\citenamefont {Greif}\ \emph {et~al.}(2013)\citenamefont {Greif},
  \citenamefont {Uehlinger}, \citenamefont {Jotzu}, \citenamefont {Tarruell},\
  and\ \citenamefont {Esslinger}}]{Greif1307}%
  \BibitemOpen
  \bibfield  {author} {\bibinfo {author} {\bibfnamefont {D.}~\bibnamefont
  {Greif}}, \bibinfo {author} {\bibfnamefont {T.}~\bibnamefont {Uehlinger}},
  \bibinfo {author} {\bibfnamefont {G.}~\bibnamefont {Jotzu}}, \bibinfo
  {author} {\bibfnamefont {L.}~\bibnamefont {Tarruell}}, \ and\ \bibinfo
  {author} {\bibfnamefont {T.}~\bibnamefont {Esslinger}},\ }\href {\doibase
  10.1126/science.1236362} {\bibfield  {journal} {\bibinfo  {journal}
  {Science}\ }\textbf {\bibinfo {volume} {340}},\ \bibinfo {pages} {1307}
  (\bibinfo {year} {2013})}\BibitemShut {NoStop}%
\bibitem [{\citenamefont {Boll}\ \emph {et~al.}(2016)\citenamefont {Boll},
  \citenamefont {Hilker}, \citenamefont {Salomon}, \citenamefont {Omran},
  \citenamefont {Nespolo}, \citenamefont {Pollet}, \citenamefont {Bloch},\ and\
  \citenamefont {Gross}}]{Boll1257}%
  \BibitemOpen
  \bibfield  {author} {\bibinfo {author} {\bibfnamefont {M.}~\bibnamefont
  {Boll}}, \bibinfo {author} {\bibfnamefont {T.~A.}\ \bibnamefont {Hilker}},
  \bibinfo {author} {\bibfnamefont {G.}~\bibnamefont {Salomon}}, \bibinfo
  {author} {\bibfnamefont {A.}~\bibnamefont {Omran}}, \bibinfo {author}
  {\bibfnamefont {J.}~\bibnamefont {Nespolo}}, \bibinfo {author} {\bibfnamefont
  {L.}~\bibnamefont {Pollet}}, \bibinfo {author} {\bibfnamefont
  {I.}~\bibnamefont {Bloch}}, \ and\ \bibinfo {author} {\bibfnamefont
  {C.}~\bibnamefont {Gross}},\ }\href {\doibase 10.1126/science.aag1635}
  {\bibfield  {journal} {\bibinfo  {journal} {Science}\ }\textbf {\bibinfo
  {volume} {353}},\ \bibinfo {pages} {1257} (\bibinfo {year}
  {2016})}\BibitemShut {NoStop}%
\bibitem [{\citenamefont {Mazurenko}\ \emph {et~al.}(2017)\citenamefont
  {Mazurenko}, \citenamefont {Chiu}, \citenamefont {Ji}, \citenamefont
  {Parsons}, \citenamefont {Kan{\'a}sz-Nagy}, \citenamefont {Schmidt},
  \citenamefont {Grusdt}, \citenamefont {Demler}, \citenamefont {Greif},\ and\
  \citenamefont {Greiner}}]{mazurenko_cold-atom_2017}%
  \BibitemOpen
  \bibfield  {author} {\bibinfo {author} {\bibfnamefont {A.}~\bibnamefont
  {Mazurenko}}, \bibinfo {author} {\bibfnamefont {C.~S.}\ \bibnamefont {Chiu}},
  \bibinfo {author} {\bibfnamefont {G.}~\bibnamefont {Ji}}, \bibinfo {author}
  {\bibfnamefont {M.~F.}\ \bibnamefont {Parsons}}, \bibinfo {author}
  {\bibfnamefont {M.}~\bibnamefont {Kan{\'a}sz-Nagy}}, \bibinfo {author}
  {\bibfnamefont {R.}~\bibnamefont {Schmidt}}, \bibinfo {author} {\bibfnamefont
  {F.}~\bibnamefont {Grusdt}}, \bibinfo {author} {\bibfnamefont
  {E.}~\bibnamefont {Demler}}, \bibinfo {author} {\bibfnamefont
  {D.}~\bibnamefont {Greif}}, \ and\ \bibinfo {author} {\bibfnamefont
  {M.}~\bibnamefont {Greiner}},\ }\href {\doibase 10.1038/nature22362}
  {\bibfield  {journal} {\bibinfo  {journal} {Nature}\ }\textbf {\bibinfo
  {volume} {545}},\ \bibinfo {pages} {462} (\bibinfo {year}
  {2017})}\BibitemShut {NoStop}%
\bibitem [{\citenamefont {Dimitrova}\ \emph {et~al.}(2020)\citenamefont
  {Dimitrova}, \citenamefont {Jepsen}, \citenamefont {Buyskikh}, \citenamefont
  {Venegas-Gomez}, \citenamefont {Amato-Grill}, \citenamefont {Daley},\ and\
  \citenamefont {Ketterle}}]{PhysRevLett.124.043204}%
  \BibitemOpen
  \bibfield  {author} {\bibinfo {author} {\bibfnamefont {I.}~\bibnamefont
  {Dimitrova}}, \bibinfo {author} {\bibfnamefont {N.}~\bibnamefont {Jepsen}},
  \bibinfo {author} {\bibfnamefont {A.}~\bibnamefont {Buyskikh}}, \bibinfo
  {author} {\bibfnamefont {A.}~\bibnamefont {Venegas-Gomez}}, \bibinfo {author}
  {\bibfnamefont {J.}~\bibnamefont {Amato-Grill}}, \bibinfo {author}
  {\bibfnamefont {A.}~\bibnamefont {Daley}}, \ and\ \bibinfo {author}
  {\bibfnamefont {W.}~\bibnamefont {Ketterle}},\ }\href {\doibase
  10.1103/PhysRevLett.124.043204} {\bibfield  {journal} {\bibinfo  {journal}
  {Phys. Rev. Lett.}\ }\textbf {\bibinfo {volume} {124}},\ \bibinfo {pages}
  {043204} (\bibinfo {year} {2020})}\BibitemShut {NoStop}%
\bibitem [{\citenamefont {Sebby-Strabley}\ \emph {et~al.}(2006)\citenamefont
  {Sebby-Strabley}, \citenamefont {Anderlini}, \citenamefont {Jessen},\ and\
  \citenamefont {Porto}}]{sebby-strabley_lattice_2006}%
  \BibitemOpen
  \bibfield  {author} {\bibinfo {author} {\bibfnamefont {J.}~\bibnamefont
  {Sebby-Strabley}}, \bibinfo {author} {\bibfnamefont {M.}~\bibnamefont
  {Anderlini}}, \bibinfo {author} {\bibfnamefont {P.~S.}\ \bibnamefont
  {Jessen}}, \ and\ \bibinfo {author} {\bibfnamefont {J.~V.}\ \bibnamefont
  {Porto}},\ }\href {\doibase 10.1103/PhysRevA.73.033605} {\bibfield  {journal}
  {\bibinfo  {journal} {Phys. Rev. A}\ }\textbf {\bibinfo {volume} {73}},\
  \bibinfo {pages} {033605} (\bibinfo {year} {2006})}\BibitemShut {NoStop}%
\bibitem [{\citenamefont {F{\"o}lling}\ \emph {et~al.}(2007)\citenamefont
  {F{\"o}lling}, \citenamefont {Trotzky}, \citenamefont {Cheinet},
  \citenamefont {Feld}, \citenamefont {Saers}, \citenamefont {Widera},
  \citenamefont {M{\"u}ller},\ and\ \citenamefont
  {Bloch}}]{folling_direct_2007}%
  \BibitemOpen
  \bibfield  {author} {\bibinfo {author} {\bibfnamefont {S.}~\bibnamefont
  {F{\"o}lling}}, \bibinfo {author} {\bibfnamefont {S.}~\bibnamefont
  {Trotzky}}, \bibinfo {author} {\bibfnamefont {P.}~\bibnamefont {Cheinet}},
  \bibinfo {author} {\bibfnamefont {M.}~\bibnamefont {Feld}}, \bibinfo {author}
  {\bibfnamefont {R.}~\bibnamefont {Saers}}, \bibinfo {author} {\bibfnamefont
  {A.}~\bibnamefont {Widera}}, \bibinfo {author} {\bibfnamefont
  {T.}~\bibnamefont {M{\"u}ller}}, \ and\ \bibinfo {author} {\bibfnamefont
  {I.}~\bibnamefont {Bloch}},\ }\href {\doibase 10.1038/nature06112} {\bibfield
   {journal} {\bibinfo  {journal} {Nature}\ }\textbf {\bibinfo {volume}
  {448}},\ \bibinfo {pages} {1029} (\bibinfo {year} {2007})}\BibitemShut
  {NoStop}%
\bibitem [{\citenamefont {Trotzky}\ \emph {et~al.}(2012)\citenamefont
  {Trotzky}, \citenamefont {Chen}, \citenamefont {Flesch}, \citenamefont
  {McCulloch}, \citenamefont {Schollw{\"o}ck}, \citenamefont {Eisert},\ and\
  \citenamefont {Bloch}}]{trotzky_probing_2012}%
  \BibitemOpen
  \bibfield  {author} {\bibinfo {author} {\bibfnamefont {S.}~\bibnamefont
  {Trotzky}}, \bibinfo {author} {\bibfnamefont {Y.-A.}\ \bibnamefont {Chen}},
  \bibinfo {author} {\bibfnamefont {A.}~\bibnamefont {Flesch}}, \bibinfo
  {author} {\bibfnamefont {I.~P.}\ \bibnamefont {McCulloch}}, \bibinfo {author}
  {\bibfnamefont {U.}~\bibnamefont {Schollw{\"o}ck}}, \bibinfo {author}
  {\bibfnamefont {J.}~\bibnamefont {Eisert}}, \ and\ \bibinfo {author}
  {\bibfnamefont {I.}~\bibnamefont {Bloch}},\ }\href {\doibase
  10.1038/nphys2232} {\bibfield  {journal} {\bibinfo  {journal} {Nature Phys}\
  }\textbf {\bibinfo {volume} {8}},\ \bibinfo {pages} {325} (\bibinfo {year}
  {2012})}\BibitemShut {NoStop}%
\bibitem [{\citenamefont {F{\"o}lling}\ \emph {et~al.}(2005)\citenamefont
  {F{\"o}lling}, \citenamefont {Gerbier}, \citenamefont {Widera}, \citenamefont
  {Mandel}, \citenamefont {Gericke},\ and\ \citenamefont
  {Bloch}}]{folling_spatial_2005}%
  \BibitemOpen
  \bibfield  {author} {\bibinfo {author} {\bibfnamefont {S.}~\bibnamefont
  {F{\"o}lling}}, \bibinfo {author} {\bibfnamefont {F.}~\bibnamefont
  {Gerbier}}, \bibinfo {author} {\bibfnamefont {A.}~\bibnamefont {Widera}},
  \bibinfo {author} {\bibfnamefont {O.}~\bibnamefont {Mandel}}, \bibinfo
  {author} {\bibfnamefont {T.}~\bibnamefont {Gericke}}, \ and\ \bibinfo
  {author} {\bibfnamefont {I.}~\bibnamefont {Bloch}},\ }\href {\doibase
  10.1038/nature03500} {\bibfield  {journal} {\bibinfo  {journal} {Nature}\
  }\textbf {\bibinfo {volume} {434}},\ \bibinfo {pages} {481} (\bibinfo {year}
  {2005})}\BibitemShut {NoStop}%
\bibitem [{\citenamefont {Rom}\ \emph {et~al.}(2006)\citenamefont {Rom},
  \citenamefont {Best}, \citenamefont {van Oosten}, \citenamefont {Schneider},
  \citenamefont {F{\"o}lling}, \citenamefont {Paredes},\ and\ \citenamefont
  {Bloch}}]{rom_free_2006}%
  \BibitemOpen
  \bibfield  {author} {\bibinfo {author} {\bibfnamefont {T.}~\bibnamefont
  {Rom}}, \bibinfo {author} {\bibfnamefont {T.}~\bibnamefont {Best}}, \bibinfo
  {author} {\bibfnamefont {D.}~\bibnamefont {van Oosten}}, \bibinfo {author}
  {\bibfnamefont {U.}~\bibnamefont {Schneider}}, \bibinfo {author}
  {\bibfnamefont {S.}~\bibnamefont {F{\"o}lling}}, \bibinfo {author}
  {\bibfnamefont {B.}~\bibnamefont {Paredes}}, \ and\ \bibinfo {author}
  {\bibfnamefont {I.}~\bibnamefont {Bloch}},\ }\href {\doibase
  10.1038/nature05319} {\bibfield  {journal} {\bibinfo  {journal} {Nature}\
  }\textbf {\bibinfo {volume} {444}},\ \bibinfo {pages} {733} (\bibinfo {year}
  {2006})}\BibitemShut {NoStop}%
\bibitem [{\citenamefont {Trotzky}\ \emph {et~al.}(2010)\citenamefont
  {Trotzky}, \citenamefont {Chen}, \citenamefont {Schnorrberger}, \citenamefont
  {Cheinet},\ and\ \citenamefont {Bloch}}]{trotzky_controlling_2010}%
  \BibitemOpen
  \bibfield  {author} {\bibinfo {author} {\bibfnamefont {S.}~\bibnamefont
  {Trotzky}}, \bibinfo {author} {\bibfnamefont {Y.-A.}\ \bibnamefont {Chen}},
  \bibinfo {author} {\bibfnamefont {U.}~\bibnamefont {Schnorrberger}}, \bibinfo
  {author} {\bibfnamefont {P.}~\bibnamefont {Cheinet}}, \ and\ \bibinfo
  {author} {\bibfnamefont {I.}~\bibnamefont {Bloch}},\ }\href {\doibase
  10.1103/PhysRevLett.105.265303} {\bibfield  {journal} {\bibinfo  {journal}
  {Phys. Rev. Lett.}\ }\textbf {\bibinfo {volume} {105}},\ \bibinfo {pages}
  {265303} (\bibinfo {year} {2010})}\BibitemShut {NoStop}%
\bibitem [{\citenamefont {Bakr}\ \emph {et~al.}(2010)\citenamefont {Bakr},
  \citenamefont {Peng}, \citenamefont {Tai}, \citenamefont {Ma}, \citenamefont
  {Simon}, \citenamefont {Gillen}, \citenamefont {F{\"o}lling}, \citenamefont
  {Pollet},\ and\ \citenamefont {Greiner}}]{bakr_probing_2010}%
  \BibitemOpen
  \bibfield  {author} {\bibinfo {author} {\bibfnamefont {W.~S.}\ \bibnamefont
  {Bakr}}, \bibinfo {author} {\bibfnamefont {A.}~\bibnamefont {Peng}}, \bibinfo
  {author} {\bibfnamefont {M.~E.}\ \bibnamefont {Tai}}, \bibinfo {author}
  {\bibfnamefont {R.}~\bibnamefont {Ma}}, \bibinfo {author} {\bibfnamefont
  {J.}~\bibnamefont {Simon}}, \bibinfo {author} {\bibfnamefont {J.~I.}\
  \bibnamefont {Gillen}}, \bibinfo {author} {\bibfnamefont {S.}~\bibnamefont
  {F{\"o}lling}}, \bibinfo {author} {\bibfnamefont {L.}~\bibnamefont {Pollet}},
  \ and\ \bibinfo {author} {\bibfnamefont {M.}~\bibnamefont {Greiner}},\ }\href
  {\doibase 10.1126/science.1192368} {\bibfield  {journal} {\bibinfo  {journal}
  {Science}\ }\textbf {\bibinfo {volume} {329}},\ \bibinfo {pages} {547}
  (\bibinfo {year} {2010})},\ \bibinfo {note} {publisher: American Association
  for the Advancement of Science Section: Report}\BibitemShut {NoStop}%
\bibitem [{\citenamefont {Sherson}\ \emph {et~al.}(2010)\citenamefont
  {Sherson}, \citenamefont {Weitenberg}, \citenamefont {Endres}, \citenamefont
  {Cheneau}, \citenamefont {Bloch},\ and\ \citenamefont
  {Kuhr}}]{sherson_single-atom-resolved_2010}%
  \BibitemOpen
  \bibfield  {author} {\bibinfo {author} {\bibfnamefont {J.~F.}\ \bibnamefont
  {Sherson}}, \bibinfo {author} {\bibfnamefont {C.}~\bibnamefont {Weitenberg}},
  \bibinfo {author} {\bibfnamefont {M.}~\bibnamefont {Endres}}, \bibinfo
  {author} {\bibfnamefont {M.}~\bibnamefont {Cheneau}}, \bibinfo {author}
  {\bibfnamefont {I.}~\bibnamefont {Bloch}}, \ and\ \bibinfo {author}
  {\bibfnamefont {S.}~\bibnamefont {Kuhr}},\ }\href {\doibase
  10.1038/nature09378} {\bibfield  {journal} {\bibinfo  {journal} {Nature}\
  }\textbf {\bibinfo {volume} {467}},\ \bibinfo {pages} {68} (\bibinfo {year}
  {2010})}\BibitemShut {NoStop}%
\bibitem [{\citenamefont {Koepsell}\ \emph {et~al.}(2020)\citenamefont
  {Koepsell}, \citenamefont {Hirthe}, \citenamefont {Bourgund}, \citenamefont
  {Sompet}, \citenamefont {Vijayan}, \citenamefont {Salomon}, \citenamefont
  {Gross},\ and\ \citenamefont {Bloch}}]{Koepsell_2020}%
  \BibitemOpen
  \bibfield  {author} {\bibinfo {author} {\bibfnamefont {J.}~\bibnamefont
  {Koepsell}}, \bibinfo {author} {\bibfnamefont {S.}~\bibnamefont {Hirthe}},
  \bibinfo {author} {\bibfnamefont {D.}~\bibnamefont {Bourgund}}, \bibinfo
  {author} {\bibfnamefont {P.}~\bibnamefont {Sompet}}, \bibinfo {author}
  {\bibfnamefont {J.}~\bibnamefont {Vijayan}}, \bibinfo {author} {\bibfnamefont
  {G.}~\bibnamefont {Salomon}}, \bibinfo {author} {\bibfnamefont
  {C.}~\bibnamefont {Gross}}, \ and\ \bibinfo {author} {\bibfnamefont
  {I.}~\bibnamefont {Bloch}},\ }\href {\doibase 10.1103/PhysRevLett.125.010403}
  {\bibfield  {journal} {\bibinfo  {journal} {Phys. Rev. Lett.}\ }\textbf
  {\bibinfo {volume} {125}},\ \bibinfo {pages} {010403} (\bibinfo {year}
  {2020})}\BibitemShut {NoStop}%
\bibitem [{\citenamefont {{Hartke}}\ \emph {et~al.}(2020)\citenamefont
  {{Hartke}}, \citenamefont {{Oreg}}, \citenamefont {{Jia}},\ and\
  \citenamefont {{Zwierlein}}}]{Hartke_2020}%
  \BibitemOpen
  \bibfield  {author} {\bibinfo {author} {\bibfnamefont {T.}~\bibnamefont
  {{Hartke}}}, \bibinfo {author} {\bibfnamefont {B.}~\bibnamefont {{Oreg}}},
  \bibinfo {author} {\bibfnamefont {N.}~\bibnamefont {{Jia}}}, \ and\ \bibinfo
  {author} {\bibfnamefont {M.}~\bibnamefont {{Zwierlein}}},\ }\href@noop {}
  {\bibfield  {journal} {\bibinfo  {journal} {arXiv e-prints}\ ,\ \bibinfo
  {eid} {arXiv:2003.11669}} (\bibinfo {year} {2020})},\ \Eprint
  {http://arxiv.org/abs/2003.11669} {arXiv:2003.11669 [cond-mat.quant-gas]}
  \BibitemShut {NoStop}%
\bibitem [{\citenamefont {Yang}\ \emph
  {et~al.}(2020{\natexlab{b}})\citenamefont {Yang}, \citenamefont {Sun},
  \citenamefont {Huang}, \citenamefont {Wang}, \citenamefont {Deng},
  \citenamefont {Dai}, \citenamefont {Yuan},\ and\ \citenamefont
  {Pan}}]{yang_cooling_2020}%
  \BibitemOpen
  \bibfield  {author} {\bibinfo {author} {\bibfnamefont {B.}~\bibnamefont
  {Yang}}, \bibinfo {author} {\bibfnamefont {H.}~\bibnamefont {Sun}}, \bibinfo
  {author} {\bibfnamefont {C.-J.}\ \bibnamefont {Huang}}, \bibinfo {author}
  {\bibfnamefont {H.-Y.}\ \bibnamefont {Wang}}, \bibinfo {author}
  {\bibfnamefont {Y.}~\bibnamefont {Deng}}, \bibinfo {author} {\bibfnamefont
  {H.-N.}\ \bibnamefont {Dai}}, \bibinfo {author} {\bibfnamefont {Z.-S.}\
  \bibnamefont {Yuan}}, \ and\ \bibinfo {author} {\bibfnamefont {J.-W.}\
  \bibnamefont {Pan}},\ }\href {\doibase 10.1126/science.aaz6801} {\bibfield
  {journal} {\bibinfo  {journal} {Science}\ } (\bibinfo {year}
  {2020}{\natexlab{b}}),\ 10.1126/science.aaz6801}\BibitemShut {NoStop}%
\bibitem [{\citenamefont {Coleman}\ and\ \citenamefont
  {Witten}(1980)}]{Coleman_1980}%
  \BibitemOpen
  \bibfield  {author} {\bibinfo {author} {\bibfnamefont {S.}~\bibnamefont
  {Coleman}}\ and\ \bibinfo {author} {\bibfnamefont {E.}~\bibnamefont
  {Witten}},\ }\href {\doibase 10.1103/PhysRevLett.45.100} {\bibfield
  {journal} {\bibinfo  {journal} {Phys. Rev. Lett.}\ }\textbf {\bibinfo
  {volume} {45}},\ \bibinfo {pages} {100} (\bibinfo {year} {1980})}\BibitemShut
  {NoStop}%
\bibitem [{\citenamefont {Wilczek}(1999)}]{wilczek_1999}%
  \BibitemOpen
  \bibfield  {author} {\bibinfo {author} {\bibfnamefont {F.}~\bibnamefont
  {Wilczek}},\ }\href {\doibase 10.1063/1.882879} {\bibfield  {journal}
  {\bibinfo  {journal} {Physics Today}\ }\textbf {\bibinfo {volume} {52}},\
  \bibinfo {pages} {11} (\bibinfo {year} {1999})}\BibitemShut {NoStop}%
\bibitem [{\citenamefont {Rapp}\ and\ \citenamefont
  {Wambach}(2000)}]{Rapp_2000}%
  \BibitemOpen
  \bibfield  {author} {\bibinfo {author} {\bibfnamefont {R.}~\bibnamefont
  {Rapp}}\ and\ \bibinfo {author} {\bibfnamefont {J.}~\bibnamefont {Wambach}},\
  }\enquote {\bibinfo {title} {Chiral symmetry restoration and dileptons in
  relativistic heavy-ion collisions},}\ in\ \href {\doibase
  10.1007/0-306-47101-9_1} {\emph {\bibinfo {booktitle} {Advances in Nuclear
  Physics}}},\ \bibinfo {editor} {edited by\ \bibinfo {editor} {\bibfnamefont
  {J.~W.}\ \bibnamefont {Negele}}\ and\ \bibinfo {editor} {\bibfnamefont
  {E.}~\bibnamefont {Vogt}}}\ (\bibinfo  {publisher} {Springer US},\ \bibinfo
  {address} {Boston, MA},\ \bibinfo {year} {2000})\ pp.\ \bibinfo {pages}
  {1--205}\BibitemShut {NoStop}%
\bibitem [{\citenamefont {Nambu}\ and\ \citenamefont
  {Jona-Lasinio}(1961)}]{Nambu_1961}%
  \BibitemOpen
  \bibfield  {author} {\bibinfo {author} {\bibfnamefont {Y.}~\bibnamefont
  {Nambu}}\ and\ \bibinfo {author} {\bibfnamefont {G.}~\bibnamefont
  {Jona-Lasinio}},\ }\href {\doibase 10.1103/PhysRev.122.345} {\bibfield
  {journal} {\bibinfo  {journal} {Phys. Rev.}\ }\textbf {\bibinfo {volume}
  {122}},\ \bibinfo {pages} {345} (\bibinfo {year} {1961})}\BibitemShut
  {NoStop}%
\bibitem [{\citenamefont {Brambilla}\ \emph {et~al.}(2014)\citenamefont
  {Brambilla}, \citenamefont {Eidelman}, \citenamefont {Foka}, \citenamefont
  {Gardner}, \citenamefont {Kronfeld}, \citenamefont {Alford}, \citenamefont
  {Alkofer}, \citenamefont {Butenschoen}, \citenamefont {Cohen}, \citenamefont
  {Erdmenger}, \citenamefont {Fabbietti}, \citenamefont {Faber}, \citenamefont
  {Goity}, \citenamefont {Ketzer}, \citenamefont {Lin}, \citenamefont
  {Llanes-Estrada}, \citenamefont {Meyer}, \citenamefont {Pakhlov},
  \citenamefont {Pallante}, \citenamefont {Polikarpov}, \citenamefont
  {Sazdjian}, \citenamefont {Schmitt}, \citenamefont {Snow}, \citenamefont
  {Vairo}, \citenamefont {Vogt}, \citenamefont {Vuorinen}, \citenamefont
  {Wittig}, \citenamefont {Arnold}, \citenamefont {Christakoglou},
  \citenamefont {Di~Nezza}, \citenamefont {Fodor}, \citenamefont {Garcia~i
  Tormo}, \citenamefont {H{\"o}llwieser}, \citenamefont {Janik}, \citenamefont
  {Kalweit}, \citenamefont {Keane}, \citenamefont {Kiritsis}, \citenamefont
  {Mischke}, \citenamefont {Mizuk}, \citenamefont {Odyniec}, \citenamefont
  {Papadodimas}, \citenamefont {Pich}, \citenamefont {Pittau}, \citenamefont
  {Qiu}, \citenamefont {Ricciardi}, \citenamefont {Salgado}, \citenamefont
  {Schwenzer}, \citenamefont {Stefanis}, \citenamefont {von Hippel},\ and\
  \citenamefont {Zakharov}}]{Brambilla_2014}%
  \BibitemOpen
  \bibfield  {author} {\bibinfo {author} {\bibfnamefont {N.}~\bibnamefont
  {Brambilla}}, \bibinfo {author} {\bibfnamefont {S.}~\bibnamefont {Eidelman}},
  \bibinfo {author} {\bibfnamefont {P.}~\bibnamefont {Foka}}, \bibinfo {author}
  {\bibfnamefont {S.}~\bibnamefont {Gardner}}, \bibinfo {author} {\bibfnamefont
  {A.~S.}\ \bibnamefont {Kronfeld}}, \bibinfo {author} {\bibfnamefont {M.~G.}\
  \bibnamefont {Alford}}, \bibinfo {author} {\bibfnamefont {R.}~\bibnamefont
  {Alkofer}}, \bibinfo {author} {\bibfnamefont {M.}~\bibnamefont
  {Butenschoen}}, \bibinfo {author} {\bibfnamefont {T.~D.}\ \bibnamefont
  {Cohen}}, \bibinfo {author} {\bibfnamefont {J.}~\bibnamefont {Erdmenger}},
  \bibinfo {author} {\bibfnamefont {L.}~\bibnamefont {Fabbietti}}, \bibinfo
  {author} {\bibfnamefont {M.}~\bibnamefont {Faber}}, \bibinfo {author}
  {\bibfnamefont {J.~L.}\ \bibnamefont {Goity}}, \bibinfo {author}
  {\bibfnamefont {B.}~\bibnamefont {Ketzer}}, \bibinfo {author} {\bibfnamefont
  {H.~W.}\ \bibnamefont {Lin}}, \bibinfo {author} {\bibfnamefont {F.~J.}\
  \bibnamefont {Llanes-Estrada}}, \bibinfo {author} {\bibfnamefont {H.~B.}\
  \bibnamefont {Meyer}}, \bibinfo {author} {\bibfnamefont {P.}~\bibnamefont
  {Pakhlov}}, \bibinfo {author} {\bibfnamefont {E.}~\bibnamefont {Pallante}},
  \bibinfo {author} {\bibfnamefont {M.~I.}\ \bibnamefont {Polikarpov}},
  \bibinfo {author} {\bibfnamefont {H.}~\bibnamefont {Sazdjian}}, \bibinfo
  {author} {\bibfnamefont {A.}~\bibnamefont {Schmitt}}, \bibinfo {author}
  {\bibfnamefont {W.~M.}\ \bibnamefont {Snow}}, \bibinfo {author}
  {\bibfnamefont {A.}~\bibnamefont {Vairo}}, \bibinfo {author} {\bibfnamefont
  {R.}~\bibnamefont {Vogt}}, \bibinfo {author} {\bibfnamefont {A.}~\bibnamefont
  {Vuorinen}}, \bibinfo {author} {\bibfnamefont {H.}~\bibnamefont {Wittig}},
  \bibinfo {author} {\bibfnamefont {P.}~\bibnamefont {Arnold}}, \bibinfo
  {author} {\bibfnamefont {P.}~\bibnamefont {Christakoglou}}, \bibinfo {author}
  {\bibfnamefont {P.}~\bibnamefont {Di~Nezza}}, \bibinfo {author}
  {\bibfnamefont {Z.}~\bibnamefont {Fodor}}, \bibinfo {author} {\bibfnamefont
  {X.}~\bibnamefont {Garcia~i Tormo}}, \bibinfo {author} {\bibfnamefont
  {R.}~\bibnamefont {H{\"o}llwieser}}, \bibinfo {author} {\bibfnamefont
  {M.~A.}\ \bibnamefont {Janik}}, \bibinfo {author} {\bibfnamefont
  {A.}~\bibnamefont {Kalweit}}, \bibinfo {author} {\bibfnamefont
  {D.}~\bibnamefont {Keane}}, \bibinfo {author} {\bibfnamefont
  {E.}~\bibnamefont {Kiritsis}}, \bibinfo {author} {\bibfnamefont
  {A.}~\bibnamefont {Mischke}}, \bibinfo {author} {\bibfnamefont
  {R.}~\bibnamefont {Mizuk}}, \bibinfo {author} {\bibfnamefont
  {G.}~\bibnamefont {Odyniec}}, \bibinfo {author} {\bibfnamefont
  {K.}~\bibnamefont {Papadodimas}}, \bibinfo {author} {\bibfnamefont
  {A.}~\bibnamefont {Pich}}, \bibinfo {author} {\bibfnamefont {R.}~\bibnamefont
  {Pittau}}, \bibinfo {author} {\bibfnamefont {J.-W.}\ \bibnamefont {Qiu}},
  \bibinfo {author} {\bibfnamefont {G.}~\bibnamefont {Ricciardi}}, \bibinfo
  {author} {\bibfnamefont {C.~A.}\ \bibnamefont {Salgado}}, \bibinfo {author}
  {\bibfnamefont {K.}~\bibnamefont {Schwenzer}}, \bibinfo {author}
  {\bibfnamefont {N.~G.}\ \bibnamefont {Stefanis}}, \bibinfo {author}
  {\bibfnamefont {G.~M.}\ \bibnamefont {von Hippel}}, \ and\ \bibinfo {author}
  {\bibfnamefont {V.~I.}\ \bibnamefont {Zakharov}},\ }\href {\doibase
  10.1140/epjc/s10052-014-2981-5} {\bibfield  {journal} {\bibinfo  {journal}
  {The European Physical Journal C}\ }\textbf {\bibinfo {volume} {74}},\
  \bibinfo {pages} {2981} (\bibinfo {year} {2014})}\BibitemShut {NoStop}%
\bibitem [{\citenamefont {McLerran}\ and\ \citenamefont
  {Pisarski}(2007)}]{McLerran_2007}%
  \BibitemOpen
  \bibfield  {author} {\bibinfo {author} {\bibfnamefont {L.}~\bibnamefont
  {McLerran}}\ and\ \bibinfo {author} {\bibfnamefont {R.~D.}\ \bibnamefont
  {Pisarski}},\ }\href {\doibase
  https://doi.org/10.1016/j.nuclphysa.2007.08.013} {\bibfield  {journal}
  {\bibinfo  {journal} {Nuclear Physics A}\ }\textbf {\bibinfo {volume}
  {796}},\ \bibinfo {pages} {83 } (\bibinfo {year} {2007})}\BibitemShut
  {NoStop}%
\bibitem [{\citenamefont {Bali}(2001)}]{BALI20011}%
  \BibitemOpen
  \bibfield  {author} {\bibinfo {author} {\bibfnamefont {G.~S.}\ \bibnamefont
  {Bali}},\ }\href {\doibase https://doi.org/10.1016/S0370-1573(00)00079-X}
  {\bibfield  {journal} {\bibinfo  {journal} {Physics Reports}\ }\textbf
  {\bibinfo {volume} {343}},\ \bibinfo {pages} {1 } (\bibinfo {year}
  {2001})}\BibitemShut {NoStop}%
\bibitem [{\citenamefont {Coleman}\ \emph {et~al.}(1975)\citenamefont
  {Coleman}, \citenamefont {Jackiw},\ and\ \citenamefont
  {Susskind}}]{COLEMAN1975267}%
  \BibitemOpen
  \bibfield  {author} {\bibinfo {author} {\bibfnamefont {S.}~\bibnamefont
  {Coleman}}, \bibinfo {author} {\bibfnamefont {R.}~\bibnamefont {Jackiw}}, \
  and\ \bibinfo {author} {\bibfnamefont {L.}~\bibnamefont {Susskind}},\ }\href
  {\doibase https://doi.org/10.1016/0003-4916(75)90212-2} {\bibfield  {journal}
  {\bibinfo  {journal} {Annals of Physics}\ }\textbf {\bibinfo {volume} {93}},\
  \bibinfo {pages} {267 } (\bibinfo {year} {1975})}\BibitemShut {NoStop}%
\bibitem [{\citenamefont {Buyens}\ \emph {et~al.}(2016)\citenamefont {Buyens},
  \citenamefont {Haegeman}, \citenamefont {Verschelde}, \citenamefont
  {Verstraete},\ and\ \citenamefont {Van~Acoleyen}}]{PhysRevX.6.041040}%
  \BibitemOpen
  \bibfield  {author} {\bibinfo {author} {\bibfnamefont {B.}~\bibnamefont
  {Buyens}}, \bibinfo {author} {\bibfnamefont {J.}~\bibnamefont {Haegeman}},
  \bibinfo {author} {\bibfnamefont {H.}~\bibnamefont {Verschelde}}, \bibinfo
  {author} {\bibfnamefont {F.}~\bibnamefont {Verstraete}}, \ and\ \bibinfo
  {author} {\bibfnamefont {K.}~\bibnamefont {Van~Acoleyen}},\ }\href {\doibase
  10.1103/PhysRevX.6.041040} {\bibfield  {journal} {\bibinfo  {journal} {Phys.
  Rev. X}\ }\textbf {\bibinfo {volume} {6}},\ \bibinfo {pages} {041040}
  (\bibinfo {year} {2016})}\BibitemShut {NoStop}%
\bibitem [{\citenamefont {Chodos}\ \emph {et~al.}(1974)\citenamefont {Chodos},
  \citenamefont {Jaffe}, \citenamefont {Johnson}, \citenamefont {Thorn},\ and\
  \citenamefont {Weisskopf}}]{PhysRevD.9.3471}%
  \BibitemOpen
  \bibfield  {author} {\bibinfo {author} {\bibfnamefont {A.}~\bibnamefont
  {Chodos}}, \bibinfo {author} {\bibfnamefont {R.~L.}\ \bibnamefont {Jaffe}},
  \bibinfo {author} {\bibfnamefont {K.}~\bibnamefont {Johnson}}, \bibinfo
  {author} {\bibfnamefont {C.~B.}\ \bibnamefont {Thorn}}, \ and\ \bibinfo
  {author} {\bibfnamefont {V.~F.}\ \bibnamefont {Weisskopf}},\ }\href {\doibase
  10.1103/PhysRevD.9.3471} {\bibfield  {journal} {\bibinfo  {journal} {Phys.
  Rev. D}\ }\textbf {\bibinfo {volume} {9}},\ \bibinfo {pages} {3471} (\bibinfo
  {year} {1974})}\BibitemShut {NoStop}%
\bibitem [{\citenamefont {Bardeen}\ \emph {et~al.}(1975)\citenamefont
  {Bardeen}, \citenamefont {Chanowitz}, \citenamefont {Drell}, \citenamefont
  {Weinstein},\ and\ \citenamefont {Yan}}]{PhysRevD.11.1094}%
  \BibitemOpen
  \bibfield  {author} {\bibinfo {author} {\bibfnamefont {W.~A.}\ \bibnamefont
  {Bardeen}}, \bibinfo {author} {\bibfnamefont {M.~S.}\ \bibnamefont
  {Chanowitz}}, \bibinfo {author} {\bibfnamefont {S.~D.}\ \bibnamefont
  {Drell}}, \bibinfo {author} {\bibfnamefont {M.}~\bibnamefont {Weinstein}}, \
  and\ \bibinfo {author} {\bibfnamefont {T.~M.}\ \bibnamefont {Yan}},\ }\href
  {\doibase 10.1103/PhysRevD.11.1094} {\bibfield  {journal} {\bibinfo
  {journal} {Phys. Rev. D}\ }\textbf {\bibinfo {volume} {11}},\ \bibinfo
  {pages} {1094} (\bibinfo {year} {1975})}\BibitemShut {NoStop}%
\bibitem [{\citenamefont {Friedberg}\ and\ \citenamefont
  {Lee}(1977{\natexlab{a}})}]{PhysRevD.15.1694}%
  \BibitemOpen
  \bibfield  {author} {\bibinfo {author} {\bibfnamefont {R.}~\bibnamefont
  {Friedberg}}\ and\ \bibinfo {author} {\bibfnamefont {T.~D.}\ \bibnamefont
  {Lee}},\ }\href {\doibase 10.1103/PhysRevD.15.1694} {\bibfield  {journal}
  {\bibinfo  {journal} {Phys. Rev. D}\ }\textbf {\bibinfo {volume} {15}},\
  \bibinfo {pages} {1694} (\bibinfo {year} {1977}{\natexlab{a}})}\BibitemShut
  {NoStop}%
\bibitem [{\citenamefont {Friedberg}\ and\ \citenamefont
  {Lee}(1977{\natexlab{b}})}]{PhysRevD.16.1096}%
  \BibitemOpen
  \bibfield  {author} {\bibinfo {author} {\bibfnamefont {R.}~\bibnamefont
  {Friedberg}}\ and\ \bibinfo {author} {\bibfnamefont {T.~D.}\ \bibnamefont
  {Lee}},\ }\href {\doibase 10.1103/PhysRevD.16.1096} {\bibfield  {journal}
  {\bibinfo  {journal} {Phys. Rev. D}\ }\textbf {\bibinfo {volume} {16}},\
  \bibinfo {pages} {1096} (\bibinfo {year} {1977}{\natexlab{b}})}\BibitemShut
  {NoStop}%
\bibitem [{\citenamefont {Su}\ \emph {et~al.}(1979)\citenamefont {Su},
  \citenamefont {Schrieffer},\ and\ \citenamefont
  {Heeger}}]{PhysRevLett.42.1698}%
  \BibitemOpen
  \bibfield  {author} {\bibinfo {author} {\bibfnamefont {W.~P.}\ \bibnamefont
  {Su}}, \bibinfo {author} {\bibfnamefont {J.~R.}\ \bibnamefont {Schrieffer}},
  \ and\ \bibinfo {author} {\bibfnamefont {A.~J.}\ \bibnamefont {Heeger}},\
  }\href {\doibase 10.1103/PhysRevLett.42.1698} {\bibfield  {journal} {\bibinfo
   {journal} {Phys. Rev. Lett.}\ }\textbf {\bibinfo {volume} {42}},\ \bibinfo
  {pages} {1698} (\bibinfo {year} {1979})}\BibitemShut {NoStop}%
\bibitem [{\citenamefont {Campbell}\ and\ \citenamefont
  {Bishop}(1982)}]{CAMPBELL1982297}%
  \BibitemOpen
  \bibfield  {author} {\bibinfo {author} {\bibfnamefont {D.}~\bibnamefont
  {Campbell}}\ and\ \bibinfo {author} {\bibfnamefont {A.}~\bibnamefont
  {Bishop}},\ }\href {\doibase https://doi.org/10.1016/0550-3213(82)90089-X}
  {\bibfield  {journal} {\bibinfo  {journal} {Nuclear Physics B}\ }\textbf
  {\bibinfo {volume} {200}},\ \bibinfo {pages} {297 } (\bibinfo {year}
  {1982})}\BibitemShut {NoStop}%
\bibitem [{\citenamefont {Gonz\'alez-Cuadra}\ \emph {et~al.}(2019)\citenamefont
  {Gonz\'alez-Cuadra}, \citenamefont {Dauphin}, \citenamefont {Grzybowski},
  \citenamefont {W\'ojcik}, \citenamefont {Lewenstein},\ and\ \citenamefont
  {Bermudez}}]{PhysRevB.99.045139}%
  \BibitemOpen
  \bibfield  {author} {\bibinfo {author} {\bibfnamefont {D.}~\bibnamefont
  {Gonz\'alez-Cuadra}}, \bibinfo {author} {\bibfnamefont {A.}~\bibnamefont
  {Dauphin}}, \bibinfo {author} {\bibfnamefont {P.~R.}\ \bibnamefont
  {Grzybowski}}, \bibinfo {author} {\bibfnamefont {P.}~\bibnamefont
  {W\'ojcik}}, \bibinfo {author} {\bibfnamefont {M.}~\bibnamefont
  {Lewenstein}}, \ and\ \bibinfo {author} {\bibfnamefont {A.}~\bibnamefont
  {Bermudez}},\ }\href {\doibase 10.1103/PhysRevB.99.045139} {\bibfield
  {journal} {\bibinfo  {journal} {Phys. Rev. B}\ }\textbf {\bibinfo {volume}
  {99}},\ \bibinfo {pages} {045139} (\bibinfo {year} {2019})}\BibitemShut
  {NoStop}%
\bibitem [{\citenamefont {Gonz{\'a}lez-Cuadra}\ \emph
  {et~al.}(2019)\citenamefont {Gonz{\'a}lez-Cuadra}, \citenamefont {Bermudez},
  \citenamefont {Grzybowski}, \citenamefont {Lewenstein},\ and\ \citenamefont
  {Dauphin}}]{Gonzalez-Cuadra2019}%
  \BibitemOpen
  \bibfield  {author} {\bibinfo {author} {\bibfnamefont {D.}~\bibnamefont
  {Gonz{\'a}lez-Cuadra}}, \bibinfo {author} {\bibfnamefont {A.}~\bibnamefont
  {Bermudez}}, \bibinfo {author} {\bibfnamefont {P.~R.}\ \bibnamefont
  {Grzybowski}}, \bibinfo {author} {\bibfnamefont {M.}~\bibnamefont
  {Lewenstein}}, \ and\ \bibinfo {author} {\bibfnamefont {A.}~\bibnamefont
  {Dauphin}},\ }\href {\doibase 10.1038/s41467-019-10796-8} {\bibfield
  {journal} {\bibinfo  {journal} {Nature Communications}\ }\textbf {\bibinfo
  {volume} {10}},\ \bibinfo {pages} {2694} (\bibinfo {year}
  {2019})}\BibitemShut {NoStop}%
\bibitem [{\citenamefont {Gonz\'alez-Cuadra}\ \emph {et~al.}(2019)\citenamefont
  {Gonz\'alez-Cuadra}, \citenamefont {Dauphin}, \citenamefont {Grzybowski},
  \citenamefont {Lewenstein},\ and\ \citenamefont {Bermudez}}]{1908.02186}%
  \BibitemOpen
  \bibfield  {author} {\bibinfo {author} {\bibfnamefont {D.}~\bibnamefont
  {Gonz\'alez-Cuadra}}, \bibinfo {author} {\bibfnamefont {A.}~\bibnamefont
  {Dauphin}}, \bibinfo {author} {\bibfnamefont {P.~R.}\ \bibnamefont
  {Grzybowski}}, \bibinfo {author} {\bibfnamefont {M.}~\bibnamefont
  {Lewenstein}}, \ and\ \bibinfo {author} {\bibfnamefont {A.}~\bibnamefont
  {Bermudez}},\ }\href@noop {} {\enquote {\bibinfo {title} {$\mathbb{Z}_n$
  solitons in intertwined topological phases},}\ } (\bibinfo {year} {2019}),\
  \Eprint {http://arxiv.org/abs/arXiv:1908.02186} {arXiv:1908.02186}
  \BibitemShut {NoStop}%
\bibitem [{\citenamefont {Gonz\'alez-Cuadra}\ \emph
  {et~al.}(2020{\natexlab{b}})\citenamefont {Gonz\'alez-Cuadra}, \citenamefont
  {Dauphin}, \citenamefont {Grzybowski}, \citenamefont {Lewenstein},\ and\
  \citenamefont {Bermudez}}]{2003.10994}%
  \BibitemOpen
  \bibfield  {author} {\bibinfo {author} {\bibfnamefont {D.}~\bibnamefont
  {Gonz\'alez-Cuadra}}, \bibinfo {author} {\bibfnamefont {A.}~\bibnamefont
  {Dauphin}}, \bibinfo {author} {\bibfnamefont {P.~R.}\ \bibnamefont
  {Grzybowski}}, \bibinfo {author} {\bibfnamefont {M.}~\bibnamefont
  {Lewenstein}}, \ and\ \bibinfo {author} {\bibfnamefont {A.}~\bibnamefont
  {Bermudez}},\ }\href@noop {} {\enquote {\bibinfo {title} {Dynamical solitons
  and boson fractionalization in cold-atom topological insulators},}\ }
  (\bibinfo {year} {2020}{\natexlab{b}}),\ \Eprint
  {http://arxiv.org/abs/arXiv:2003.10994} {arXiv:2003.10994} \BibitemShut
  {NoStop}%
\bibitem [{\citenamefont {Bermudez}\ \emph {et~al.}(2018)\citenamefont
  {Bermudez}, \citenamefont {Tirrito}, \citenamefont {Rizzi}, \citenamefont
  {Lewenstein},\ and\ \citenamefont {Hands}}]{BERMUDEZ2018149}%
  \BibitemOpen
  \bibfield  {author} {\bibinfo {author} {\bibfnamefont {A.}~\bibnamefont
  {Bermudez}}, \bibinfo {author} {\bibfnamefont {E.}~\bibnamefont {Tirrito}},
  \bibinfo {author} {\bibfnamefont {M.}~\bibnamefont {Rizzi}}, \bibinfo
  {author} {\bibfnamefont {M.}~\bibnamefont {Lewenstein}}, \ and\ \bibinfo
  {author} {\bibfnamefont {S.}~\bibnamefont {Hands}},\ }\href {\doibase
  https://doi.org/10.1016/j.aop.2018.10.007} {\bibfield  {journal} {\bibinfo
  {journal} {Annals of Physics}\ }\textbf {\bibinfo {volume} {399}},\ \bibinfo
  {pages} {149 } (\bibinfo {year} {2018})}\BibitemShut {NoStop}%
\bibitem [{\citenamefont {Tirrito}\ \emph {et~al.}(2019)\citenamefont
  {Tirrito}, \citenamefont {Rizzi}, \citenamefont {Sierra}, \citenamefont
  {Lewenstein},\ and\ \citenamefont {Bermudez}}]{PhysRevB.99.125106}%
  \BibitemOpen
  \bibfield  {author} {\bibinfo {author} {\bibfnamefont {E.}~\bibnamefont
  {Tirrito}}, \bibinfo {author} {\bibfnamefont {M.}~\bibnamefont {Rizzi}},
  \bibinfo {author} {\bibfnamefont {G.}~\bibnamefont {Sierra}}, \bibinfo
  {author} {\bibfnamefont {M.}~\bibnamefont {Lewenstein}}, \ and\ \bibinfo
  {author} {\bibfnamefont {A.}~\bibnamefont {Bermudez}},\ }\href {\doibase
  10.1103/PhysRevB.99.125106} {\bibfield  {journal} {\bibinfo  {journal} {Phys.
  Rev. B}\ }\textbf {\bibinfo {volume} {99}},\ \bibinfo {pages} {125106}
  (\bibinfo {year} {2019})}\BibitemShut {NoStop}%
\bibitem [{\citenamefont {J\"unemann}\ \emph {et~al.}(2017)\citenamefont
  {J\"unemann}, \citenamefont {Piga}, \citenamefont {Ran}, \citenamefont
  {Lewenstein}, \citenamefont {Rizzi},\ and\ \citenamefont
  {Bermudez}}]{PhysRevX.7.031057}%
  \BibitemOpen
  \bibfield  {author} {\bibinfo {author} {\bibfnamefont {J.}~\bibnamefont
  {J\"unemann}}, \bibinfo {author} {\bibfnamefont {A.}~\bibnamefont {Piga}},
  \bibinfo {author} {\bibfnamefont {S.-J.}\ \bibnamefont {Ran}}, \bibinfo
  {author} {\bibfnamefont {M.}~\bibnamefont {Lewenstein}}, \bibinfo {author}
  {\bibfnamefont {M.}~\bibnamefont {Rizzi}}, \ and\ \bibinfo {author}
  {\bibfnamefont {A.}~\bibnamefont {Bermudez}},\ }\href {\doibase
  10.1103/PhysRevX.7.031057} {\bibfield  {journal} {\bibinfo  {journal} {Phys.
  Rev. X}\ }\textbf {\bibinfo {volume} {7}},\ \bibinfo {pages} {031057}
  (\bibinfo {year} {2017})}\BibitemShut {NoStop}%
\bibitem [{\citenamefont {Magnifico}\ \emph
  {et~al.}(2019{\natexlab{a}})\citenamefont {Magnifico}, \citenamefont
  {Vodola}, \citenamefont {Ercolessi}, \citenamefont {Kumar}, \citenamefont
  {M\"uller},\ and\ \citenamefont {Bermudez}}]{PhysRevD.99.014503}%
  \BibitemOpen
  \bibfield  {author} {\bibinfo {author} {\bibfnamefont {G.}~\bibnamefont
  {Magnifico}}, \bibinfo {author} {\bibfnamefont {D.}~\bibnamefont {Vodola}},
  \bibinfo {author} {\bibfnamefont {E.}~\bibnamefont {Ercolessi}}, \bibinfo
  {author} {\bibfnamefont {S.~P.}\ \bibnamefont {Kumar}}, \bibinfo {author}
  {\bibfnamefont {M.}~\bibnamefont {M\"uller}}, \ and\ \bibinfo {author}
  {\bibfnamefont {A.}~\bibnamefont {Bermudez}},\ }\href {\doibase
  10.1103/PhysRevD.99.014503} {\bibfield  {journal} {\bibinfo  {journal} {Phys.
  Rev. D}\ }\textbf {\bibinfo {volume} {99}},\ \bibinfo {pages} {014503}
  (\bibinfo {year} {2019}{\natexlab{a}})}\BibitemShut {NoStop}%
\bibitem [{\citenamefont {Magnifico}\ \emph
  {et~al.}(2019{\natexlab{b}})\citenamefont {Magnifico}, \citenamefont
  {Vodola}, \citenamefont {Ercolessi}, \citenamefont {Kumar}, \citenamefont
  {M\"uller},\ and\ \citenamefont {Bermudez}}]{PhysRevB.100.115152}%
  \BibitemOpen
  \bibfield  {author} {\bibinfo {author} {\bibfnamefont {G.}~\bibnamefont
  {Magnifico}}, \bibinfo {author} {\bibfnamefont {D.}~\bibnamefont {Vodola}},
  \bibinfo {author} {\bibfnamefont {E.}~\bibnamefont {Ercolessi}}, \bibinfo
  {author} {\bibfnamefont {S.~P.}\ \bibnamefont {Kumar}}, \bibinfo {author}
  {\bibfnamefont {M.}~\bibnamefont {M\"uller}}, \ and\ \bibinfo {author}
  {\bibfnamefont {A.}~\bibnamefont {Bermudez}},\ }\href {\doibase
  10.1103/PhysRevB.100.115152} {\bibfield  {journal} {\bibinfo  {journal}
  {Phys. Rev. B}\ }\textbf {\bibinfo {volume} {100}},\ \bibinfo {pages}
  {115152} (\bibinfo {year} {2019}{\natexlab{b}})}\BibitemShut {NoStop}%
\bibitem [{\citenamefont {Simoni}\ \emph {et~al.}(2008)\citenamefont {Simoni},
  \citenamefont {Zaccanti}, \citenamefont {D{\textquoteright}Errico},
  \citenamefont {Fattori}, \citenamefont {Roati}, \citenamefont {Inguscio},\
  and\ \citenamefont {Modugno}}]{simoni_nearthreshold_2008}%
  \BibitemOpen
  \bibfield  {author} {\bibinfo {author} {\bibfnamefont {A.}~\bibnamefont
  {Simoni}}, \bibinfo {author} {\bibfnamefont {M.}~\bibnamefont {Zaccanti}},
  \bibinfo {author} {\bibfnamefont {C.}~\bibnamefont
  {D{\textquoteright}Errico}}, \bibinfo {author} {\bibfnamefont
  {M.}~\bibnamefont {Fattori}}, \bibinfo {author} {\bibfnamefont
  {G.}~\bibnamefont {Roati}}, \bibinfo {author} {\bibfnamefont
  {M.}~\bibnamefont {Inguscio}}, \ and\ \bibinfo {author} {\bibfnamefont
  {G.}~\bibnamefont {Modugno}},\ }\href {\doibase 10.1103/PhysRevA.77.052705}
  {\bibfield  {journal} {\bibinfo  {journal} {Phys. Rev. A}\ }\textbf {\bibinfo
  {volume} {77}} (\bibinfo {year} {2008}),\
  10.1103/PhysRevA.77.052705}\BibitemShut {NoStop}%
\bibitem [{\citenamefont {Johansen}\ \emph {et~al.}(2017)\citenamefont
  {Johansen}, \citenamefont {DeSalvo}, \citenamefont {Patel},\ and\
  \citenamefont {Chin}}]{johansen_testing_2017}%
  \BibitemOpen
  \bibfield  {author} {\bibinfo {author} {\bibfnamefont {J.}~\bibnamefont
  {Johansen}}, \bibinfo {author} {\bibfnamefont {B.~J.}\ \bibnamefont
  {DeSalvo}}, \bibinfo {author} {\bibfnamefont {K.}~\bibnamefont {Patel}}, \
  and\ \bibinfo {author} {\bibfnamefont {C.}~\bibnamefont {Chin}},\ }\href
  {\doibase 10.1038/nphys4130} {\bibfield  {journal} {\bibinfo  {journal}
  {Nature Physics}\ }\textbf {\bibinfo {volume} {13}},\ \bibinfo {pages} {731}
  (\bibinfo {year} {2017})}\BibitemShut {NoStop}%
\bibitem [{\citenamefont {Kawamoto}\ and\ \citenamefont
  {Smit}(1981)}]{KAWAMOTO1981100}%
  \BibitemOpen
  \bibfield  {author} {\bibinfo {author} {\bibfnamefont {N.}~\bibnamefont
  {Kawamoto}}\ and\ \bibinfo {author} {\bibfnamefont {J.}~\bibnamefont
  {Smit}},\ }\href {\doibase https://doi.org/10.1016/0550-3213(81)90196-6}
  {\bibfield  {journal} {\bibinfo  {journal} {Nuclear Physics B}\ }\textbf
  {\bibinfo {volume} {192}},\ \bibinfo {pages} {100 } (\bibinfo {year}
  {1981})}\BibitemShut {NoStop}%
\end{thebibliography}%

\end{document}